\documentclass[reprint,aps]{revtex4-1}

\usepackage{amsmath}  
\usepackage{amsfonts} 
\usepackage{subfig}
\usepackage{graphicx} 
\usepackage{booktabs}
\usepackage{threeparttable}
\usepackage[font=footnotesize]{caption}

\newcommand{\mbf}[1]{\mathbf{#1}}
\newcommand{\al}{\alpha}

\begin{document}

\author{James Shee}
\email{js4564@columbia.edu}
\affiliation{Department of Chemistry, Columbia University, 3000 Broadway, New York, NY, 10027}
\author{Shiwei Zhang}
\affiliation{Department of Physics, College of William and Mary, Williamsburg, Virginia 23187-8795}
\author{David R. Reichman}
\author{Richard A. Friesner}
\affiliation{Department of Chemistry, Columbia University, 3000 Broadway, New York, NY, 10027}

\title{Chemical Transformations Approaching Chemical Accuracy via Correlated Sampling in Auxiliary-Field Quantum Monte Carlo}


\begin{abstract}
The exact and phaseless variants of Auxiliary-Field Quantum Monte Carlo (AFQMC) have been shown to be capable of producing accurate ground-state energies for a wide variety of systems including those which exhibit substantial electron correlation effects.  The computational cost of performing these calculations has to date been relatively high, impeding many important applications of these approaches.  Here we present a correlated sampling methodology for AFQMC which relies on error cancellation to dramatically accelerate the calculation of energy differences of relevance to chemical transformations.  In particular, we show that our correlated sampling-based AFQMC approach is capable of calculating redox properties, deprotonation free-energies, and hydrogen abstraction energies in an efficient manner without sacrificing accuracy.  We validate the computational protocol by calculating the ionization potentials and electron affinities of the atoms contained in the G2 Test Set, and then proceed to utilize a composite method, which treats fixed-geometry processes with correlated sampling-based AFQMC and relaxation energies via MP2, to compute the ionization potential, deprotonation free-energy, and the O-H bond disocciation energy of methanol, all to within chemical accuracy.  We show that the efficiency of correlated sampling relative to uncorrelated calculations increases with system and basis set size, and that correlated sampling greatly reduces the required number of random walkers to achieve a target statistical error.  This translates to CPU-time speed-up factors of 55, 25, and 24 for the the ionization potential of the K atom, the deprotonation of methanol, and hydrogen abstraction from the O-H bond of methanol, respectively.  We conclude with a discussion of further efficiency improvements that may open the door to the accurate description of chemical processes in complex systems.
\end{abstract}

\maketitle 

\section{Introduction}

	The pursuit of chemical accuracy (as defined by errors not exceeding 1 kcal/mol) in the \emph{ab initio} computation of the energetic properties of \emph{generic} many-electron systems is a long-standing goal that has yet to be reached.\cite{peterson2012chemical,friesner2005ab,langhoff2012quantum,sherrill2010frontiers,harvey2006accuracy}  Traditional methods such as full configuration interaction (FCI)\cite{szabo1989modern} and single-reference coupled cluster with single, double, and perturbative triple excitations (CCSD(T))\cite{purvis1982full,bartlett2007coupled} scale exponentially and with the seventh power of the system size, respectively, and the latter can break down for systems which exhibit sufficiently strong electron correlation.\cite{purwanto2015auxiliary,dutta2003full}  Multi-reference methods such as CASSCF,\cite{roos1987complete,olsen2011casscf,schmidt1998construction} often supplemented with second order perturbation theory,\cite{andersson1990second,andersson1992second} and FCI-QMC\cite{booth2009fermion, booth2011breaking, cleland2011study,cleland2012taming} have been shown to yield benchmark-quality results even for strongly correlated electronic systems, an import class of problems that includes many transition metal (TM)-containing systems.\cite{pierloot2011transition,thomas2015accurate}  However, while a judicious choice of the active space can make such calculations tractable for small systems, exponential scaling prohibits the use of these methods in studying most realistic systems of interest in biology, materials science, and chemical catalysis.  As a result, with the exception of isolated studies using more accurate, specialized methods,\cite{chan2011density,sharma2014low,kurashige2013entangled} these systems can only be investigated feasibly with less accurate but more economical approaches such as Density Functional Theory (DFT).\cite{kohn1996density}  Given the plethora of investigations using this tool to generate hypotheses concerning reaction mechanisms, equilibrium structures, etc., 
the importance of having a systematically improvable \emph{ab initio} method which is both accurate and feasible becomes readily apparent. 

	Among the class of Quantum Monte Carlo (QMC) methods used in electronic structure theory,\cite{lester1990quantum,hammond1994monte,RevModPhys.73.33} Auxiliary-Field QMC, and in particular its ``phaseless" variant (ph-AFQMC)\cite{zhang2013, zhang2003quantum} which controls the sign problem at the cost of introducing a bias which in principle can be systematically reduced, has produced state-of-the-art, benchmark-quality results for systems such as first- and second- row $d$ elements,\cite{al2006d} the chromium dimer,\cite{purwanto2015auxiliary} cobalt adatoms on graphene,\cite{virgus2014stability} and various transition metal oxides,\cite{al2006auxiliary} while exhibiting low-polynomial scaling ($M^4$ for Gaussian basis sets).  However, the widespread use of ph-AFQMC in quantum chemistry has not yet taken place.  One major reason is undoubtedly its relatively high computational cost, as the favorable scaling is masked by a large prefactor.  
	
	In this work we present an approach to greatly reducing this prefactor which involves the use of correlated sampling for a particular class of important processes.  The general idea is that for sufficiently similar systems, energy \emph{differences} are expected to converge more rapidly, \emph{i.e.} with smaller error bars, than total energies when the errors or statistical fluctuations in the calculations are biased in the same direction.  Indeed, error cancellation is largely responsible for the success of many approximate methods such as DFT in the computation of energy differences.  Correlated sampling has previously been adapted to reduce the statistical errors in QMC approaches via the use of the same set of configurations sampled for both the primary and secondary systems.\cite{lester1990quantum}  This technique is often referred to in the literature as ``Differential'' QMC, and the details of its implementation can vary depending on the type of QMC being used.  The potential energy curves of H$_2$ and BH have been calculated using correlated sampling with Variational QMC,\cite{umrigar1989two} and similarly that of the $H_3$ cation with Differential Green's Function QMC.\cite{traynor1988parallel}  The latter method has been used to compute the dipole moment of LiH, \cite{wells1985differential} and to calculate infinitesimal energy differences from which forces and various polarizabilities have been obtained.\cite{vrbik1990infinitesimal}  This idea has also been extended to Diffusion Monte Carlo, which has been used to compute forces and potential energy surfaces for the first row diatomics.\cite{PhysRevB.61.R16291}  Correlated sampling has also been used to calculate energy differences between ground and excited states of the same Hamiltonian, as illustrated by a Variational QMC study of particle-hole excitations in the two-dimensional electron gas.\cite{kwon1994quantum}  In addition, the concept has been extended to enable concerted propagation of a system with different time steps, in order to  extrapolate the Trotter error in Differential Diffusion QMC.\cite{garmer1989extrapolation}
  
	Correlated sampling is, in fact, also well-suited to model the energetics of myriad chemical reactions, since \emph{only} energy differences, as opposed to total energies, are relevant.  In this paper we present a novel correlated sampling-based AFQMC approach, and show that it is capable of computing ground-state energy differences corresponding to redox, deprotonation, and hydrogen abstraction reactions to a given statistical error in a fraction of the time previously required, without any loss of accuracy.  Redox reactions (often involving TMs) abound, for example, in metabolic and photosynthetic processes,\cite{nelson2008lehninger,siegbahn2000transition,meunier2004mechanism} battery chemistry,\cite{palacin2009recent} and catalysis (\emph{e.g.} CO$_2$ and O$_2$ reduction).\cite{gratzel2012energy,costentin2013catalysis}  A reliable \emph{ab initio} method to calculate deprotonation free energies would provide an improvement upon existing computational approaches to determining pKa's and protonation states,\cite{alexov2011progress,jerome2014accurate,bochevarov2016multiconformation} which would have significant ramifications for drug discovery,\cite{rajapakse2010sar,sprous2010qsar} materials science,\cite{cheng2010acidity,gallus2015new,bryantsev2013predicting} and the structural determination of biological complexes.\cite{ames2011theoretical,el2000protonation}  In the context of chemical catalysis, proton removal is known to be the rate-limiting step in many important reactions, \emph{e.g.} oxygen reduction on the surface of TiO$_2$.\cite{2013chemical}  Hydrogen abstraction reactions are ubiquitous and play a major role in combustion and the oxidation of hydrocarbons,\cite{olah2003hydrocarbon,wang1994calculations} diamond growth via chemical vapor deposition,\cite{page1991hydrogen} biochemical processes involving \emph{e.g.} the antioxidant vitamin E \cite{burton1989vitamin} and various metalloenzymes,\cite{mayer1998hydrogen,blomberg1997quantum} industrial processes,\cite{grasselli1999advances} and organic synthesis.\cite{snider1996manganese} 
AFQMC would be a useful benchmark for previous \emph{ab initio} studies\cite{blomberg1997quantum,basch1997ab,coote2004reliable,pu2005benchmark,carvalho2008dft,carvalho2008theoretical,fracchia2013barrier,kanai2009toward,kollias2004quantum}   in predicting thermodynamic properties of this difficult class of chemical reactions.  Thus the class of applications we target is large and important.  We highlight the fact that the cost of our correlated sampling approach, relative to the uncorrelated method, decreases with increasing system and basis set size, opening the door to the treatment of realistic large chemical systems with correlated sampling-based AFQMC in the near future.      
	
	This work is organized as follows:  Section II.A will review the exact and phaseless variants of AFQMC, while  Section II.B will present our correlated sampling approach.  We justify this approach for modeling molecular systems in Section II.C, and Section II.D will disclose further computational details.  Section III.A will present calculations of the ionization potentials (IPs) and electron affinities (EAs) of the 1st row atoms in the G2 Ion Test Set,\cite{curtiss1998assessment} while Section III.B will consider adiabatic molecular properties, taking the IP, deprotonation energy, and O-H bond dissociation energy of methanol as examples.  The efficacy of correlated sampling as a function of both basis set size and number of random walkers will be explored in Sections III.C.  The latter subsection will provide an assessment of the reduction in CPU-time afforded by the use of correlated sampling.  In Section IV we conclude, emphasizing opportunities for further gains in computational efficiency that may be possible and future targets of investigation.

\section{Methods}

\subsection{Overview of AFQMC - Exact Method and the Phaseless Constraint}

The excited states of a many-body state $| \Phi \rangle$ can be projected out via imaginary time propagation, \emph{i.e.}  $\lim_{N\rightarrow\infty}(e^{\Delta\tau(E_0 - \hat{H})})^N |\Phi\rangle \rightarrow |\Phi_0 \rangle$, with $\langle \Phi_0 | \Phi \rangle \ne 0$ and the general electronic Hamiltonian 
\begin{equation}
\hat{H} = \sum\limits_{ij}^M \ T_{ij} \sum\limits_\sigma c_{i \sigma}^\dagger c_{j \sigma} + \frac{1}{2}\sum\limits_{ijkl}^M V_{ijkl} \sum\limits_{\sigma,\tau} c_{i \sigma}^\dagger c_{j \tau}^\dagger c_{l\tau} c_{k\sigma},
\label{Hamiltonian}
\end{equation}
where $M$ is the size of the orthonormal one-particle basis, and $c_{i\sigma}^\dag$ and $c_{i\sigma}$ are the second-quantized fermionic creation and annihilation operators with particle and spin labels.  The two-body matrix elements, $V_{ijkl}$, can be expressed in terms of Cholesky vectors as $V_{ijkl} = \sum_\al L_{ik}^\al L_{jl}^\al$.\cite{purwanto2011Ca}   Defining the one-body operator $\hat{v}_\alpha \equiv i\sum_{ik} L_{ik}^\al \sum_\sigma c_{i\sigma}^\dagger c_{k\sigma}$, and subtracting the expectation value with respect to the trial wavefunction $\langle \hat{v}_\alpha\rangle$ from $\hat{v}_\alpha$, the Hamiltonian can be written as the sum of all one-body operators, $\hat{H}_1$, plus the following two-body operator
\begin{equation}
\hat{H}_2 = -\frac{1}{2}\sum_\alpha(\hat{v}_\alpha - \langle \hat{v}_\alpha \rangle )^2.
\label{H_2}
\end{equation}

Use of the Trotter-Suzuki decomposition\cite{trotter1959product,suzuki1976generalized} gives
\begin{equation}
\label{Trotter}
e^{-\Delta\tau\hat{H}} = e^{-\Delta\tau\hat{H}_1/2}e^{-\Delta\tau\hat{H}_2}e^{-\Delta\tau\hat{H}_1/2} + O(\Delta\tau^3).
\end{equation}
The exponential terms involving $\hat{H}_2$ may be decomposed using a Hubbard-Stratonovich (HS) transformation\cite{stratonovich1957method, hubbard1959calculation}
\begin{equation}
e^{\frac{1}{2}\Delta\tau(\hat{v}_\alpha - \langle \hat{v}_\alpha \rangle)^2} = \int_{-\infty}^\infty dx_\alpha \bigg(\frac{e^{-\frac{1}{2}x_\alpha^2}}{\sqrt{2\pi}}\bigg)e^{\sqrt{\Delta\tau}x_\alpha(\hat{v}_\alpha - \langle \hat{v}_\alpha \rangle)}, 
\label{HS}
\end{equation}
which expresses the exponential of a two-body operator as the exponential of a one-body operator integrated over auxiliary-fields (AFs).  This transformation allows for practical propagation in terms of the Thouless theorem, which states that the application of an exponential of a one-body operator on a Slater determinant produces another Slater determinant,\cite{hamann1990energy} which can be implemented via a simple matrix multiplication.\cite{HuyThesis, rubenstein2012finite}  

The propagator \eqref{Trotter} now takes the form of a multi-dimensional integral
\begin{equation}
e^{-\Delta\tau \hat{H}} = \int d\mbf{x} P(\mbf{x}) \hat{B}(\mbf{x}),
\label{prop}
\end{equation}
where $\mbf{x} = (x_1, x_2, \ldots , x_\al)$, $P(\mbf{x})$ is a normal distribution with unit variance, and $\hat{B}(\mbf{x}) = e^{-\Delta\tau \hat{H}_1} e^{\sqrt{\Delta\tau} \mbf{x} \cdot (\mbf{ \hat{v} } - \mbf{\langle \hat{v} \rangle})} e^{-\Delta\tau \hat{H}_1} $.  

The integral in \eqref{prop} may be approximated using a Monte Carlo scheme, with walkers whose propagation in the space of Slater determinants is guided by the complex importance function $\langle \phi_T | \phi\rangle$ which is proportional to the walker weights.  The representation of the total wavefunction is thus a weighted sum over walker determinants
\begin{equation}
|\Phi \rangle = \sum_k \frac{w_k | \phi_k \rangle}{\langle \phi_T | \phi_k \rangle},
\end{equation}
yielding essentially a multi-reference description.  The energy is calculated at intervals using the mixed-estimator
\begin{equation}
\frac{\langle \phi_T | \hat{H} | \Phi \rangle}{\langle \phi_T | \Phi \rangle} = \frac{\sum_k w_k E_L(\phi_k)}{\sum_k w_k},
\end{equation}
where the ``local energy'' is given by $E_L(\phi_k) = \frac{\langle \phi_T|\hat{H}|\phi_k \rangle}{\langle \phi_T | \phi_k \rangle }$, in analogy to Diffusion QMC.\cite{reynolds1982fixed}  

The method as described above is called the ``Free Projection" (FP) approach.\cite{shi2013symmetry}   While it is formally exact and can yield excellent results for small system sizes, this method suffers from the ``phase problem," which is a generalization of the Fermionic ``sign problem'' to the complex plane.\cite{zhang1997constrained,zhang2013,loh1990sign,troyer2005computational}  For the standard Coulomb interaction the $\hat{v}_\alpha$ operators are purely imaginary, and each application of $e^{\sqrt{\Delta\tau} \mbf{x} \cdot (\mbf{ \hat{v} } - \mbf{\langle \hat{v} \rangle})}$ can be thought of as a rotation of the Slater determinant $|\phi\rangle$, causing an evolution of the overlap $\langle \phi_T | \phi\rangle$ in the complex plane.  Over the course of the random walk, a determinant accumulates a phase $e^{i\theta}$, and the infinitely many possible values of $\theta \ \epsilon \ [0, 2\pi)$ result in the possibility of infinitely many indistinguishable determinants.  Furthermore, over the course of the propagation, walkers will populate the origin where $\langle \phi_T | \phi \rangle = 0$, and subsequent propagation yields only noise from signal cancellation effects and divergences in the weights and local energies.  

The ph-AFQMC employs a multifaceted strategy to control this problem.  The weights are initialized to a positive real constant, and after each propagation step are projected back onto the real axis, \emph{i.e.} the rotated weights are multiplied by $max\{ 0, cos(\Delta\theta)\}$, where we have defined the phase of the overlap ratio $\Delta\theta \equiv Im\{\ln \frac{\langle \phi_T | \phi^{(\tau + 1)} \rangle}{\langle \phi_T | \phi^{(\tau)} \rangle} \}$.\cite{motta2014imaginary}  For this phase projection to work, the AFs are shifted by a force bias (FB) $\mbf{\bar{x}}$,\cite{zhang2003quantum,purwanto2004quantum} the optimal choice of which is obtained by minimizing the fluctuations of the weights with respect to the AFs at their average value:
\begin{equation}
\frac{\partial}{\partial x_\alpha}\bigg[\frac{\langle\phi_T|e^{\sqrt{\Delta\tau}(x_\alpha - \bar{x}_\alpha)(\hat{v}_\alpha - \langle \hat{v}_\alpha \rangle)}|\phi_k\rangle}{\langle \phi_T | \phi_k \rangle} e^{-\bar{x}_\alpha^2/2 + x_\alpha \bar{x}} \bigg]_{x_\alpha = 0} = 0.
\label{FB}
\end{equation}
This is, in essence, a stationary-phase approximation.  Expanding the expression inside the brackets to $O(\sqrt{\Delta\tau})$ and taking the derivative gives
\begin{equation}
\bar{x}_\alpha = -\sqrt{\Delta\tau}\big[\bar{v}_\alpha - \langle \hat{v}_\alpha \rangle \big],
\label{optFB}
\end{equation} 
where $\bar{v}_\alpha \equiv \frac{\langle\phi_T|\hat{v}_\alpha|\phi\rangle}{\langle \phi_T | \phi \rangle}$.  The introduction of the FB does \emph{not} add any additional approximations, as the integration variable in \eqref{HS} is merely shifted by a constant, yet it is crucial for two reasons.  First, it diverges when the ``nodal surface" (as defined in the complex plane of overlaps) is approached, pushing the walker away from the origin.  Second,   since $\hat{v}_\alpha$ is complex, $Im[{\mathbf{\bar{x}}}]$ reduces the amount of physical information discarded in the phase projection.  Similarly, the subtraction of $\langle \hat{v} \rangle$ from $\hat{v}$ in the propagator also greatly reduces the severity of this projection, as the smaller diagonal matrix elements of the resulting propagator cause milder rotations of the phases of the orbitals.\cite{al2006gaussian}   

The choice of FB allows the weight factor which multiplies the previous weight after a propagation step to be written as $W(\phi) = e^{-\Delta\tau E_L(\phi)}$.  The second approximation in ph-AFQMC, which is much milder than the first, takes the real part of the local energy in the weight above:
\begin{equation}
E_L(\phi) \equiv Re\{ \frac{\langle \phi_T | \hat{H} | \phi \rangle} {\langle \phi_T | \phi \rangle} \}.
\label{phaseless}
\end{equation}
The severity of the phaseless constraint in ph-AFQMC can be reduced with the use of more accurate trial wavefunctions, especially those with the correct symmetry properties.\cite{shi2013symmetry}  This can be seen from \eqref{phaseless}, for when $| \phi_T \rangle = | \Phi_0 \rangle$, the local energy which determines the weights and energy measurements is a real constant (equal to the exact ground state energy).

\subsection{Correlated Sampling Methods for AFQMC}
Given that the statistical error in an AFQMC calculation arises solely from the MC evaluation of the integral over AFs, a natural way to implement correlated sampling in the calculation of an energy difference is to pair the walkers of the two systems, and use the same set of AFs to propagate each pair of walkers.  To be precise, the correlation is established on the level of each term in \eqref{H_2}, for each value of $\al$.  This, in essence, matches Cholesky vectors of the same iteration, and becomes more effective the more similar the interactions are in the two systems.

Sampling in this way, however, requires one to relinquish optimal importance sampling, which is usually implemented via a population control (PC) scheme in which walkers with large (small) weights are replicated (stochastically purged), since performing independent PC for each system separately would quickly destroy the walker-pair correlation.  Alternatively, one can implement PC with respect to the weights of the primary system \emph{for both systems}.  This, however, will only be effective if the two systems are essentially identical.  In the absence of a PC scheme, the noise from the accumulation and persistence of divergent walkers inevitably grows with propagation time.  We find that the immediate reduction in statistical error following the correlation of the AFs, augmented as needed by what we call the ``preliminary equilibration scheme" below, allows converged averages to be obtained at short and intermediate projection times.  In light of the fact that the stability of the random walks at long times, as afforded by a branching scheme, appears to be only marginally relevant when our correlated sampling approach is used, we simply choose \emph{not} to implement PC when the AFs are correlated.  In the event that a walker's weight becomes zero or negative (in the latter case the phase projection sets the weight to zero), the walker is no longer propagated and the random number stream is updated if necessary such that the correlation of the other walker-pairs is unaffected.

\begin{figure}[h!]
    \centering
    \subfloat[Averaged IPs (circles) among the repeats at each $\tau$ along the imaginary-time propagation.]{{\includegraphics[width=7.7cm, keepaspectratio]{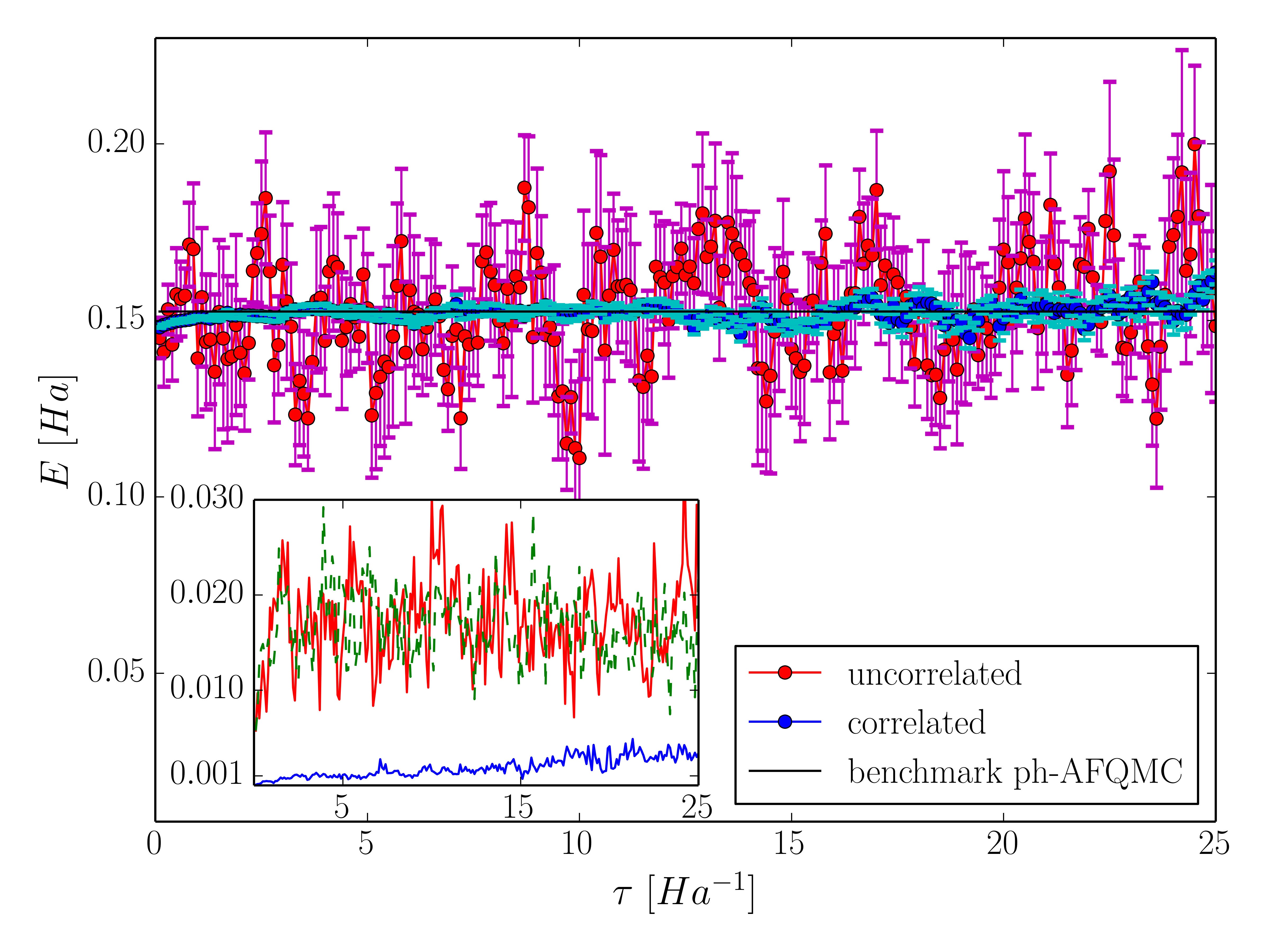} }}
    \qquad
    \subfloat[Mean values (circles) of the cumulative averages taken for $\tau > 4$.]
    {{\includegraphics[width=7.7cm, keepaspectratio]{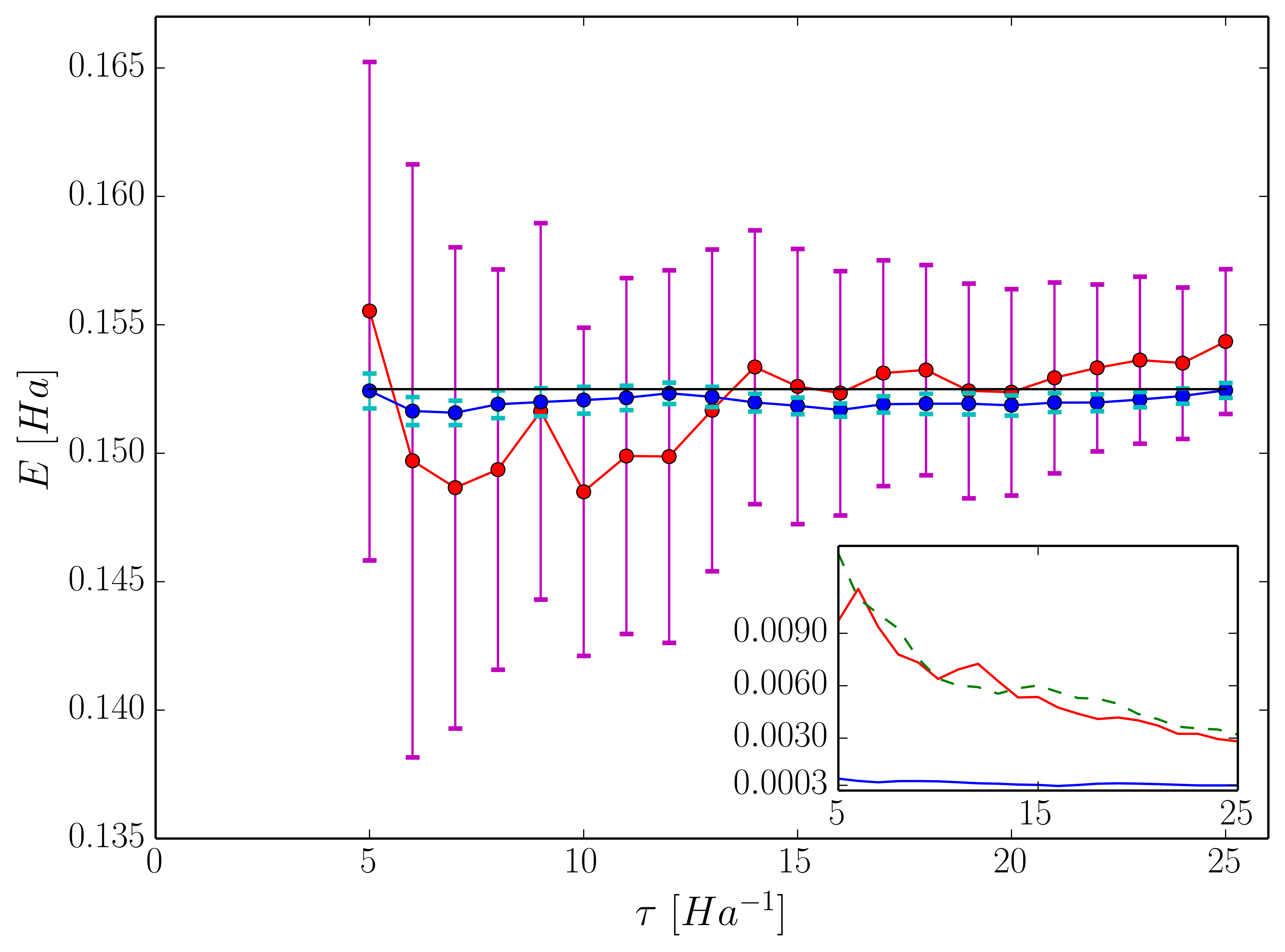} }}
    \caption{Comparison of correlated and uncorrelated sampling for the IP of the K atom in a 6-31+G* basis, with $\Delta\tau = 0.01$, 12 walkers per repeat, and a HF reference state.  The error bars give the standard errors (the standard deviation times $\frac{1}{\sqrt{N_r}}$, where $N_r$ is the number of repeats) of the mean values among the repeats at each $\tau$ point in (a), and of the cumulative averages in (b).  These standard errors are plotted in the insets, along with the scaled standard error resulting from an uncorrelated run in which PC was used with 360 walkers per repeat (dotted green).}
    \label{fig:Corr_Samp}
\end{figure}

The data shown in blue in Fig. \ref{fig:Corr_Samp} illustrates features associated with the typical correlated sampling protocol used in this work.  In this example, the propagation in imaginary time (measured in  Ha$^{-1}$) is performed using correlated AFs and repeated 11 times using different random number seeds.  The mean energy difference and associated standard error at each $\tau$ point is computed among the repeats (see left plot).  We then choose the imaginary-time at which the energy difference is seen to stabilize (in this case at $\tau \sim 4$), and for each repeat calculate the \emph{cumulative} average at each $\tau$, which represents the running average of the energy measurements taken after the end of the equilibration period up to the given value of $\tau$.  To obtain the final result, we compute the mean and standard error of the cumulative averages among the repeats (see right plot), and choose the value corresponding to the $\tau$ at which the standard error reaches a minimum (if there are multiple equivalent minima the one occurring earliest is chosen) or falls below a target error level.  We note that our choice of 11 repeats is arbitrary; a larger number could be used to reduce the standard error as necessary.  Clearly in this example correlating the AFs for $\tau \geq 0$ drastically reduces the error relative to the uncorrelated runs shown in red, and the resulting IP agrees with that obtained from independent $\tau = 80$ ph-AFQMC runs of the neutral and cationic species.  This benchmark, indicated by the solid black line, was found to have a negligible standard error of 0.2 mHa after employing a reblocking analysis which corrects for auto-correlation.\cite{flyvbjerg1989error}  
 
As it is often the case that a very small population size is sufficient to achieve a desired statistical error via the correlated sampling approach, we choose not to employ PC in the uncorrelated comparisons since this would result in a bias that typically goes as 1/$N_{wlk}$.\cite{nguyen2014cpmc}  A second reason is that over the relatively short imaginary-time scales relevant for the correlated sampling method, PC is expected to have little effect on the uncorrelated comparisons as the walker weights usually do not stray far from unity.  This is confirmed by the similarity of the error curves plotted in the insets of Fig. \ref{fig:Corr_Samp}, corresponding to uncorrelated runs with (dotted green) and without (red) PC.  The former is obtained by using a large enough population size such that the bias from the PC algorithm is negligible (360 walkers per repeat), and rescaling the resulting standard error by $\sqrt{\frac{360}{12}}$.  

\begin{figure}[h!]
    \centering
    \includegraphics[width=8cm]{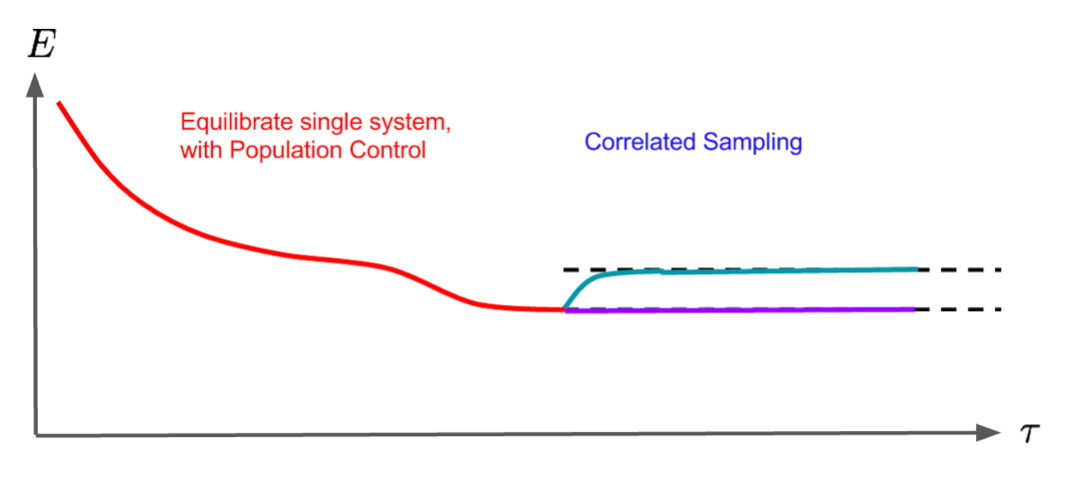} 
    \caption{Schematic of the PES version of correlated sampling.  Preliminary equilibration of the primary system is shown in red.  This phase can be as long as necessary due to the use of PC.  The resulting walkers are used to initialize the secondary system, after which the two systems are propagated with their own Hamiltonians using correlated sampling.  Equilibration of the secondary system is relatively rapid, allowing measurements to be taken before the noise growth becomes prohibitive.}   
    \label{fig:Cartoon}
\end{figure}

If one or both of the comparative systems requires a long equilibration time, which can be the case for, \emph{e.g.}, initial populations which are severely spin-contaminated or for strongly correlated systems in which the ``guiding" trial function poorly describes the true ground state, then walker pair correlation can be lost prior to convergence, and the associated noise growth can make measurements impossible.  The simplest way to overcome this problem is to use a better trial function, \emph{e.g.} a multi-determinant CASSCF wavefunction instead of the single Hartree-Fock (HF) determinant.  Alternatively, we have devised what we will refer to as the ``preliminary equilibration scheme" (PES).  First, one of the two systems is equilibrated using PC for the required interval, then the walkers of the secondary system are initialized with the resulting determinants of the equilibrated primary system, with weights scaled by $\frac{\langle \phi_T^{secondary} | \phi \rangle}{\langle \phi_T^{primary} | \phi \rangle}$ (and any resulting phases projected) to reflect the appropriate importance sampling.  Finally, both systems are propagated using correlated sampling with their respective Hamiltonians without PC for a short period of time after which energy measurements are collected.  A schematic of this procedure is shown in Fig. \ref{fig:Cartoon}. This protocol is repeated with different random number seeds to obtain satisfactory statistics on the energy difference between the ground states.  We note that a similar scheme was published many years ago by Traynor et al.\cite{traynor1988parallel}  

Inspection of the ph-AFQMC propagator
\begin{equation}
\hat{B}(\mbf{x} - \mbf{\bar{x}}) = e^{-\Delta\tau\hat{H}_1/2} e^{\sqrt{\Delta\tau}(\mbf{x} - \mbf{\bar{x}})(\mbf{\hat{v}} - \langle \mbf{\hat{v}} \rangle)} e^{-\Delta\tau\hat{H}_1/2},
\label{full_propagator}
\end{equation}
demonstrates that stricter approaches to correlated sampling exist.  In one such alternative to correlating only the AFs, each pair of walkers is propagated using both the same AFs and FBs.  However taking the simple average of the FBs of the primary and secondary systems was found to cause an early onset of noise even for marginally different systems.   This issue likely arises from divergences in the local energies and weights which can occur more frequently due to the use of sub-optimal FBs, a fact which follows from the discussion centered around Eq. \eqref{optFB}.  Separately, we choose \emph{not} to correlate the $\langle \mbf{\hat{v}} \rangle$ since any gain in sampling efficiency would come at the cost of an increased bias from the phaseless constraint.\cite{al2006gaussian}  

Finally, we note that our method of correlated sampling is not limited to the phaseless version of AFQMC, and can be used in exactly the same manner for FP calculations.  In fact, using the same AFs in the latter case results in a relatively stricter form of correlated sampling, since the FB is not present in the post-HS propagator used in FP. 

\subsection{Utilizing Optimal Correlation in Molecular Applications}
\begin{figure}[h!]
    \centering
    \subfloat[Vertical IP]{{\includegraphics[width=7.7cm,keepaspectratio]{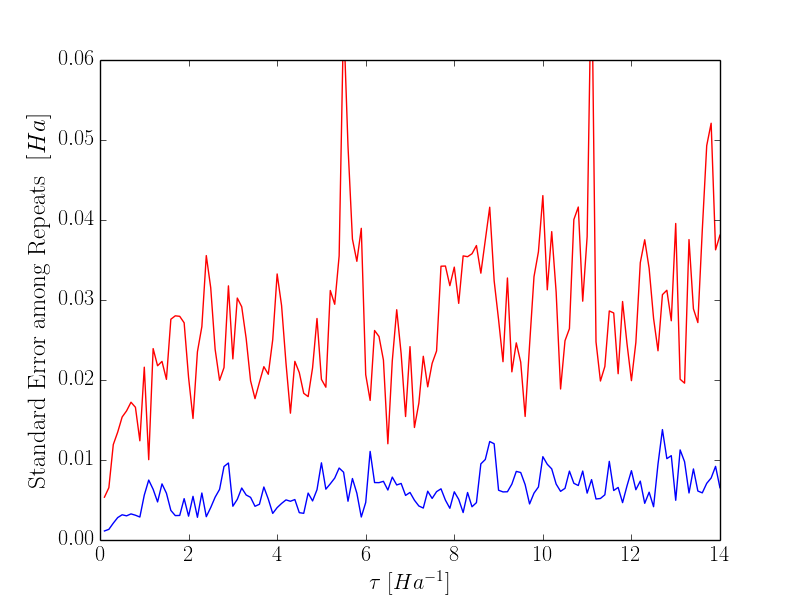} }}
    \qquad
    \subfloat[Adiabatic IP]{{\includegraphics[width=7.7cm, keepaspectratio]{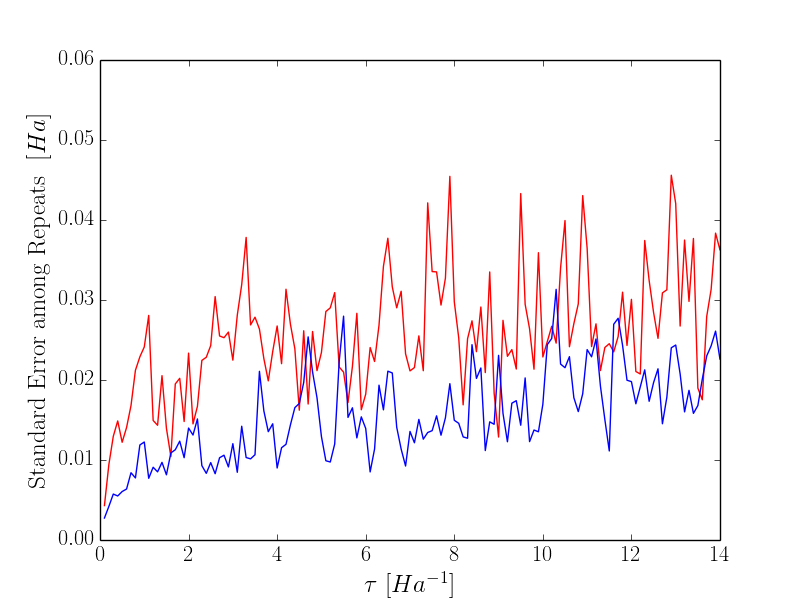} }}
    \caption{Standard errors resulting from uncorrelated (red) and correlated (blue) propagation of the repeats for methanol in the cc-pVTZ basis.  $\Delta\tau = 0.01$ and 24 walkers per repeat were used.}
    \label{fig:Vert_vs_Ad}
\end{figure}

Having justified our choice to correlate only the AFs, we claim that maximal correlation between a pair of walkers is achieved when the primary and secondary systems use the same set of basis functions.  To see this, recall the definition of the two-electron matrix elements in the Hamiltonian \eqref{Hamiltonian}:
\begin{equation}
V_{ijkl} = \int d\mbf{r}_1 d\mbf{r}_2 \phi_i(\mbf{r}_1)\phi_j(\mbf{r}_2)\frac{1}{r_{12}}\phi_k(\mbf{r}_1)\phi_l(\mbf{r}_2).
\end{equation}
When $\{\phi\}_{primary} = \{\phi\}_{secondary}$, the $V_{ijkl}$ and, in turn, the one-body operators $\mbf{\hat{v}}$ in the propagators \eqref{full_propagator} for both systems will be identical.  This ideal condition is obviously satisfied when the two systems are an atom or molecule with the same geometry but \emph{e.g.} different charges, since atomic basis functions such as Gaussian-type orbitals are usually centered on the positions of the nuclei.  In fact, for the calculation of vertical IPs and EAs, exactly the same $T_{ij}$ and $V_{ijkl}$ elements are used in the imaginary-time propagation of both systems.  For adiabatic redox processes, however, the ground-state geometries of the neutral and ionized species are, in general, different.  In most cases the differences in the $V_{ijkl}$ are slight, and correlated sampling is still found to be effective, albeit with reduced efficiency.  For example, Fig. \ref{fig:Vert_vs_Ad} compares the correlated and uncorrelated standard errors  corresponding to the vertical and adiabatic IPs of methanol.  While the errors from the uncorrelated runs are roughly similar in magnitude, that from the correlated  vertical case is significantly smaller and more constant compared to that from the correlated adiabatic case.  This suggests the use of a two-step process to compute adiabatic energy differences, in which the fixed-geometry transition is calculated with correlated sampling-based AFQMC, and the geometry relaxation energy is obtained from a lower level of theory.  Further details will be presented in Section III.B.    

\begin{figure}[h!]
    \centering
    \subfloat[Averaged deprotonation energy (circles) among the repeats at each $\tau$.]{{\includegraphics[width=7.8cm, keepaspectratio]{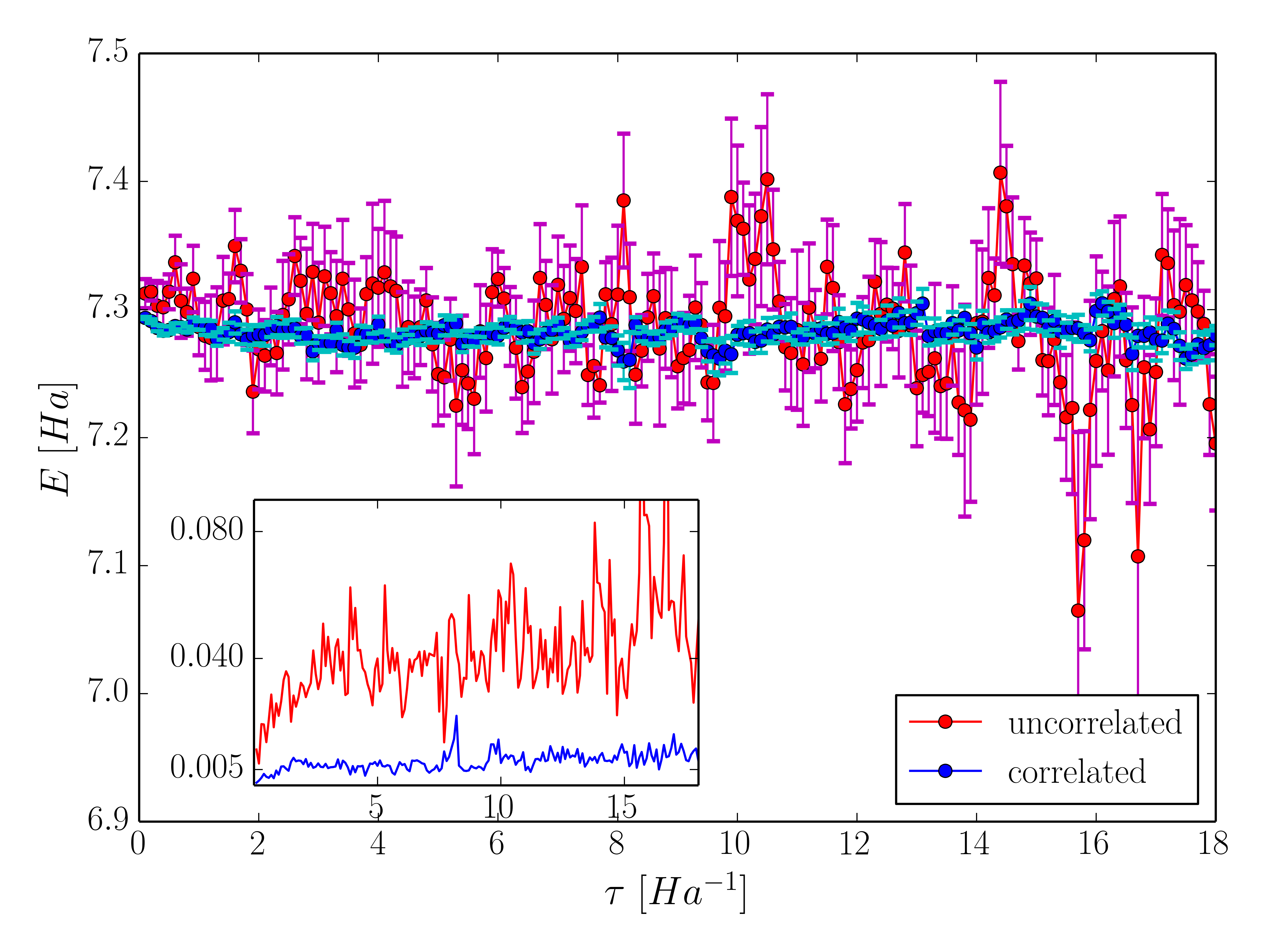} }}
    \qquad
    \subfloat[Mean values (circles) of the cumulative averages taken at $\tau > 4$.]{{\includegraphics[width=7.6cm, keepaspectratio]{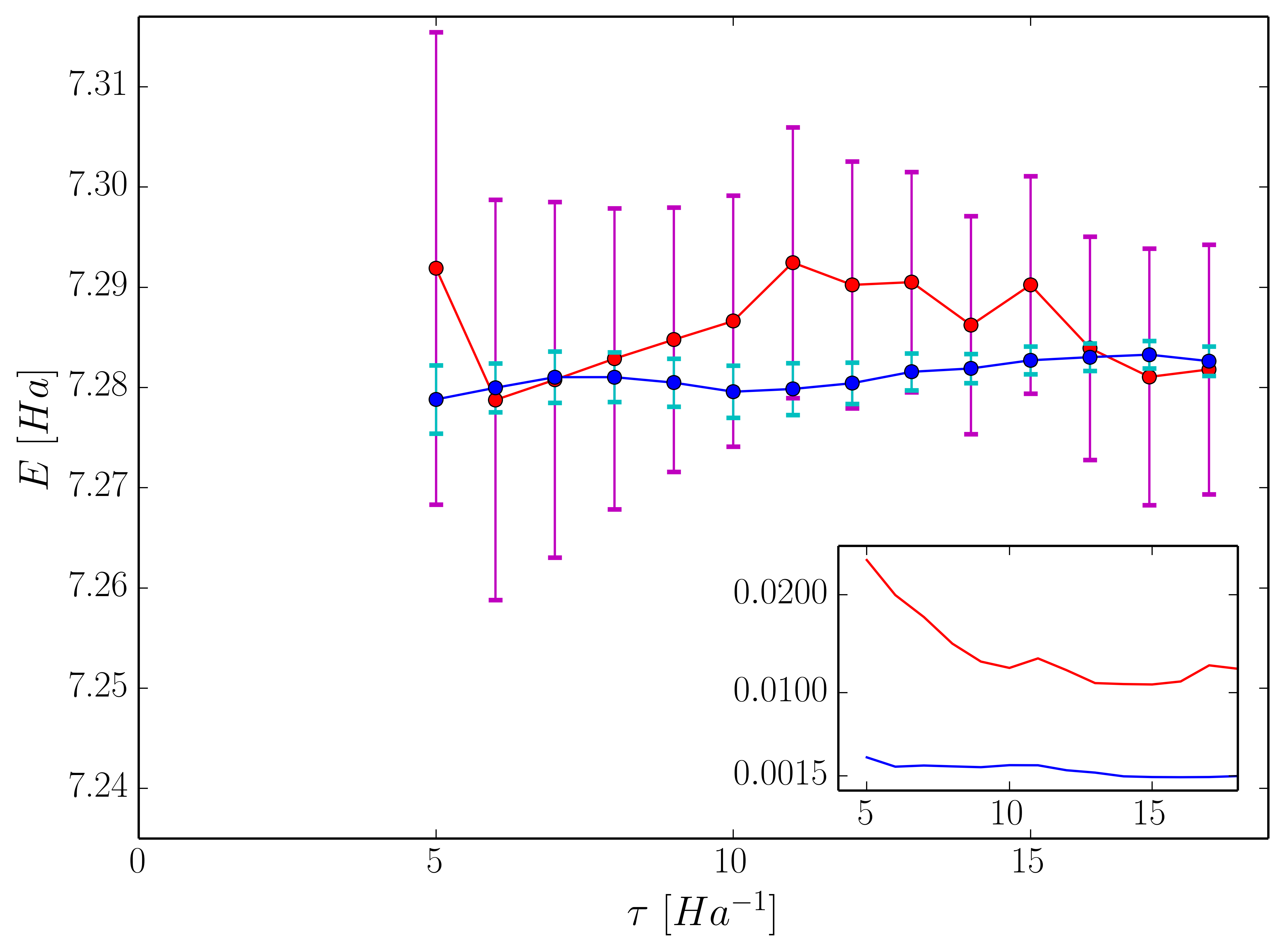} }}
    \caption{Comparison of correlated and uncorrelated sampling for the fixed-geometry deprotonation of CH$_3$OH in the cc-pVTZ basis with ``ghost" basis functions.  We use $\Delta\tau = 0.01$, 12 walkers per repeat, and a HF reference state.  The error bars, plotted in the insets, in (a) represent the standard error among the repeats at each $\tau$; those in (b) give the standard error of the cumulative averages.}
    \label{fig:Corr_Samp_deprot}
\end{figure}

\begin{figure}[h!]
    \centering
    \subfloat[Averaged H abstraction energy (circles) among the repeats at each $\tau$.]{{\includegraphics[width=7.8cm, keepaspectratio]{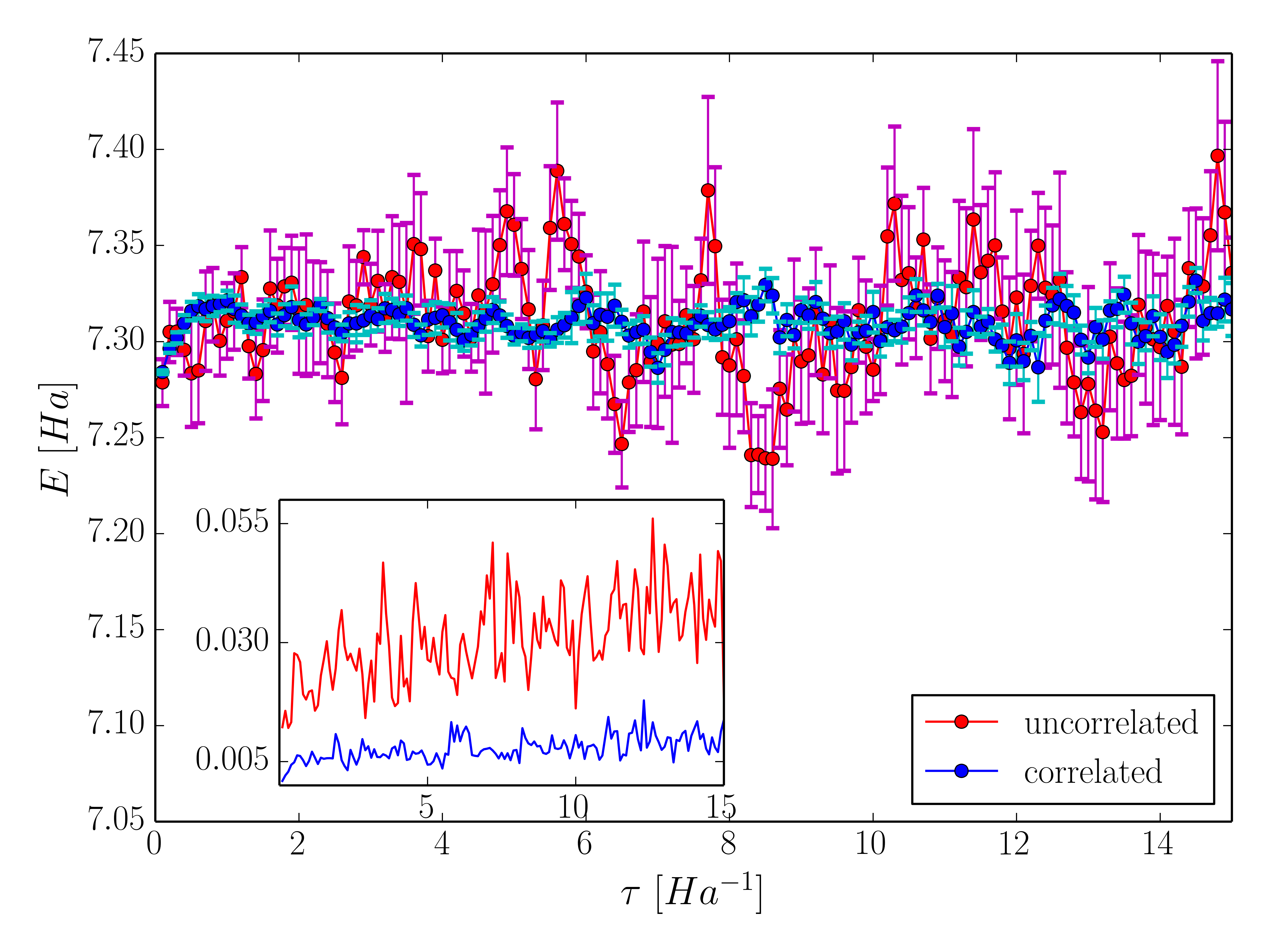} }}
    \qquad
    \subfloat[Mean values (circles) of the cumulative averages taken at $\tau > 2$.]
    {{\includegraphics[width=7.6cm, keepaspectratio]{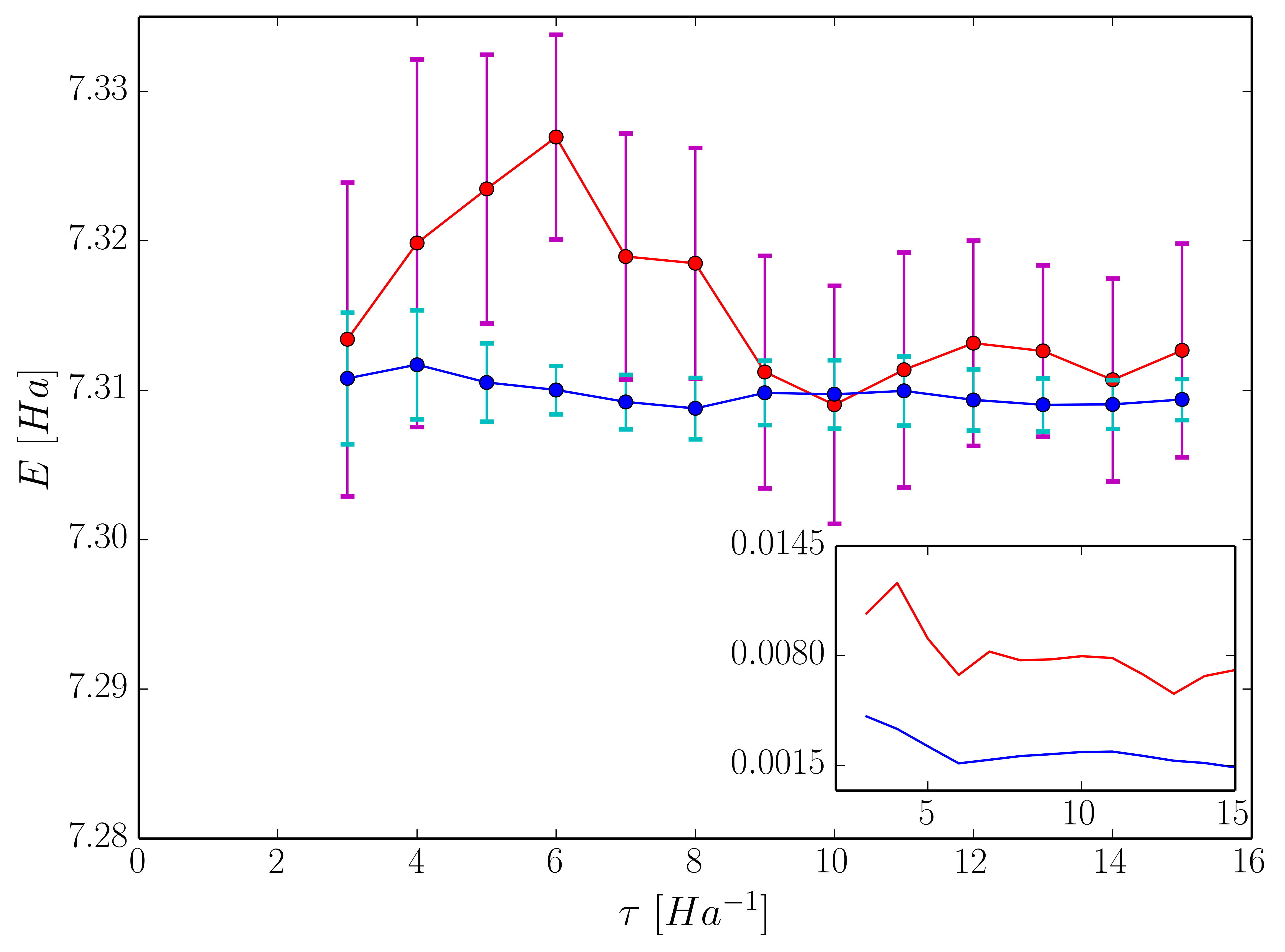} }}
    \caption{Comparison of correlated and uncorrelated sampling for the energy difference corresponding to the removal of H$\cdot$ from the O-H bond of CH$_3$OH, with the rest of the geometry fixed.  The cc-pVTZ basis is used, with $\Delta\tau = 0.01$, 12 walkers per repeat, and a CASSCF reference state.  The error bars, plotted in the insets, in (a) represent the standard error among the repeats at each $\tau$; those in (b) give the standard error of the cumulative averages.}
    \label{fig:H_abstr}
\end{figure}

In the same spirit, we have devised a protocol which enables the calculation of the change in energy corresponding to the removal of a proton while the rest of the geometry remains fixed, in which the optimal condition mentioned above is realized with the use of so-called ``ghost" basis functions.  In this scheme, the deprotonated species uses the same set of basis functions as the acid, \emph{i.e.} the position of the removed proton still serves as a center for hydrogen basis functions but \emph{not} as a center of nuclear charge.  As a result, the number of basis functions remains the same for the primary and secondary systems as required by our correlated sampling method.  Moreover the $V_{ijkl}$ and thus the $\mbf{\hat{v}}$ are, by construction, identical.  We note that even though the $T_{ij}$ now differ due to the altered electron-nucleus attraction terms, the statistical error in the calculated energy difference is \emph{not} exacerbated since it results only from the MC evaluation of the integral over AFs.  Fig. \ref{fig:Corr_Samp_deprot} illustrates the efficacy of this procedure in a calculation of the deprotonation energy for methanol, showing a drastic reduction in the statistical errors when the AFs are correlated.
The ``ghost" basis function strategy can also be directly applied effectively to hydrogen abstraction reactions.  Fig. \ref{fig:H_abstr} demonstrates the error reduction afforded in a calculation of the MeOH $\rightarrow$ MeO$\cdot$ reaction energy.

Thus we have presented a correlated sampling approach which has the potential to reduce the statistical error for redox, deprotonation, and hydrogen abstraction reactions.  The extension of this protocol to obtain adiabatic energy differences will be described in Section III.B.

\subsection{Computational Details}

One-electron and overlap integrals, Cholesky vectors, restricted open-shell and unrestricted HF trial wavefunctions were all obtained from a modified version of NWChem.\cite{valiev2010nwchem, purwanto2011Ca}  The maximum residual error in the Cholesky decomposition was chosen to be $1\times10^{-6}$ Ha.  This cutoff has a negligible effect on the QMC energies (\emph{i.e.} orders of magnitude smaller than the statistical error bars), and is utilized to decrease the number of AFs required, which in turn decreases both the error from the random noise and the computational cost of the calculations.
 
For calculations using a single-determinant trial wavefunction with unrestricted reference, spin-contamination can occur due to the presence of higher multiplicity spin configurations in the trial function,  which can prolong equilibration times and reduce accuracy.  We use a spin-projection technique\cite{purwanto2008eliminating} in AFQMC to minimize such effects.  In this scheme, the walkers are initialized with a restricted open-shell HF determinant, which is an eigenfunction of $\hat{S}^2$, such that propagation with $e^{\sqrt{\Delta\tau} \mbf{x} \cdot ( \mbf{\hat{v}} - \langle \mbf{\hat{v}} \rangle)}$ preserves spin symmetry.  The unrestricted HF determinant, while \emph{not} an eigenfunction of the total spin operator, is known in most cases to provide a better description of the ground-state energy, and is therefore used to implement the phaseless constraint and to estimate the energy.    

Multi-determinant trial states were obtained from CASSCF calculations performed with PySCF.\cite{pyscf}  The resulting expansions of the wavefunctions in CI space are truncated such that determinants associated with coefficients below a specified threshold are discarded.  In what follows, CASSCF($X$,$Y$) will denote an active space with $X$ electrons and $Y$ orbitals; for cases in which an energy difference is calculated we use the notation $X \equiv N_{neutral}/N_{ion}$ and $Y \equiv M_{neutral}/M_{ion}$ to denote the numbers of electrons and orbitals in the neutral and charged systems, respectively.  

In all AFQMC calculations the spin -up and -down sectors of the walker determinants are separately orthonormalized  via a modified Gram-Schmidt procedure\cite{white1989numerical} every 0.05 Ha$^{-1}$ to preserve the anti-symmetry of the walker determinants.  When required, we use a PC algorithm in which the number of walkers is fixed throughout the entire calculation.\cite{nguyen2014cpmc} PC (when used) and energy measurements are performed every 0.1 Ha$^{-1}$.  

Use of the ``hybrid'' formulation\cite{purwanto2009pressure} of ph-AFQMC allows for the calculation of the local energy, the bottleneck in the algorithm's scaling, at intervals rather than at every step since $E_L$ is not required to compute the weight factors, which are now calculated explicitly as $\frac{\langle \phi_T | \phi^{(\tau+1)} \rangle}{\langle \phi_T | \phi^{(\tau)} \rangle} e^{\mbf{x} \cdot \mbf{\bar{x}} - \mbf{\bar{x}}^2/2 }$.  We chose to implement the hybrid method primarily because it is much faster than the local energy method, and offers more flexibility when devising correlated sampling strategies.

The comparison of QMC energies with experimental results in general requires extrapolations to mitigate errors associated with the use of a finite time step and basis set.  The Trotter error, due to the decomposition in \eqref{Trotter} and three bounding conditions,\cite{purwanto2009pressure} can be estimated from a linear extrapolation of the energy differences from three independent simulations at $\Delta\tau = 0.02, \ 0.01, \ \text{and} \ 0.005$ Ha$^{-1}$.  For the IPs and EAs of the G2 atomic test set, the energies resulting from the 0.01 Ha$^{-1}$ time step were found in all cases to be well within 1 mHa of the $\Delta\tau$-extrapolated value.  Hence, we use only the $\Delta\tau = 0.01$ value to produce the results in Section III.   Extrapolation to the complete basis set (CBS) limit was performed following the second order M{\o}ller-Plesset perturbation theory (MP2)-assisted protocol detailed in Ref. \cite{purwanto2011Ca}  AFQMC calculations were performed at $x=$ 3 (where $x$ is the cardinal number of the basis set); UHF calculations were run at $x =$ 2-5 and MP2 calculations at $x=$ 3, 4 using GAMESS\cite{schmidt1993general} or, for the reactions which utilize ``ghost'' basis functions, NWChem.  IPs were computed using the cc-pVxZ basis,\cite{dunning1989gaussian} while EAs and the deprotonation energies to be compared with experimental data were computed using the aug-cc-pVxZ basis sets.\cite{kendall1992electron}

QMC statistical error bars were propagated through the data analysis procedure.  In assessing the statistical error on the deviation of QMC from experiment, we assume that there are negligible uncertainties associated with 1) the experimental measurements, 2) the exponential (HF) and linear (MP2) fits in the CBS extrapolation procedure where we found that the former error is an order of magnitude or more smaller than the QMC error in representative cases, and 3) the scaling factor for the zero-point energies.  With regard to 1), experimental uncertainties can be found in the NIST database,\cite{johnsonnist} and are ignored in order to isolate the statistical error associated with the QMC measurements, since the quantities presented in this work are differences between calculated energies and experimental measurements.  

Our Fortran 90 code is linked with OpenBLAS\cite{Wang_OpenBLAS,Xianyi_OpenBLAS} and Expokit.\cite{sidje1998expokit}  Random numbers were generated with the 48 bit Linear Congruential Generator with Prime Addend, as implemented in SPRNG5.\cite{mascagni2000algorithm}

\section{Results}

\subsection{Atomic IPs and EAs}

In this section, we use our correlated sampling-based AFQMC approach to calculate the IPs and EAs of the 1st row atoms included in the G2 Ion Test set, which have experimental uncertainties of less than 0.05 eV.\cite{curtiss1998assessment}  The deviations from experiment within calculated AFQMC results are presented in Tables \ref{table:G2 IP} and \ref{table:G2 EA} alongside the deviations resulting from G2 theory,\cite{curtiss1991gaussian} and DFT with the B3LYP exchange-correlation  functional\cite{becke1993density,lee1988development} in the 6-311+G(3df,2p) basis.\cite{curtiss1998assessment}  

\begin{table*}[ht]
\centering
\begin{threeparttable}
\caption{Experimental IPs and the deviations of various calculated results (theory - experiment) for atoms in the G2 Test Set in eV.  QMC statistical errors in the two right-most digits are shown in parenthesis.  QMC calculations using single-determinant trial functions (phaseless and FP) use 5040 walkers per repeat, while those using multi-determinant trial functions use 1056 walkers per repeat.}
\begin{tabular}{c c c c c c c}
\hline\hline
Atom & Expt. & $\Delta$ph-HF/QMC & $\Delta$FP & $\Delta$ph-CAS/QMC & $\Delta$G2 & $\Delta$B3LYP \\ [0.5ex] 
\hline
B  & 8.2980   & -0.156(10)\tnote{*}  & -0.0012(50) & -0.0162(26)\tnote{a}  &  0.10 & -0.44 \\
C  & 11.2603  &  0.0342(90)\tnote{*} &             &  0.0045(43)\tnote{b}  &  0.08 & -0.29 \\
N  & 14.5341  &  0.1214(61)\tnote{*} &  0.0050(84) &  0.0100(38)\tnote{c}  &  0.06 & -0.14 \\
O  & 13.6181  & -0.0830(22)\tnote{*} &             &  -0.0360(24)\tnote{d} &  0.08 & -0.55 \\
F  & 17.4228  &  0.0010(35)          &             &  0.0015(46)\tnote{e}  &  0.03 & -0.34 \\ 
Ne & 21.5645  &  0.0775(51)          &             &  0.0159(34)\tnote{f}  & -0.05 & -0.21 \\ [1ex]
\hline
\end{tabular}
\begin{tablenotes}
\footnotesize 
\item[*] PES
\item[a] QMC trial from CASSCF(5/4,8) with minimum CI coefficient of 0.034 (20/14 determinants)
\item[b] QMC trial from CASSCF(4/3,8) with minimum CI coefficient of 0.034 (14/7 determinants)
\item[c] QMC trial from CASSCF(5/4,8) with minimum CI coefficient of 0.01 (29/32 determinants)
\item[d] QMC trial from CASSCF(6/5,13) with minimum CI coefficient of 0.01 (67/62 determinants)
\item[e] QMC trial from CASSCF(7/6,8) with minimum CI coefficient of 0.033 (7/2 determinants)
\item[f] QMC trial from CASSCF(8/7,16) with minimum CI coefficient of 0.0085 (101/104 determinants)
\end{tablenotes}
\label{table:G2 IP}
\end{threeparttable}
\end{table*}

\begin{table*}[ht]
\centering
\begin{threeparttable}
\caption{Experimental EAs and the deviations of various calculated results (theory - experiment) for atoms in the G2 Test Set in eV.  QMC statistical errors in the two right-most digits are shown in parenthesis.  QMC calculations using single-determinant trial functions (phaseless and FP) use 5040 walkers per repeat, while those using multi-determinant trial functions use 1056 walkers per repeat.}
\begin{tabular}{c c c c c c c}
\hline\hline
Atom & Expt. & $\Delta$ph-HF/QMC & $\Delta$FP & $\Delta$ph-CAS/QMC & $\Delta$G2 & $\Delta$B3LYP \\ [0.5ex] 
\hline
B  & 0.2797 & -0.0422(31)           &            & -0.0090(30)\tnote{a}   &  0.09 & -0.18 \\
C  & 1.2621 &  0.0762(84)\tnote{*}  &            & 0.0031(43)\tnote{b}    &  0.07 & -0.11 \\
O  & 1.4620 &  0.0505(86)           &            & 0.0327(53)\tnote{c}    &  0.06 & -0.22 \\
F  & 3.4013 &  0.222(18)\tnote{*}   & -0.033(20) & 0.0474(49)\tnote{d}    & -0.08 & -0.13 \\ [1ex] 
\hline
\end{tabular}
\begin{tablenotes}
\footnotesize 
\item[*] PES
\item[a] QMC trial from CASSCF(3/4,8) with minimum CI coefficient of 0.01 (34/46 determinants) 
\item[b] QMC trial from CASSCF(4/5,8) with minimum CI coefficient of 0.01 (30/53 determinants) 
\item[c] QMC trial from CASSCF(6/7,8) with minimum CI coefficient of 0.01 (39/77 determinants) 
\item[d] QMC trial from CASSCF(7/8,16) with minimum CI coefficient of 0.0075 (145/243 determinants) 
\end{tablenotes}
\label{table:G2 EA}
\end{threeparttable}
\end{table*}

For the calculations which used the unrestricted HF determinant as the trial function, PES was used for the IPs of B, C, N, and O with equilibration times of 35, 20, 15, and 15 $Ha^{-1}$, respectively, and similarly for the EAs of C and F with 15 and 10 $Ha^{-1}$, respectively.  For the remaining species, equilibration was facile and thus the AFs were correlated from the beginning of the imaginary-time propagation.  Notable deviations from the experimental values are found for the IPs of B and N, and the EA of F when the UHF state is used as the trial function (nearly identical errors have been previously reported within ph-AFQMC\cite{al2006gaussian}).  These discrepancies are resolved in the correlated sampling-based FP results.  Since the energies from FP are not biased by the phaseless constraint and thus insensitive to the trial function used, (any) small residual errors can be attributed to the MP2-assisted CBS extrapolation scheme.    

The long equilibration times and inaccuracies encountered in the above cases are manifestations of the fact that single-determinant trial functions obtained from mean-field calculations are generally not well-suited to describe the open-shell systems involved in redox reactions.  For instance in the IP of B, comparing FP and ph-AFQMC calculations for both the neutral and cationic species exposes the fact that the error in the IP stems from inaccuracies in the computation of the total energy of the neutral B atom, which has a single unpaired electron in the triply-degenerate $p$ orbital manifold.  The use of trial wavefunctions with proper symmetry properties has previously been shown to lead to improved accuracy within the phaseless approximation,\cite{shi2013symmetry} so we now consider trial functions which are eigenfunctions of $\hat{S}^2$ for use with the phaseless constraint.  While employing a restricted open-shell HF trial affords no appreciable gain in accuracy, the use of a truncated CASSCF trial with a very modest active space size and a small number of determinants is sufficient to eliminate most of the error for the IPs of both B and N. 

The accurate description of the EA of F proves to be more demanding.  It is possible that the phase problem is particularly problematic for the case of F- due to the small atomic radius which may lead to very strong electronic correlations.  Such sizable correlations manifest themselves in the algorithm as a large imaginary component of the propagator, which in turn leads to the elimination of crucial physical information during the projection to the real axis if the trial function does not adequately provide the gauge information on the Slater determinants of the ground state.  In such cases, excitations into a large number of virtual orbitals will make significant contributions to the correlation energy, and therefore a large active space is required to generate the CASSCF trial function so that the resulting phase projections are sufficiently benign.  

In general, the implementation and efficacy of the correlated sampling scheme presented in this work remain unaltered in the case of a multi-determinant trial function since walker determinants are propagated in exactly the same manner.  We do find a reduction in the required equilibration times for all multi-determinant ph-AFQMC calculations that we have performed, relative to single-determinant calculations of the same systems, rendering the use of PES unnecessary.  In addition, the use of multi-determinant trial functions results in drastically smaller error bars, even when more than 4x \emph{fewer} walkers are used.  Increasing the quality of the CASSCF trial function is a promising way to systematically reduce the error from the phaseless constraint, and we find that in all cases the use of a reasonable number of determinants produces redox energies within the maximum experimental error of 0.05 eV for the G2 Test Set.

\subsection{The Case of Methanol}

Motivated by the arguments set forth in Section II.C, we now describe the details of a composite method for calculating the \emph{adiabatic} IPs, deprotonation free-energies, and hydrogen-dissociation energies of molecular systems, and illustrate the accuracy of our approach on the case of methanol.  We utilize a stepwise process consisting of:  1) the fixed-geometry process calculated with correlated sampling-based ph-AFQMC, and 2) a geometry-relaxation step calculated within MP2.  Optimal correlation can be achieved in 1) since the same set of basis functions is used, while in 2) we expect large error cancellation resulting from the fact that the initial and final ground-state geometries are typically very similar in redox, deprotonation, and hydrogen abstraction reactions.  The calculations are performed in a triple-zeta basis with additional diffuse basis functions for all species involved in the deprotonation reaction.  The resulting energy differences from these two steps are added together, and the endpoints are extrapolated to the CBS limit.  

For all reaction types, zero-point energies are calculated at the level of HF/6-31G* and scaled by a factor of 0.899 to account for anharmonicity and the known shortcomings of HF theory, following the G2 protocol.  Deprotonation free-energy results incorporate the value of -6.28 kcal/mol as the free-energy of a proton at 298 K,\cite{tawa1998calculation} and we use the exact ground-state energy of H$\cdot$ (-0.5 Ha) to calculate the bond dissociation enthalpy of the O-H bond. 

Table \ref{table:MeOH} illustrates the accuracy of our correlated sampling protocol with respect to experimental results for all three reaction types.  The quality of our IP and deprotonation free-energy results surpass that of the more costly G2 method, while the O-H bond dissociation result has an error of comparable magnitude.  We note that the single-determinant trial wavefunction is sufficient to produce a near-exact deprotonation free-energy, which we attribute to the fact that in this reaction type both the protonated and deprotonated species are closed-shell.  Moreover, our correlated sampling-based ph-AFQMC calculations, using an extremely modest number of walkers, predicts the energy differences corresponding to all of these reaction types to within chemical accuracy, which is not the case for the G2 method.        

\begin{table*}[ht]
\centering
\begin{threeparttable}
\caption{Adiabatic Reaction Energies for Methanol.  Experimental values and the deviations of the correlated sampling-based ph-AFQMC and G2 results (theory - experiment) in eV.  QMC statistical errors in the two right-most digits are shown in parenthesis.  All QMC calculations use 192 walkers per repeat.}
\begin{tabular}{c @{\hskip 0.1in} c c @{\hskip 0.2in} c}
\hline\hline
Reaction & Expt. & $\Delta$ph-QMC & $\Delta$G2 \\ [0.5ex] 
\hline
Ionization Potential   &   10.84             &  0.034(27)\tnote{$\dagger$}   &   -0.11      \\
Deprotonation Free-Energy & 16.2695 &  -0.005(21)\tnote{$\dagger\dagger$}  & 0.0484\tnote{a} \\ 
O-H Bond Dissociation Energy & 4.5359 &  0.039(14)\tnote{$\dagger$}  & 0.0166\tnote{b}     \\ [1ex] 
\hline
\end{tabular}
\begin{tablenotes}
\footnotesize 
\item[$\dagger$] QMC trial from CASSCF(10/11, 10/11) with minimum CI coefficient of 0.02 (27/32 determinants)
\item[$\dagger\dagger$] HF trial
\item[a] Ref. \cite{merrill1996calculated}
\item[b] Ref. \cite{curtiss1991energies}
\end{tablenotes}
\label{table:MeOH}
\end{threeparttable}
\end{table*}

\subsection{Basis Set Size, Number of Random Walkers, and CPU-time Reduction}
\subsubsection{Basis Set Size}
The statistical noise of individual AFQMC runs is expected to increase with the number of AFs, $\alpha$, yet the precise scaling is subtle.  Each AF contributes additional noise.  On the other hand, the magnitude of the contribution is moderated by a degree which depends on the form of the interaction and the decomposition procedure leading to Eq. \eqref{prop}.  For example, the error bar is seen to change little beyond a modest cutoff in plane-wave AFQMC.\cite{suewattana2007phaseless}  In principle $\al$ grows as $M^2$, but our (rather conservative) truncation of the Cholesky decomposition via the cutoff mentioned in Section II.D results in $\al \sim10M$.  As a result, we expect the statistical error in an uncorrelated calculation to increase with $M$ before saturating toward the CBS limit.  This could be problematic given that large basis sets containing functions associated with high angular momenta are frequently required to accurately describe the correlation energy,\cite{dunning1989gaussian} and also given the fact that most interesting applications involve large systems.  

\begin{figure}[h!]
\centering
{\includegraphics[width=0.45\textwidth]{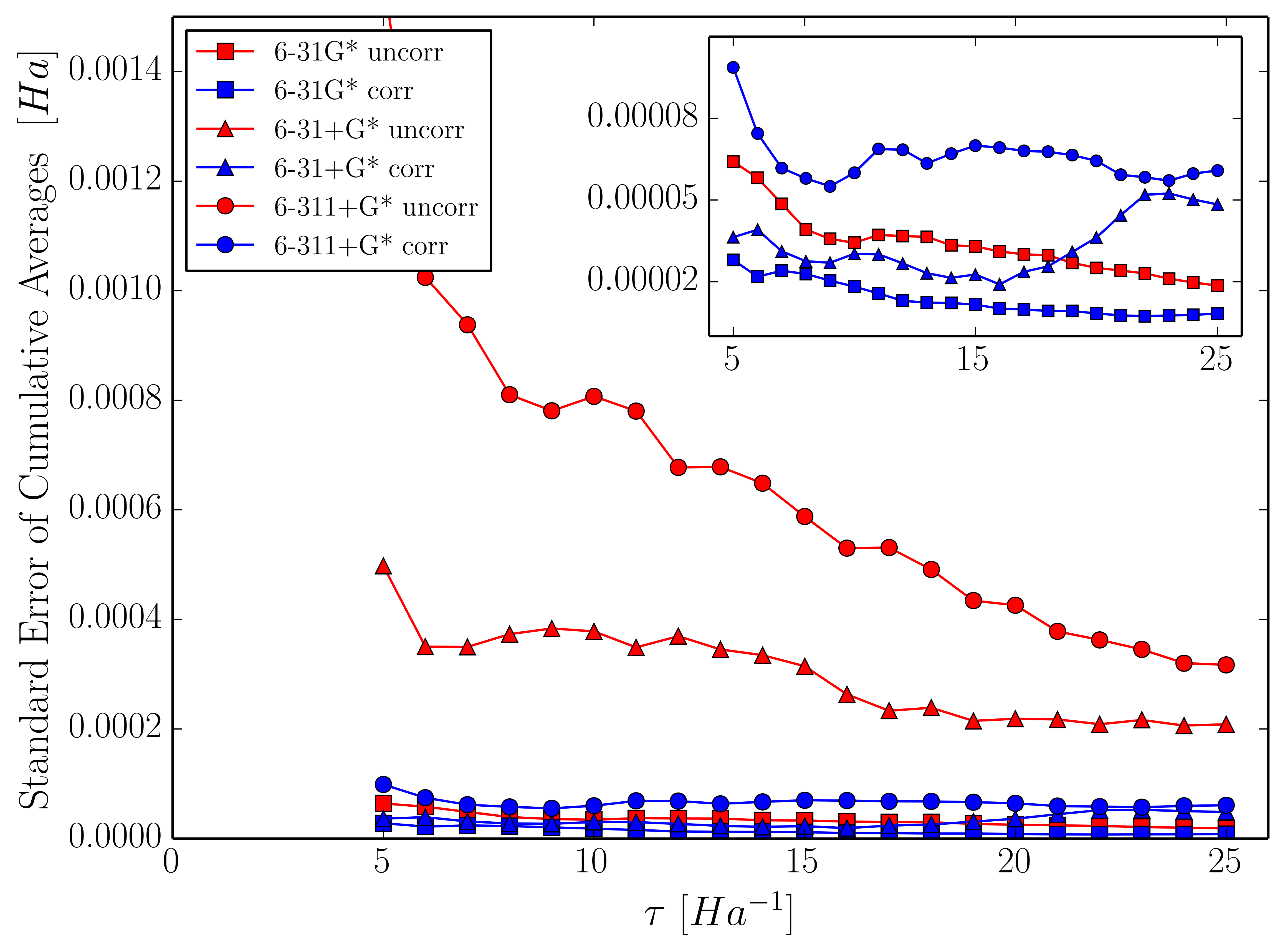}}
\caption{Comparison of the standard errors of the cumulative averages resulting from correlated and uncorrelated sampling in computing the IP of the K atom in three different basis sets with 5040 walkers per repeat.  The inset zooms in on the lower region, for clarity.}
\label{fig:basis_IP}
\end{figure}

\begin{figure}[h!]
\centering
{\includegraphics[width=0.45\textwidth]{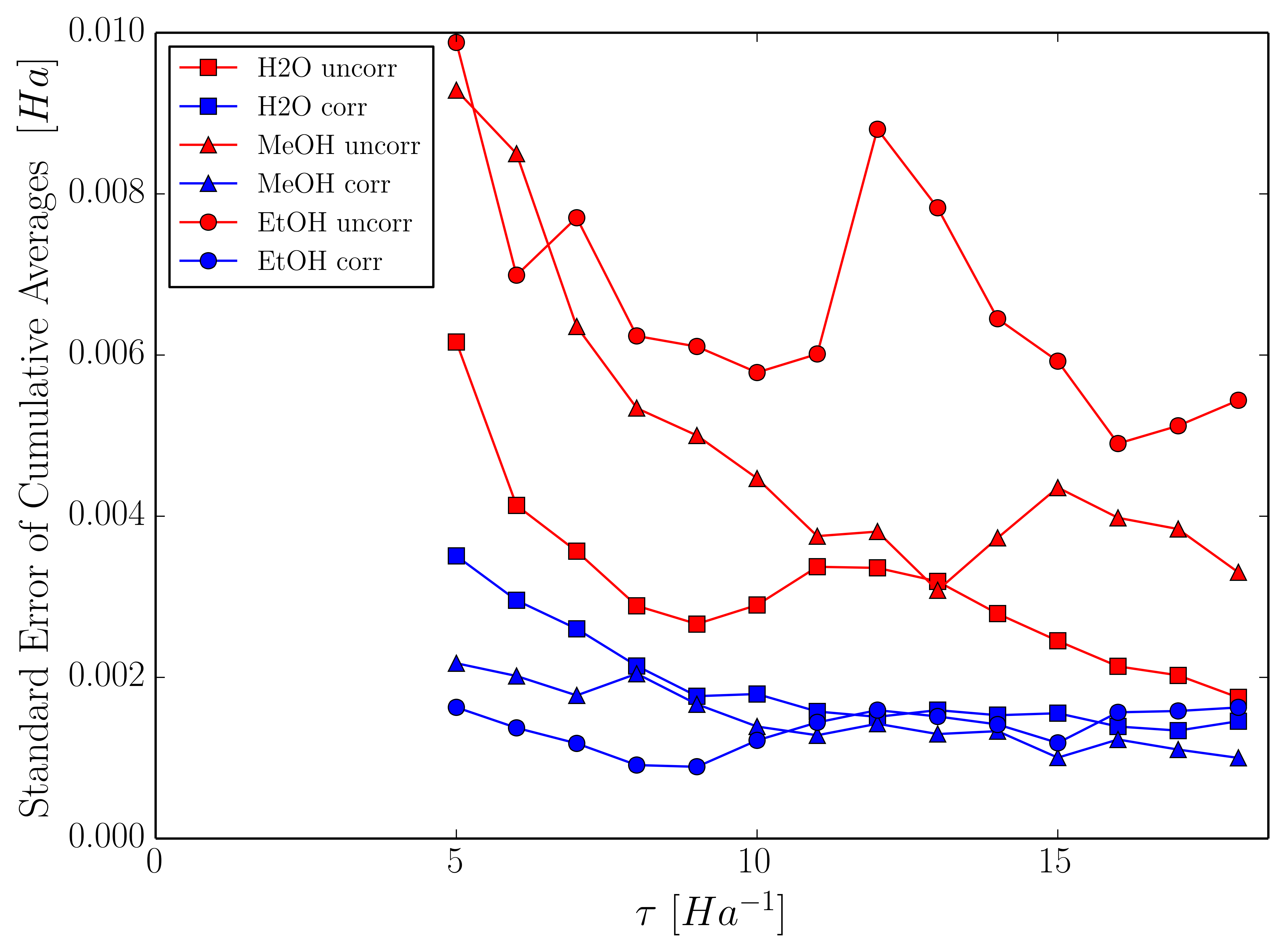}}
\caption{Comparison of the errors resulting from correlated and uncorrelated sampling in computing the fixed-geometry deprotonation energies of H$_2$O (M=58), CH$_3$OH (M=116), and C$_2$H$_5$OH (M=174) with 192 walkers per repeat. }
\label{fig:basis_deprot}
\end{figure}

Here we show that the magnitude of the reduction in statistical error enabled by the use of correlated sampling grows with $M$, such that the error is nearly independent of $M$.  Fig. \ref{fig:basis_IP} shows the standard errors resulting from calculations of the IP of the K in the 6-31G*, 6-31+G*, and 6-311+G* basis sets (which consist of 23, 35, and 45 basis functions, respectively); Fig. \ref{fig:basis_deprot} illustrates the same effect for fixed-geometry deprotonation reactions of water, methanol, and ethanol (58, 116, and 174 basis functions).  For both reaction types, while the errors from the uncorrelated calculations increase significantly with $M$, we find that those resulting from correlated sampling remained roughly constant.  We anticipate that this finding will hold in general, and will be of crucial importance in future applications of correlated sampling-based AFQMC to larger molecules.    
 
\subsubsection{Reduction in the Number of Random Walkers and CPU-Time}  
In an uncorrelated QMC calculation for a fixed length of propagation time, the resulting standard error can be reduced by increasing the number of random walkers, $N_{wlk}$, used in the MC evaluation of the HS integral in Eq. \eqref{prop}.  However, given that the required computational expense increases linearly with $N_{wlk}$, using a brute-force approach that simply increases $N_{wlk}$ is less practical for many systems.  For energy differences, correlated sampling provides a much cheaper alternative as it allows for a dramatic reduction in the $N_{wlk}$ required to achieve a given statistical error.

\begin{figure}[h!]
\centering
{\includegraphics[width=0.45\textwidth]{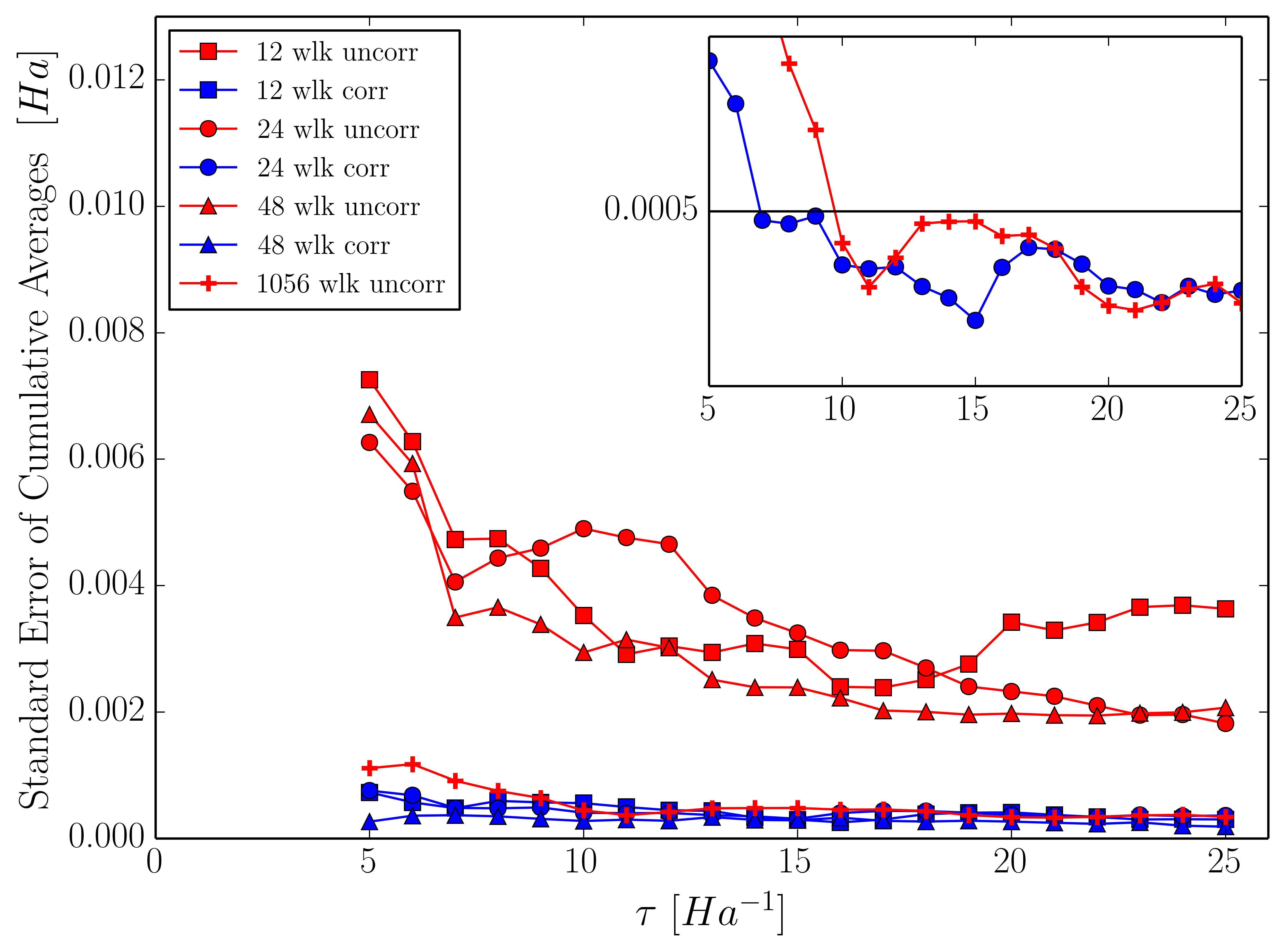}}
\caption{Dependence of standard error on the number of random walkers per repeat for the IP of K in the 6-31+G* basis.  The inset highlights the errors of the uncorrelated run with 1056 walkers and the correlated run with 24 walkers, compared with the 0.5 mHa error target.}
\label{fig:Nwlk_K}
\end{figure}

\begin{figure}[h!]
\centering
{\includegraphics[width=0.45\textwidth]{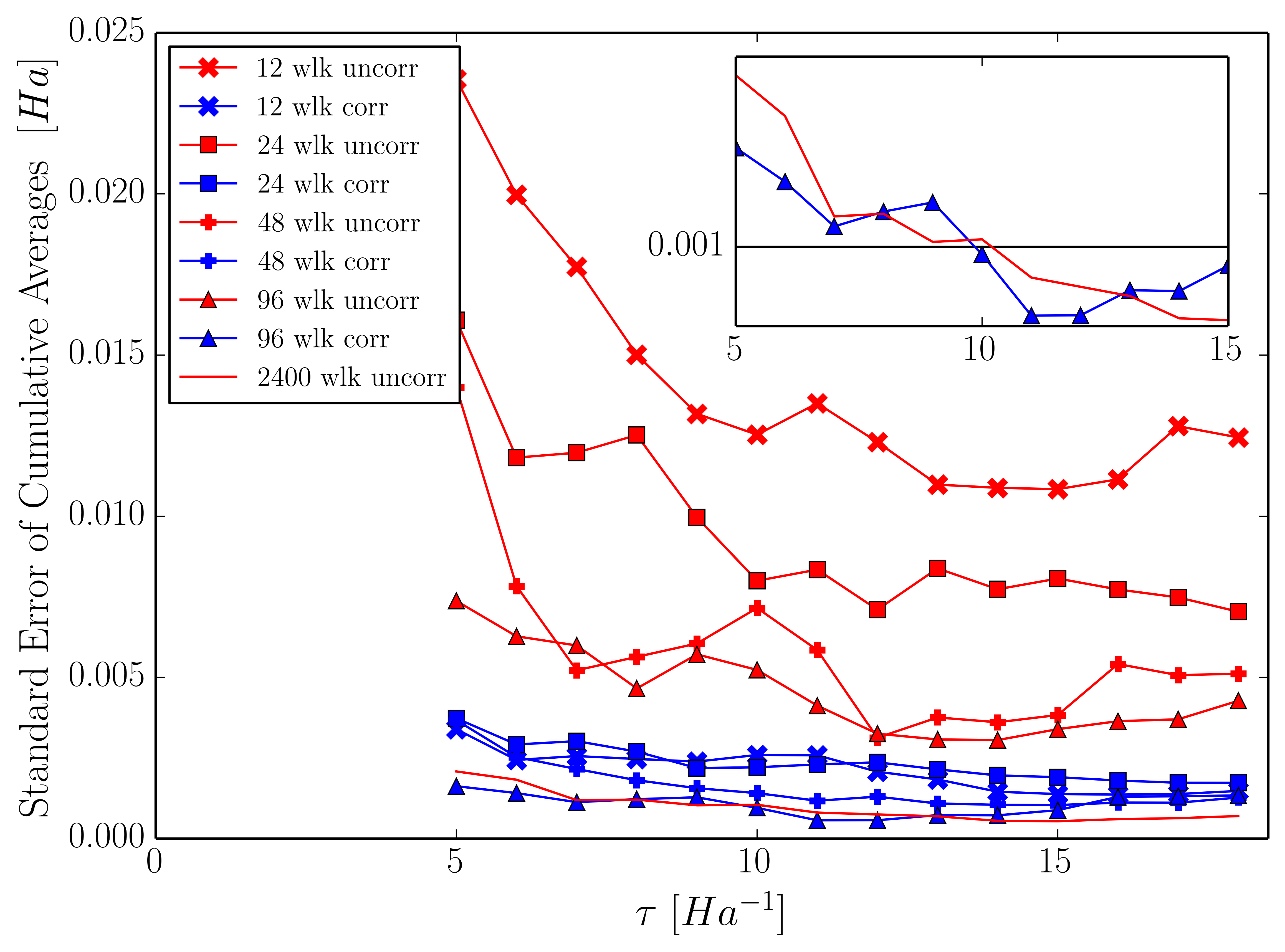}}
\caption{Dependence of standard error on the number of random walkers per repeat for the fixed-geometry deprotonation of methanol in the cc-pVTZ basis.  The inset highlights the errors of the uncorrelated run with 2400 walkers and the correlated run with 96 walkers, compared with the 1 mHa error target.}
\label{fig:Nwlk_MeOH}
\end{figure}

The errors of the cumulative averages of the IP of K in the 6-31+G* basis (M=35) and the fixed-geometry deprotonation of methanol in the cc-pVTZ basis (M=116) are shown for different values of $N_{wlk}$ in Figs. \ref{fig:Nwlk_K} and \ref{fig:Nwlk_MeOH}, respectively.  For both reaction types, we make the following observations:  First, for a given $N_{wlk}$, the standard error is significantly lower in the correlated sampling case.  Second, the magnitude of this reduction is greater for smaller $N_{wlk}$.  Finally, while the standard errors of the uncorrelated runs increase as $N_{wlk}$ is reduced, in the correlated sampling runs the error is relatively insensitive to $N_{wlk}$.

In light of these findings, we are in a position to understand how a reduction in the standard error due to correlating the AFs translates into a reduction in CPU-time.  Considering first the IP of K in the 6-31+G* basis, we choose a target standard error of 0.5 mHa on the QMC energy (which defines the 99$\%$ confidence interval as the cumulative average of the energy $\pm$ 1 kcal/mol), and compare the total CPU-time required to achieve this via correlated and uncorrelated approaches.  In the former case, we find that using only 24 walkers in each of the 11 repeat calculations is sufficient to achieve the target error and a resulting energy in agreement (\emph{i.e.} within the 99$\%$ confidence interval) with a benchmark result obtained with 5040 walkers.  In fact, as few as 6 walkers produced the same level of accuracy in some cases.  The inset of Fig. \ref{fig:Nwlk_K} shows that the statistical error falls below the target at $\tau \sim 7$. The total CPU-time required to propagate 11 repeats for this length of imaginary-time is 41.2 minutes on a single 2.60 GHz Intel Xeon processor.  Using the same number of walkers \emph{without} correlating the AFs, we find that the target error is not reached even after 200 $Ha^{-1}$.  Using 1056 walkers gives rise to a standard error that falls below 0.5 mHa after 10 $Ha^{-1}$, as shown in the inset of Fig. \ref{fig:Nwlk_K}, and a resulting QMC energy that is in agreement with the benchmark result.  This calculation takes 2262.6 minutes on a single processor, and we thus conclude that our correlated sampling approach reduces the CPU-time by a factor of approximately 55.  

\begin{figure}[h!]
\centering
{\includegraphics[width=0.45\textwidth]{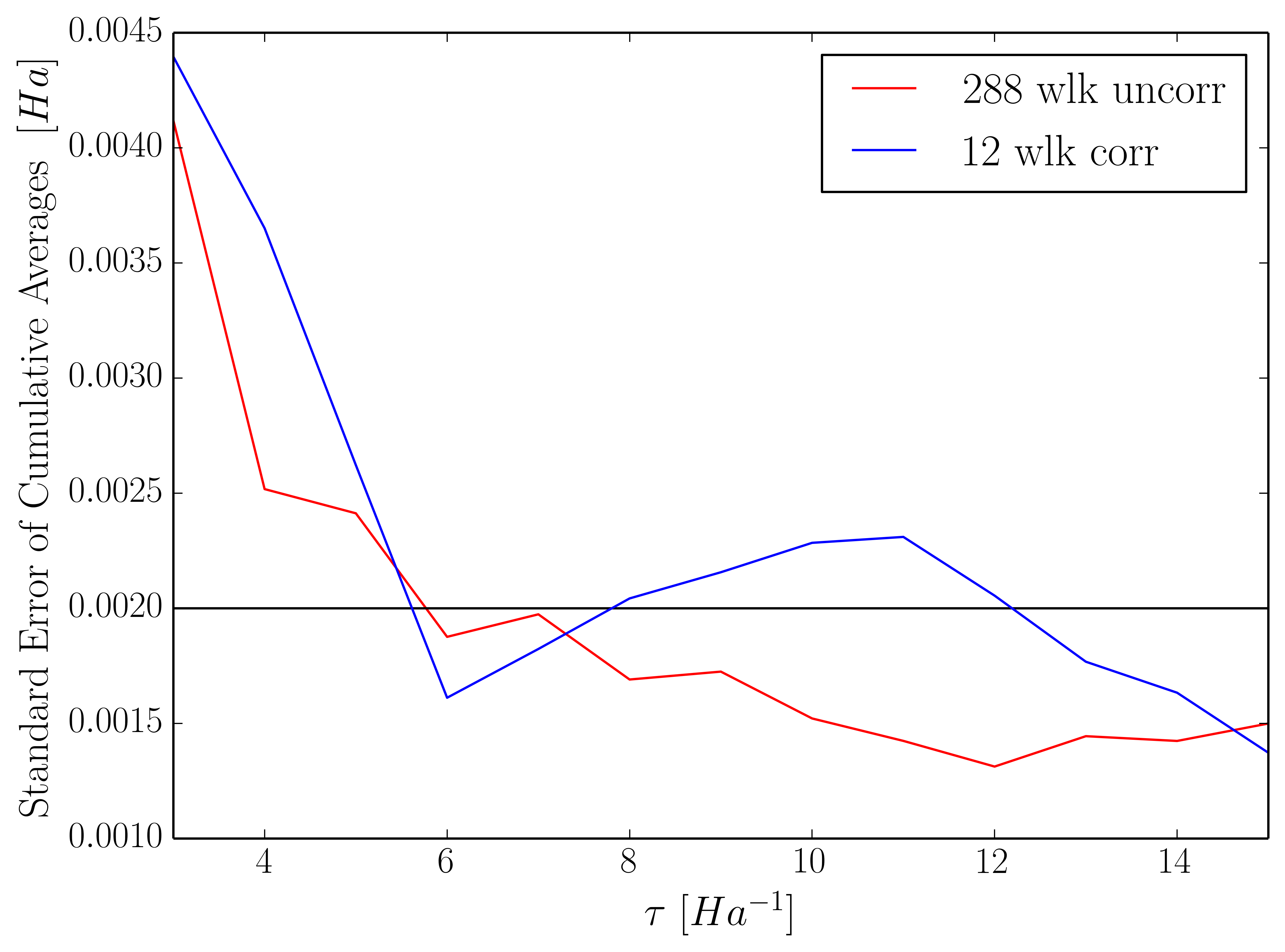}}
\caption{Comparison of the standard errors resulting from the use of correlated sampling with 12 walkers per repeat and uncorrelated sampling with 288 walkers per repeat to calculate the energy difference associated with the fixed-geometry removal of H$\cdot$ from the O-H bond of methanol in the cc-pVTZ basis.  The 2 mHa error target is shown in black.   }
\label{fig:MeOH_H_abstr_timing}
\end{figure}

For the deprotonation of methanol we use a target error of 1 mHa.  As shown in the inset of Fig. \ref{fig:Nwlk_MeOH}, both the uncorrelated run with 2400 walkers and the correlated run with 96 walkers yield results that fall below our target error at $\tau \sim 10$.  We use the fact that the total CPU-time is proportional to the product of the number of walkers and the propagation time to estimate that correlating the AFs reduces the CPU-time by a factor of approximately 25.  Currently the total calculation, including all 11 repeats, requires $\sim$154 hours on a single CPU core.  We perform a similar analysis for the dissociation of H$\cdot$ from the O-H bond of methanol.  Due to the relatively large computational cost of using a CASSCF trial function (with the same active space and truncation scheme described in Table III), we increase our target error to 2 mHa.  Figure \ref{fig:MeOH_H_abstr_timing} shows that the errors on the cumulative averages from both an uncorrelated run with 288 walkers and a correlated run with 12 walkers fall below our target error at $\tau \sim 6$.  Thus we deduce a speed-up factor of 24 for this H abstraction reaction.  On a single CPU core this requires $\sim$157 hours. 

	As the previous sections have shown, the magnitude of the reduction in the statistical error, and consequently the CPU-time, as compared with an uncorrelated calculation depends on the number of walkers used and the size of the basis set.  In addition, one intuitively expects our correlated sampling approach to work better for systems that more closely resemble each other.  This is crudely the case, yet the subtleties involved in such a claim warrant further discussion.  Indeed, referring to the propagator in \eqref{full_propagator}, while our correlated sampling method ensures that the $\mbf{\hat{v}}$ operators and the AFs, $\mbf{x}$, are the same for both the primary and secondary systems, the FBs, $\mbf{\bar{x}}$, as defined in \eqref{optFB} and the expectation values with respect to the trial functions $\langle \mbf{\hat{v}} \rangle$ will in general be different.  In the limit that the primary and secondary systems are identical, the entire propagator in \eqref{full_propagator} is identical for both systems, and the statistical error in the energy difference will be exactly and trivially zero as a result of perfect walker-pair correlation.  Otherwise, the reduction in statistical error afforded by our correlated sampling approach becomes less pronounced the more the trial wavefunctions of the primary and secondary systems differ, since the the FBs and $\langle \mbf{\hat{v}} \rangle$ are expectation values that depend explicitly on the trial wavefunctions.  It is encouraging to note that despite the differences in trial functions correlating only the AFs yields such large speed-ups in CPU-time.  While additionally correlating the $\langle \mbf{\hat{v}} \rangle$, possibly by using some combination of the trial functions for both systems, would compromise the accuracy of the phaseless approximations, future studies will explore optimal ways to pair walkers such that the similarity in the FBs of walker pairs is maximized.  It is encouraging that even for systems for which the \emph{electronic} energies of the primary and secondary systems differ by some 200 eV, as is the case in the deprotonation of methanol, our correlated sampling approach still yields dramatic efficiency gains.  

\section{Conclusions and Outlook} 

	In this work we have devised a correlated sampling protocol for the calculation of chemically relevant energy differences within the exact and phaseless variants of AFQMC.  For molecules we utilize a two step strategy in which optimal walker-pair correlation is achieved in the ph-AFQMC description of the fixed-geometry process,  while the geometry relaxation energy is calculated with the confines of MP2.  Together with an MP2-assisted CBS extrapolation method we obtain calculated IPs, EAs, deprotonation free-energies, and bond dissociation energies that are in excellent agreement with experiments.  Moreover, our correlated sampling approach yields large reductions in the statistical errors relative to those obtained from uncorrelated approaches.  In contrast to uncorrelated AFQMC, where the error bars are found to increase with system size and/or the number of basis functions, correlating the AFs keeps the statistical error relatively constant as chemical complexity increases.  In addition, our approach drastically reduces the number of walkers required to achieve a given statistical error target, which translates into large reductions in CPU-time.  

	We utilize a ``ghost" basis function strategy that enables the application of a correlated sampling-based approach to processes involving large energetic changes. Indeed, given that the correlated sampling scheme outlined here is successful for electron, proton, and H$\cdot$ transfer reactions, we are optimistic that other chemical changes are within reach.  In a future work we will systematically investigate chemical reactions which involve substantial changes in geometry, including the addition/removal of larger, more complex functional groups.  Along the same lines we are optimistic about the savings that our correlated sampling approach may yield when basis functions that are independent of the nuclear coordinates, such as plane-waves, are used.  We anticipate that the insensitivity of the statistical error to basis set size that we observe in this work may partially or totally offset the relatively large number of plane-waves typically required for convergence.
    
	Additional improvements in computational efficiency are necessary before our correlated sampling-based AFQMC approach can be routinely applied to the large, complex systems that we ultimately hope to investigate.  In addition to incorporating orbital localization techniques and experimenting with more compact basis sets, we will pursue a variety of algorithmic (\emph{e.g.} the implementation of mixed-precision and sparse linear algebra routines) and hardware (\emph{e.g.} porting the code to run on GPUs) optimizations.  Future targets of investigation include the IP of B$_2$F$_4$,\cite{chan2012comment,li2002b2f4} redox potentials and pKa's of TM clusters and battery electrolytes,\cite{coskun2016evaluation,jerome2014accurate,roy2009calculation,konezny2012reduction,baik2002computing,borodin2013oxidative,shao2012oxidation,qu2015electrolyte} and the relative energies of low-lying states of interacting TM centers such as Fe-S complexes and the active site of Photosystem II.\cite{sharma2014low,ames2011theoretical,siegbahn2000transition}

\section{Acknowledgements}
JS gratefully acknowledge Fengjie Ma, Mario Motta, Hao Shi, and James Dama for their generous help and guidance in implementing and understanding AFQMC methods.  DRR acknowledges funding from grant NSF-CHE-1464802.  The work of SZ was supported by grants DE-SC0001303 and DE-SC0008627.

\bibliographystyle{apsrev4-1}
\bibliography{References}

\begin{thebibliography}{121}%
\makeatletter
\providecommand \@ifxundefined [1]{%
 \@ifx{#1\undefined}
}%
\providecommand \@ifnum [1]{%
 \ifnum #1\expandafter \@firstoftwo
 \else \expandafter \@secondoftwo
 \fi
}%
\providecommand \@ifx [1]{%
 \ifx #1\expandafter \@firstoftwo
 \else \expandafter \@secondoftwo
 \fi
}%
\providecommand \natexlab [1]{#1}%
\providecommand \enquote  [1]{``#1''}%
\providecommand \bibnamefont  [1]{#1}%
\providecommand \bibfnamefont [1]{#1}%
\providecommand \citenamefont [1]{#1}%
\providecommand \href@noop [0]{\@secondoftwo}%
\providecommand \href [0]{\begingroup \@sanitize@url \@href}%
\providecommand \@href[1]{\@@startlink{#1}\@@href}%
\providecommand \@@href[1]{\endgroup#1\@@endlink}%
\providecommand \@sanitize@url [0]{\catcode `\\12\catcode `\$12\catcode
  `\&12\catcode `\#12\catcode `\^12\catcode `\_12\catcode `\%12\relax}%
\providecommand \@@startlink[1]{}%
\providecommand \@@endlink[0]{}%
\providecommand \url  [0]{\begingroup\@sanitize@url \@url }%
\providecommand \@url [1]{\endgroup\@href {#1}{\urlprefix }}%
\providecommand \urlprefix  [0]{URL }%
\providecommand \Eprint [0]{\href }%
\providecommand \doibase [0]{http://dx.doi.org/}%
\providecommand \selectlanguage [0]{\@gobble}%
\providecommand \bibinfo  [0]{\@secondoftwo}%
\providecommand \bibfield  [0]{\@secondoftwo}%
\providecommand \translation [1]{[#1]}%
\providecommand \BibitemOpen [0]{}%
\providecommand \bibitemStop [0]{}%
\providecommand \bibitemNoStop [0]{.\EOS\space}%
\providecommand \EOS [0]{\spacefactor3000\relax}%
\providecommand \BibitemShut  [1]{\csname bibitem#1\endcsname}%
\let\auto@bib@innerbib\@empty
\bibitem [{\citenamefont {Peterson}\ \emph {et~al.}(2012)\citenamefont
  {Peterson}, \citenamefont {Feller},\ and\ \citenamefont
  {Dixon}}]{peterson2012chemical}%
  \BibitemOpen
  \bibfield  {author} {\bibinfo {author} {\bibfnamefont {K.~A.}\ \bibnamefont
  {Peterson}}, \bibinfo {author} {\bibfnamefont {D.}~\bibnamefont {Feller}}, \
  and\ \bibinfo {author} {\bibfnamefont {D.~A.}\ \bibnamefont {Dixon}},\
  }\href@noop {} {\bibfield  {journal} {\bibinfo  {journal} {Theor. Chem.
  Acc.}\ }\textbf {\bibinfo {volume} {131}},\ \bibinfo {pages} {1} (\bibinfo
  {year} {2012})}\BibitemShut {NoStop}%
\bibitem [{\citenamefont {Friesner}(2005)}]{friesner2005ab}%
  \BibitemOpen
  \bibfield  {author} {\bibinfo {author} {\bibfnamefont {R.~A.}\ \bibnamefont
  {Friesner}},\ }\href@noop {} {\bibfield  {journal} {\bibinfo  {journal}
  {Proc. Natl. Acad. Sci. USA}\ }\textbf {\bibinfo {volume} {102}},\ \bibinfo
  {pages} {6648} (\bibinfo {year} {2005})}\BibitemShut {NoStop}%
\bibitem [{\citenamefont {Langhoff}(2012)}]{langhoff2012quantum}%
  \BibitemOpen
  \bibfield  {author} {\bibinfo {author} {\bibfnamefont {S.}~\bibnamefont
  {Langhoff}},\ }\href@noop {} {\emph {\bibinfo {title} {Quantum mechanical
  electronic structure calculations with chemical accuracy}}},\ Vol.~\bibinfo
  {volume} {13}\ (\bibinfo  {publisher} {Springer Science \& Business Media},\
  \bibinfo {year} {2012})\BibitemShut {NoStop}%
\bibitem [{\citenamefont {Sherrill}(2010)}]{sherrill2010frontiers}%
  \BibitemOpen
  \bibfield  {author} {\bibinfo {author} {\bibfnamefont {C.~D.}\ \bibnamefont
  {Sherrill}},\ }\href@noop {} {\bibfield  {journal} {\bibinfo  {journal} {J.
  Chem. Phys.}\ }\textbf {\bibinfo {volume} {132}},\ \bibinfo {pages} {110902}
  (\bibinfo {year} {2010})}\BibitemShut {NoStop}%
\bibitem [{\citenamefont {Harvey}(2006)}]{harvey2006accuracy}%
  \BibitemOpen
  \bibfield  {author} {\bibinfo {author} {\bibfnamefont {J.~N.}\ \bibnamefont
  {Harvey}},\ }\href@noop {} {\bibfield  {journal} {\bibinfo  {journal} {Annu.
  Rep. Prog. Chem., Sect. C: Phys. Chem.}\ }\textbf {\bibinfo {volume} {102}},\
  \bibinfo {pages} {203} (\bibinfo {year} {2006})}\BibitemShut {NoStop}%
\bibitem [{\citenamefont {Szabo}\ and\ \citenamefont
  {Ostlund}(1989)}]{szabo1989modern}%
  \BibitemOpen
  \bibfield  {author} {\bibinfo {author} {\bibfnamefont {A.}~\bibnamefont
  {Szabo}}\ and\ \bibinfo {author} {\bibfnamefont {N.~S.}\ \bibnamefont
  {Ostlund}},\ }\href@noop {} {\emph {\bibinfo {title} {Modern quantum
  chemistry: introduction to advanced electronic structure theory}}}\ (\bibinfo
   {publisher} {Courier Corporation},\ \bibinfo {year} {1989})\BibitemShut
  {NoStop}%
\bibitem [{\citenamefont {Purvis~III}\ and\ \citenamefont
  {Bartlett}(1982)}]{purvis1982full}%
  \BibitemOpen
  \bibfield  {author} {\bibinfo {author} {\bibfnamefont {G.~D.}\ \bibnamefont
  {Purvis~III}}\ and\ \bibinfo {author} {\bibfnamefont {R.~J.}\ \bibnamefont
  {Bartlett}},\ }\href@noop {} {\bibfield  {journal} {\bibinfo  {journal} {J.
  Chem. Phys.}\ }\textbf {\bibinfo {volume} {76}},\ \bibinfo {pages} {1910}
  (\bibinfo {year} {1982})}\BibitemShut {NoStop}%
\bibitem [{\citenamefont {Bartlett}\ and\ \citenamefont
  {Musia\l{}}(2007)}]{bartlett2007coupled}%
  \BibitemOpen
  \bibfield  {author} {\bibinfo {author} {\bibfnamefont {R.~J.}\ \bibnamefont
  {Bartlett}}\ and\ \bibinfo {author} {\bibfnamefont {M.}~\bibnamefont
  {Musia\l{}}},\ }\href {\doibase 10.1103/RevModPhys.79.291} {\bibfield
  {journal} {\bibinfo  {journal} {Rev. Mod. Phys.}\ }\textbf {\bibinfo {volume}
  {79}},\ \bibinfo {pages} {291} (\bibinfo {year} {2007})}\BibitemShut
  {NoStop}%
\bibitem [{\citenamefont {Purwanto}\ \emph {et~al.}(2015)\citenamefont
  {Purwanto}, \citenamefont {Zhang},\ and\ \citenamefont
  {Krakauer}}]{purwanto2015auxiliary}%
  \BibitemOpen
  \bibfield  {author} {\bibinfo {author} {\bibfnamefont {W.}~\bibnamefont
  {Purwanto}}, \bibinfo {author} {\bibfnamefont {S.}~\bibnamefont {Zhang}}, \
  and\ \bibinfo {author} {\bibfnamefont {H.}~\bibnamefont {Krakauer}},\
  }\href@noop {} {\bibfield  {journal} {\bibinfo  {journal} {J. Chem. Phys.}\
  }\textbf {\bibinfo {volume} {142}},\ \bibinfo {pages} {064302} (\bibinfo
  {year} {2015})}\BibitemShut {NoStop}%
\bibitem [{\citenamefont {Dutta}\ and\ \citenamefont
  {Sherrill}(2003)}]{dutta2003full}%
  \BibitemOpen
  \bibfield  {author} {\bibinfo {author} {\bibfnamefont {A.}~\bibnamefont
  {Dutta}}\ and\ \bibinfo {author} {\bibfnamefont {C.~D.}\ \bibnamefont
  {Sherrill}},\ }\href@noop {} {\bibfield  {journal} {\bibinfo  {journal} {J.
  Chem. Phys.}\ }\textbf {\bibinfo {volume} {118}},\ \bibinfo {pages} {1610}
  (\bibinfo {year} {2003})}\BibitemShut {NoStop}%
\bibitem [{\citenamefont {Roos}(1987)}]{roos1987complete}%
  \BibitemOpen
  \bibfield  {author} {\bibinfo {author} {\bibfnamefont {B.~O.}\ \bibnamefont
  {Roos}},\ }\href@noop {} {\bibfield  {journal} {\bibinfo  {journal} {Advances
  in Chemical Physics: Ab Initio Methods in Quantum Chemistry Part 2, Volume
  69}\ ,\ \bibinfo {pages} {399}} (\bibinfo {year} {1987})}\BibitemShut
  {NoStop}%
\bibitem [{\citenamefont {Olsen}(2011)}]{olsen2011casscf}%
  \BibitemOpen
  \bibfield  {author} {\bibinfo {author} {\bibfnamefont {J.}~\bibnamefont
  {Olsen}},\ }\href@noop {} {\bibfield  {journal} {\bibinfo  {journal} {Int. J.
  Quantum Chem.}\ }\textbf {\bibinfo {volume} {111}},\ \bibinfo {pages} {3267}
  (\bibinfo {year} {2011})}\BibitemShut {NoStop}%
\bibitem [{\citenamefont {Schmidt}\ and\ \citenamefont
  {Gordon}(1998)}]{schmidt1998construction}%
  \BibitemOpen
  \bibfield  {author} {\bibinfo {author} {\bibfnamefont {M.~W.}\ \bibnamefont
  {Schmidt}}\ and\ \bibinfo {author} {\bibfnamefont {M.~S.}\ \bibnamefont
  {Gordon}},\ }\href@noop {} {\bibfield  {journal} {\bibinfo  {journal} {Annu.
  Rev. Phys. Chem.}\ }\textbf {\bibinfo {volume} {49}},\ \bibinfo {pages} {233}
  (\bibinfo {year} {1998})}\BibitemShut {NoStop}%
\bibitem [{\citenamefont {Andersson}\ \emph {et~al.}(1990)\citenamefont
  {Andersson}, \citenamefont {Malmqvist}, \citenamefont {Roos}, \citenamefont
  {Sadlej},\ and\ \citenamefont {Wolinski}}]{andersson1990second}%
  \BibitemOpen
  \bibfield  {author} {\bibinfo {author} {\bibfnamefont {K.}~\bibnamefont
  {Andersson}}, \bibinfo {author} {\bibfnamefont {P.~A.}\ \bibnamefont
  {Malmqvist}}, \bibinfo {author} {\bibfnamefont {B.~O.}\ \bibnamefont {Roos}},
  \bibinfo {author} {\bibfnamefont {A.~J.}\ \bibnamefont {Sadlej}}, \ and\
  \bibinfo {author} {\bibfnamefont {K.}~\bibnamefont {Wolinski}},\ }\href@noop
  {} {\bibfield  {journal} {\bibinfo  {journal} {J. Phys. Chem.}\ }\textbf
  {\bibinfo {volume} {94}},\ \bibinfo {pages} {5483} (\bibinfo {year}
  {1990})}\BibitemShut {NoStop}%
\bibitem [{\citenamefont {Andersson}\ \emph {et~al.}(1992)\citenamefont
  {Andersson}, \citenamefont {Malmqvist},\ and\ \citenamefont
  {Roos}}]{andersson1992second}%
  \BibitemOpen
  \bibfield  {author} {\bibinfo {author} {\bibfnamefont {K.}~\bibnamefont
  {Andersson}}, \bibinfo {author} {\bibfnamefont {P.-{\AA}.}\ \bibnamefont
  {Malmqvist}}, \ and\ \bibinfo {author} {\bibfnamefont {B.~O.}\ \bibnamefont
  {Roos}},\ }\href@noop {} {\bibfield  {journal} {\bibinfo  {journal} {J. Chem.
  Phys.}\ }\textbf {\bibinfo {volume} {96}},\ \bibinfo {pages} {1218} (\bibinfo
  {year} {1992})}\BibitemShut {NoStop}%
\bibitem [{\citenamefont {Booth}\ \emph {et~al.}(2009)\citenamefont {Booth},
  \citenamefont {Thom},\ and\ \citenamefont {Alavi}}]{booth2009fermion}%
  \BibitemOpen
  \bibfield  {author} {\bibinfo {author} {\bibfnamefont {G.~H.}\ \bibnamefont
  {Booth}}, \bibinfo {author} {\bibfnamefont {A.~J.~W.}\ \bibnamefont {Thom}},
  \ and\ \bibinfo {author} {\bibfnamefont {A.}~\bibnamefont {Alavi}},\ }\href
  {\doibase 10.1063/1.3193710} {\bibfield  {journal} {\bibinfo  {journal} {J.
  Chem. Phys.}\ }\textbf {\bibinfo {volume} {131}},\ \bibinfo {pages} {054106}
  (\bibinfo {year} {2009})},\ \Eprint
  {http://arxiv.org/abs/http://aip.scitation.org/doi/pdf/10.1063/1.3193710}
  {http://aip.scitation.org/doi/pdf/10.1063/1.3193710} \BibitemShut {NoStop}%
\bibitem [{\citenamefont {Booth}\ \emph {et~al.}(2011)\citenamefont {Booth},
  \citenamefont {Cleland}, \citenamefont {Thom},\ and\ \citenamefont
  {Alavi}}]{booth2011breaking}%
  \BibitemOpen
  \bibfield  {author} {\bibinfo {author} {\bibfnamefont {G.~H.}\ \bibnamefont
  {Booth}}, \bibinfo {author} {\bibfnamefont {D.}~\bibnamefont {Cleland}},
  \bibinfo {author} {\bibfnamefont {A.~J.~W.}\ \bibnamefont {Thom}}, \ and\
  \bibinfo {author} {\bibfnamefont {A.}~\bibnamefont {Alavi}},\ }\href
  {\doibase 10.1063/1.3624383} {\bibfield  {journal} {\bibinfo  {journal} {J.
  Chem. Phys.}\ }\textbf {\bibinfo {volume} {135}},\ \bibinfo {pages} {084104}
  (\bibinfo {year} {2011})},\ \Eprint
  {http://arxiv.org/abs/http://dx.doi.org/10.1063/1.3624383}
  {http://dx.doi.org/10.1063/1.3624383} \BibitemShut {NoStop}%
\bibitem [{\citenamefont {Cleland}\ \emph {et~al.}(2011)\citenamefont
  {Cleland}, \citenamefont {Booth},\ and\ \citenamefont
  {Alavi}}]{cleland2011study}%
  \BibitemOpen
  \bibfield  {author} {\bibinfo {author} {\bibfnamefont {D.}~\bibnamefont
  {Cleland}}, \bibinfo {author} {\bibfnamefont {G.~H.}\ \bibnamefont {Booth}},
  \ and\ \bibinfo {author} {\bibfnamefont {A.}~\bibnamefont {Alavi}},\
  }\href@noop {} {\bibfield  {journal} {\bibinfo  {journal} {J. Chem. Phys.}\
  }\textbf {\bibinfo {volume} {134}},\ \bibinfo {pages} {024112} (\bibinfo
  {year} {2011})}\BibitemShut {NoStop}%
\bibitem [{\citenamefont {Cleland}\ \emph {et~al.}(2012)\citenamefont
  {Cleland}, \citenamefont {Booth}, \citenamefont {Overy},\ and\ \citenamefont
  {Alavi}}]{cleland2012taming}%
  \BibitemOpen
  \bibfield  {author} {\bibinfo {author} {\bibfnamefont {D.}~\bibnamefont
  {Cleland}}, \bibinfo {author} {\bibfnamefont {G.~H.}\ \bibnamefont {Booth}},
  \bibinfo {author} {\bibfnamefont {C.}~\bibnamefont {Overy}}, \ and\ \bibinfo
  {author} {\bibfnamefont {A.}~\bibnamefont {Alavi}},\ }\href@noop {}
  {\bibfield  {journal} {\bibinfo  {journal} {J. Chem. Theory Comput.}\
  }\textbf {\bibinfo {volume} {8}},\ \bibinfo {pages} {4138} (\bibinfo {year}
  {2012})}\BibitemShut {NoStop}%
\bibitem [{\citenamefont {Pierloot}(2011)}]{pierloot2011transition}%
  \BibitemOpen
  \bibfield  {author} {\bibinfo {author} {\bibfnamefont {K.}~\bibnamefont
  {Pierloot}},\ }\href@noop {} {\bibfield  {journal} {\bibinfo  {journal} {Int.
  J. Quantum Chem.}\ }\textbf {\bibinfo {volume} {111}},\ \bibinfo {pages}
  {3291} (\bibinfo {year} {2011})}\BibitemShut {NoStop}%
\bibitem [{\citenamefont {Thomas}\ \emph {et~al.}(2015)\citenamefont {Thomas},
  \citenamefont {Booth},\ and\ \citenamefont {Alavi}}]{thomas2015accurate}%
  \BibitemOpen
  \bibfield  {author} {\bibinfo {author} {\bibfnamefont {R.~E.}\ \bibnamefont
  {Thomas}}, \bibinfo {author} {\bibfnamefont {G.~H.}\ \bibnamefont {Booth}}, \
  and\ \bibinfo {author} {\bibfnamefont {A.}~\bibnamefont {Alavi}},\ }\href
  {\doibase 10.1103/PhysRevLett.114.033001} {\bibfield  {journal} {\bibinfo
  {journal} {Phys. Rev. Lett.}\ }\textbf {\bibinfo {volume} {114}},\ \bibinfo
  {pages} {033001} (\bibinfo {year} {2015})}\BibitemShut {NoStop}%
\bibitem [{\citenamefont {Chan}\ and\ \citenamefont
  {Sharma}(2011)}]{chan2011density}%
  \BibitemOpen
  \bibfield  {author} {\bibinfo {author} {\bibfnamefont {G.~K.-L.}\
  \bibnamefont {Chan}}\ and\ \bibinfo {author} {\bibfnamefont {S.}~\bibnamefont
  {Sharma}},\ }\href@noop {} {\bibfield  {journal} {\bibinfo  {journal} {Annu.
  Rev. Phys. Chem.}\ }\textbf {\bibinfo {volume} {62}},\ \bibinfo {pages} {465}
  (\bibinfo {year} {2011})}\BibitemShut {NoStop}%
\bibitem [{\citenamefont {Sharma}\ \emph {et~al.}(2014)\citenamefont {Sharma},
  \citenamefont {Sivalingam}, \citenamefont {Neese},\ and\ \citenamefont
  {Chan}}]{sharma2014low}%
  \BibitemOpen
  \bibfield  {author} {\bibinfo {author} {\bibfnamefont {S.}~\bibnamefont
  {Sharma}}, \bibinfo {author} {\bibfnamefont {K.}~\bibnamefont {Sivalingam}},
  \bibinfo {author} {\bibfnamefont {F.}~\bibnamefont {Neese}}, \ and\ \bibinfo
  {author} {\bibfnamefont {G.~K.-L.}\ \bibnamefont {Chan}},\ }\href@noop {}
  {\bibfield  {journal} {\bibinfo  {journal} {Nat. Chem.}\ }\textbf {\bibinfo
  {volume} {6}},\ \bibinfo {pages} {927} (\bibinfo {year} {2014})}\BibitemShut
  {NoStop}%
\bibitem [{\citenamefont {Kurashige}\ \emph {et~al.}(2013)\citenamefont
  {Kurashige}, \citenamefont {Chan},\ and\ \citenamefont
  {Yanai}}]{kurashige2013entangled}%
  \BibitemOpen
  \bibfield  {author} {\bibinfo {author} {\bibfnamefont {Y.}~\bibnamefont
  {Kurashige}}, \bibinfo {author} {\bibfnamefont {G.~K.-L.}\ \bibnamefont
  {Chan}}, \ and\ \bibinfo {author} {\bibfnamefont {T.}~\bibnamefont {Yanai}},\
  }\href@noop {} {\bibfield  {journal} {\bibinfo  {journal} {Nat. Chem.}\
  }\textbf {\bibinfo {volume} {5}},\ \bibinfo {pages} {660} (\bibinfo {year}
  {2013})}\BibitemShut {NoStop}%
\bibitem [{\citenamefont {Kohn}\ \emph {et~al.}(1996)\citenamefont {Kohn},
  \citenamefont {Becke},\ and\ \citenamefont {Parr}}]{kohn1996density}%
  \BibitemOpen
  \bibfield  {author} {\bibinfo {author} {\bibfnamefont {W.}~\bibnamefont
  {Kohn}}, \bibinfo {author} {\bibfnamefont {A.~D.}\ \bibnamefont {Becke}}, \
  and\ \bibinfo {author} {\bibfnamefont {R.~G.}\ \bibnamefont {Parr}},\
  }\href@noop {} {\bibfield  {journal} {\bibinfo  {journal} {J. Phys. Chem.}\
  }\textbf {\bibinfo {volume} {100}},\ \bibinfo {pages} {12974} (\bibinfo
  {year} {1996})}\BibitemShut {NoStop}%
\bibitem [{\citenamefont {Lester~Jr}\ and\ \citenamefont
  {Hammond}(1990)}]{lester1990quantum}%
  \BibitemOpen
  \bibfield  {author} {\bibinfo {author} {\bibfnamefont {W.~A.}\ \bibnamefont
  {Lester~Jr}}\ and\ \bibinfo {author} {\bibfnamefont {B.~L.}\ \bibnamefont
  {Hammond}},\ }\href@noop {} {\bibfield  {journal} {\bibinfo  {journal} {Annu.
  Rev. Phys. Chem.}\ }\textbf {\bibinfo {volume} {41}},\ \bibinfo {pages} {283}
  (\bibinfo {year} {1990})}\BibitemShut {NoStop}%
\bibitem [{\citenamefont {Hammond}\ \emph {et~al.}(1994)\citenamefont
  {Hammond}, \citenamefont {Lester~Jr},\ and\ \citenamefont
  {Reynolds}}]{hammond1994monte}%
  \BibitemOpen
  \bibfield  {author} {\bibinfo {author} {\bibfnamefont {B.~L.}\ \bibnamefont
  {Hammond}}, \bibinfo {author} {\bibfnamefont {W.~A.}\ \bibnamefont
  {Lester~Jr}}, \ and\ \bibinfo {author} {\bibfnamefont {P.~J.}\ \bibnamefont
  {Reynolds}},\ }\href@noop {} {\emph {\bibinfo {title} {Monte Carlo methods in
  ab initio quantum chemistry}}},\ Vol.~\bibinfo {volume} {1}\ (\bibinfo
  {publisher} {World Scientific},\ \bibinfo {year} {1994})\BibitemShut
  {NoStop}%
\bibitem [{\citenamefont {Foulkes}\ \emph {et~al.}(2001)\citenamefont
  {Foulkes}, \citenamefont {Mitas}, \citenamefont {Needs},\ and\ \citenamefont
  {Rajagopal}}]{RevModPhys.73.33}%
  \BibitemOpen
  \bibfield  {author} {\bibinfo {author} {\bibfnamefont {W.~M.~C.}\
  \bibnamefont {Foulkes}}, \bibinfo {author} {\bibfnamefont {L.}~\bibnamefont
  {Mitas}}, \bibinfo {author} {\bibfnamefont {R.~J.}\ \bibnamefont {Needs}}, \
  and\ \bibinfo {author} {\bibfnamefont {G.}~\bibnamefont {Rajagopal}},\ }\href
  {\doibase 10.1103/RevModPhys.73.33} {\bibfield  {journal} {\bibinfo
  {journal} {Rev. Mod. Phys.}\ }\textbf {\bibinfo {volume} {73}},\ \bibinfo
  {pages} {33} (\bibinfo {year} {2001})}\BibitemShut {NoStop}%
\bibitem [{\citenamefont {Zhang}(2013)}]{zhang2013}%
  \BibitemOpen
  \bibfield  {author} {\bibinfo {author} {\bibfnamefont {S.}~\bibnamefont
  {Zhang}},\ }\href@noop {} {\bibfield  {journal} {\bibinfo  {journal}
  {Emergent Phenomena in Correlated Matter}\ }\textbf {\bibinfo {volume} {3}}
  (\bibinfo {year} {2013})}\BibitemShut {NoStop}%
\bibitem [{\citenamefont {Zhang}\ and\ \citenamefont
  {Krakauer}(2003)}]{zhang2003quantum}%
  \BibitemOpen
  \bibfield  {author} {\bibinfo {author} {\bibfnamefont {S.}~\bibnamefont
  {Zhang}}\ and\ \bibinfo {author} {\bibfnamefont {H.}~\bibnamefont
  {Krakauer}},\ }\href {\doibase 10.1103/PhysRevLett.90.136401} {\bibfield
  {journal} {\bibinfo  {journal} {Phys. Rev. Lett.}\ }\textbf {\bibinfo
  {volume} {90}},\ \bibinfo {pages} {136401} (\bibinfo {year}
  {2003})}\BibitemShut {NoStop}%
\bibitem [{\citenamefont {Al-Saidi}\ \emph
  {et~al.}(2006{\natexlab{a}})\citenamefont {Al-Saidi}, \citenamefont
  {Krakauer},\ and\ \citenamefont {Zhang}}]{al2006d}%
  \BibitemOpen
  \bibfield  {author} {\bibinfo {author} {\bibfnamefont {W.}~\bibnamefont
  {Al-Saidi}}, \bibinfo {author} {\bibfnamefont {H.}~\bibnamefont {Krakauer}},
  \ and\ \bibinfo {author} {\bibfnamefont {S.}~\bibnamefont {Zhang}},\
  }\href@noop {} {\bibfield  {journal} {\bibinfo  {journal} {J. Chem. Phys.}\
  }\textbf {\bibinfo {volume} {125}},\ \bibinfo {pages} {154110} (\bibinfo
  {year} {2006}{\natexlab{a}})}\BibitemShut {NoStop}%
\bibitem [{\citenamefont {Virgus}\ \emph {et~al.}(2014)\citenamefont {Virgus},
  \citenamefont {Purwanto}, \citenamefont {Krakauer},\ and\ \citenamefont
  {Zhang}}]{virgus2014stability}%
  \BibitemOpen
  \bibfield  {author} {\bibinfo {author} {\bibfnamefont {Y.}~\bibnamefont
  {Virgus}}, \bibinfo {author} {\bibfnamefont {W.}~\bibnamefont {Purwanto}},
  \bibinfo {author} {\bibfnamefont {H.}~\bibnamefont {Krakauer}}, \ and\
  \bibinfo {author} {\bibfnamefont {S.}~\bibnamefont {Zhang}},\ }\href@noop {}
  {\bibfield  {journal} {\bibinfo  {journal} {Phys. Rev. Lett.}\ }\textbf
  {\bibinfo {volume} {113}},\ \bibinfo {pages} {175502} (\bibinfo {year}
  {2014})}\BibitemShut {NoStop}%
\bibitem [{\citenamefont {Al-Saidi}\ \emph
  {et~al.}(2006{\natexlab{b}})\citenamefont {Al-Saidi}, \citenamefont
  {Krakauer},\ and\ \citenamefont {Zhang}}]{al2006auxiliary}%
  \BibitemOpen
  \bibfield  {author} {\bibinfo {author} {\bibfnamefont {W.~A.}\ \bibnamefont
  {Al-Saidi}}, \bibinfo {author} {\bibfnamefont {H.}~\bibnamefont {Krakauer}},
  \ and\ \bibinfo {author} {\bibfnamefont {S.}~\bibnamefont {Zhang}},\ }\href
  {\doibase 10.1103/PhysRevB.73.075103} {\bibfield  {journal} {\bibinfo
  {journal} {Phys. Rev. B}\ }\textbf {\bibinfo {volume} {73}},\ \bibinfo
  {pages} {075103} (\bibinfo {year} {2006}{\natexlab{b}})}\BibitemShut
  {NoStop}%
\bibitem [{\citenamefont {Umrigar}(1989)}]{umrigar1989two}%
  \BibitemOpen
  \bibfield  {author} {\bibinfo {author} {\bibfnamefont {C.}~\bibnamefont
  {Umrigar}},\ }\href@noop {} {\bibfield  {journal} {\bibinfo  {journal} {Int.
  J. Quantum Chem.}\ }\textbf {\bibinfo {volume} {36}},\ \bibinfo {pages} {217}
  (\bibinfo {year} {1989})}\BibitemShut {NoStop}%
\bibitem [{\citenamefont {Traynor}\ and\ \citenamefont
  {Anderson}(1988)}]{traynor1988parallel}%
  \BibitemOpen
  \bibfield  {author} {\bibinfo {author} {\bibfnamefont {C.}~\bibnamefont
  {Traynor}}\ and\ \bibinfo {author} {\bibfnamefont {J.~B.}\ \bibnamefont
  {Anderson}},\ }\href@noop {} {\bibfield  {journal} {\bibinfo  {journal}
  {Chem. Phys. Lett.}\ }\textbf {\bibinfo {volume} {147}},\ \bibinfo {pages}
  {389} (\bibinfo {year} {1988})}\BibitemShut {NoStop}%
\bibitem [{\citenamefont {Wells}(1985)}]{wells1985differential}%
  \BibitemOpen
  \bibfield  {author} {\bibinfo {author} {\bibfnamefont {B.~H.}\ \bibnamefont
  {Wells}},\ }\href@noop {} {\bibfield  {journal} {\bibinfo  {journal} {Chem.
  Phys. Lett.}\ }\textbf {\bibinfo {volume} {115}},\ \bibinfo {pages} {89}
  (\bibinfo {year} {1985})}\BibitemShut {NoStop}%
\bibitem [{\citenamefont {Vrbik}\ \emph {et~al.}(1990)\citenamefont {Vrbik},
  \citenamefont {Legare},\ and\ \citenamefont
  {Rothstein}}]{vrbik1990infinitesimal}%
  \BibitemOpen
  \bibfield  {author} {\bibinfo {author} {\bibfnamefont {J.}~\bibnamefont
  {Vrbik}}, \bibinfo {author} {\bibfnamefont {D.~A.}\ \bibnamefont {Legare}}, \
  and\ \bibinfo {author} {\bibfnamefont {S.~M.}\ \bibnamefont {Rothstein}},\
  }\href@noop {} {\bibfield  {journal} {\bibinfo  {journal} {J. Chem. Phys.}\
  }\textbf {\bibinfo {volume} {92}},\ \bibinfo {pages} {1221} (\bibinfo {year}
  {1990})}\BibitemShut {NoStop}%
\bibitem [{\citenamefont {Filippi}\ and\ \citenamefont
  {Umrigar}(2000)}]{PhysRevB.61.R16291}%
  \BibitemOpen
  \bibfield  {author} {\bibinfo {author} {\bibfnamefont {C.}~\bibnamefont
  {Filippi}}\ and\ \bibinfo {author} {\bibfnamefont {C.~J.}\ \bibnamefont
  {Umrigar}},\ }\href {\doibase 10.1103/PhysRevB.61.R16291} {\bibfield
  {journal} {\bibinfo  {journal} {Phys. Rev. B}\ }\textbf {\bibinfo {volume}
  {61}},\ \bibinfo {pages} {R16291} (\bibinfo {year} {2000})}\BibitemShut
  {NoStop}%
\bibitem [{\citenamefont {Kwon}\ \emph {et~al.}(1994)\citenamefont {Kwon},
  \citenamefont {Ceperley},\ and\ \citenamefont {Martin}}]{kwon1994quantum}%
  \BibitemOpen
  \bibfield  {author} {\bibinfo {author} {\bibfnamefont {Y.}~\bibnamefont
  {Kwon}}, \bibinfo {author} {\bibfnamefont {D.~M.}\ \bibnamefont {Ceperley}},
  \ and\ \bibinfo {author} {\bibfnamefont {R.~M.}\ \bibnamefont {Martin}},\
  }\href {\doibase 10.1103/PhysRevB.50.1684} {\bibfield  {journal} {\bibinfo
  {journal} {Phys. Rev. B}\ }\textbf {\bibinfo {volume} {50}},\ \bibinfo
  {pages} {1684} (\bibinfo {year} {1994})}\BibitemShut {NoStop}%
\bibitem [{\citenamefont {Garmer}(1989)}]{garmer1989extrapolation}%
  \BibitemOpen
  \bibfield  {author} {\bibinfo {author} {\bibfnamefont {D.~R.}\ \bibnamefont
  {Garmer}},\ }\href@noop {} {\bibfield  {journal} {\bibinfo  {journal} {J.
  Comput. Chem.}\ }\textbf {\bibinfo {volume} {10}},\ \bibinfo {pages} {176}
  (\bibinfo {year} {1989})}\BibitemShut {NoStop}%
\bibitem [{\citenamefont {Nelson}\ \emph {et~al.}(2008)\citenamefont {Nelson},
  \citenamefont {Lehninger},\ and\ \citenamefont {Cox}}]{nelson2008lehninger}%
  \BibitemOpen
  \bibfield  {author} {\bibinfo {author} {\bibfnamefont {D.~L.}\ \bibnamefont
  {Nelson}}, \bibinfo {author} {\bibfnamefont {A.~L.}\ \bibnamefont
  {Lehninger}}, \ and\ \bibinfo {author} {\bibfnamefont {M.~M.}\ \bibnamefont
  {Cox}},\ }\href@noop {} {\emph {\bibinfo {title} {Lehninger principles of
  biochemistry}}}\ (\bibinfo  {publisher} {Macmillan},\ \bibinfo {year}
  {2008})\BibitemShut {NoStop}%
\bibitem [{\citenamefont {Siegbahn}\ and\ \citenamefont
  {Blomberg}(2000)}]{siegbahn2000transition}%
  \BibitemOpen
  \bibfield  {author} {\bibinfo {author} {\bibfnamefont {P.~E.}\ \bibnamefont
  {Siegbahn}}\ and\ \bibinfo {author} {\bibfnamefont {M.~R.}\ \bibnamefont
  {Blomberg}},\ }\href@noop {} {\bibfield  {journal} {\bibinfo  {journal}
  {Chem. Rev.}\ }\textbf {\bibinfo {volume} {100}},\ \bibinfo {pages} {421}
  (\bibinfo {year} {2000})}\BibitemShut {NoStop}%
\bibitem [{\citenamefont {Meunier}\ \emph {et~al.}(2004)\citenamefont
  {Meunier}, \citenamefont {De~Visser},\ and\ \citenamefont
  {Shaik}}]{meunier2004mechanism}%
  \BibitemOpen
  \bibfield  {author} {\bibinfo {author} {\bibfnamefont {B.}~\bibnamefont
  {Meunier}}, \bibinfo {author} {\bibfnamefont {S.~P.}\ \bibnamefont
  {De~Visser}}, \ and\ \bibinfo {author} {\bibfnamefont {S.}~\bibnamefont
  {Shaik}},\ }\href@noop {} {\bibfield  {journal} {\bibinfo  {journal} {Chem.
  Rev.}\ }\textbf {\bibinfo {volume} {104}},\ \bibinfo {pages} {3947} (\bibinfo
  {year} {2004})}\BibitemShut {NoStop}%
\bibitem [{\citenamefont {Palacin}(2009)}]{palacin2009recent}%
  \BibitemOpen
  \bibfield  {author} {\bibinfo {author} {\bibfnamefont {M.~R.}\ \bibnamefont
  {Palacin}},\ }\href@noop {} {\bibfield  {journal} {\bibinfo  {journal}
  {Chemical Society Reviews}\ }\textbf {\bibinfo {volume} {38}},\ \bibinfo
  {pages} {2565} (\bibinfo {year} {2009})}\BibitemShut {NoStop}%
\bibitem [{\citenamefont {Gratzel}(2012)}]{gratzel2012energy}%
  \BibitemOpen
  \bibfield  {author} {\bibinfo {author} {\bibfnamefont {M.}~\bibnamefont
  {Gratzel}},\ }\href@noop {} {\emph {\bibinfo {title} {Energy resources
  through photochemistry and catalysis}}}\ (\bibinfo  {publisher} {Elsevier},\
  \bibinfo {year} {2012})\BibitemShut {NoStop}%
\bibitem [{\citenamefont {Costentin}\ \emph {et~al.}(2013)\citenamefont
  {Costentin}, \citenamefont {Robert},\ and\ \citenamefont
  {Sav{\'e}ant}}]{costentin2013catalysis}%
  \BibitemOpen
  \bibfield  {author} {\bibinfo {author} {\bibfnamefont {C.}~\bibnamefont
  {Costentin}}, \bibinfo {author} {\bibfnamefont {M.}~\bibnamefont {Robert}}, \
  and\ \bibinfo {author} {\bibfnamefont {J.-M.}\ \bibnamefont {Sav{\'e}ant}},\
  }\href@noop {} {\bibfield  {journal} {\bibinfo  {journal} {Chemical Society
  Reviews}\ }\textbf {\bibinfo {volume} {42}},\ \bibinfo {pages} {2423}
  (\bibinfo {year} {2013})}\BibitemShut {NoStop}%
\bibitem [{\citenamefont {Alexov}\ \emph {et~al.}(2011)\citenamefont {Alexov},
  \citenamefont {Mehler}, \citenamefont {Baker}, \citenamefont {M~Baptista},
  \citenamefont {Huang}, \citenamefont {Milletti}, \citenamefont
  {Erik~Nielsen}, \citenamefont {Farrell}, \citenamefont {Carstensen},
  \citenamefont {Olsson}, \citenamefont {Shen}, \citenamefont {Warwicker},
  \citenamefont {Williams},\ and\ \citenamefont {Word}}]{alexov2011progress}%
  \BibitemOpen
  \bibfield  {author} {\bibinfo {author} {\bibfnamefont {E.}~\bibnamefont
  {Alexov}}, \bibinfo {author} {\bibfnamefont {E.~L.}\ \bibnamefont {Mehler}},
  \bibinfo {author} {\bibfnamefont {N.}~\bibnamefont {Baker}}, \bibinfo
  {author} {\bibfnamefont {A.}~\bibnamefont {M~Baptista}}, \bibinfo {author}
  {\bibfnamefont {Y.}~\bibnamefont {Huang}}, \bibinfo {author} {\bibfnamefont
  {F.}~\bibnamefont {Milletti}}, \bibinfo {author} {\bibfnamefont
  {J.}~\bibnamefont {Erik~Nielsen}}, \bibinfo {author} {\bibfnamefont
  {D.}~\bibnamefont {Farrell}}, \bibinfo {author} {\bibfnamefont
  {T.}~\bibnamefont {Carstensen}}, \bibinfo {author} {\bibfnamefont {M.~H.}\
  \bibnamefont {Olsson}}, \bibinfo {author} {\bibfnamefont {J.~K.}\
  \bibnamefont {Shen}}, \bibinfo {author} {\bibfnamefont {J.}~\bibnamefont
  {Warwicker}}, \bibinfo {author} {\bibfnamefont {S.}~\bibnamefont {Williams}},
  \ and\ \bibinfo {author} {\bibfnamefont {M.~J.}\ \bibnamefont {Word}},\
  }\href@noop {} {\bibfield  {journal} {\bibinfo  {journal} {Proteins: Struct.,
  Funct., Bioinf.}\ }\textbf {\bibinfo {volume} {79}},\ \bibinfo {pages} {3260}
  (\bibinfo {year} {2011})}\BibitemShut {NoStop}%
\bibitem [{\citenamefont {Jerome}\ \emph {et~al.}(2014)\citenamefont {Jerome},
  \citenamefont {Hughes},\ and\ \citenamefont {Friesner}}]{jerome2014accurate}%
  \BibitemOpen
  \bibfield  {author} {\bibinfo {author} {\bibfnamefont {S.~V.}\ \bibnamefont
  {Jerome}}, \bibinfo {author} {\bibfnamefont {T.~F.}\ \bibnamefont {Hughes}},
  \ and\ \bibinfo {author} {\bibfnamefont {R.~A.}\ \bibnamefont {Friesner}},\
  }\href@noop {} {\bibfield  {journal} {\bibinfo  {journal} {J. Phys. Chem. B}\
  }\textbf {\bibinfo {volume} {118}},\ \bibinfo {pages} {8008} (\bibinfo {year}
  {2014})}\BibitemShut {NoStop}%
\bibitem [{\citenamefont {Bochevarov}\ \emph {et~al.}(2016)\citenamefont
  {Bochevarov}, \citenamefont {Watson}, \citenamefont {Greenwood},\ and\
  \citenamefont {Philipp}}]{bochevarov2016multiconformation}%
  \BibitemOpen
  \bibfield  {author} {\bibinfo {author} {\bibfnamefont {A.~D.}\ \bibnamefont
  {Bochevarov}}, \bibinfo {author} {\bibfnamefont {M.~A.}\ \bibnamefont
  {Watson}}, \bibinfo {author} {\bibfnamefont {J.~R.}\ \bibnamefont
  {Greenwood}}, \ and\ \bibinfo {author} {\bibfnamefont {D.~M.}\ \bibnamefont
  {Philipp}},\ }\href {\doibase 10.1021/acs.jctc.6b00805} {\bibfield  {journal}
  {\bibinfo  {journal} {J. Chem. Theory Comput.}\ }\textbf {\bibinfo {volume}
  {12}},\ \bibinfo {pages} {6001} (\bibinfo {year} {2016})},\ \bibinfo {note}
  {pMID: 27951674},\ \Eprint
  {http://arxiv.org/abs/http://dx.doi.org/10.1021/acs.jctc.6b00805}
  {http://dx.doi.org/10.1021/acs.jctc.6b00805} \BibitemShut {NoStop}%
\bibitem [{\citenamefont {Rajapakse}\ \emph {et~al.}(2010)\citenamefont
  {Rajapakse}, \citenamefont {Nantermet}, \citenamefont {Selnick},
  \citenamefont {Barrow}, \citenamefont {McGaughey}, \citenamefont {Munshi},
  \citenamefont {Lindsley}, \citenamefont {Young}, \citenamefont {Ngo},
  \citenamefont {Holloway}, \citenamefont {Laid}, \citenamefont {Espesethd},
  \citenamefont {Shid}, \citenamefont {Colussid}, \citenamefont {Pietrakd},
  \citenamefont {Crouthameld}, \citenamefont {Tugushevad}, \citenamefont
  {Huangd}, \citenamefont {Xud}, \citenamefont {Simond}, \citenamefont {Kuoc},
  \citenamefont {Hazudad}, \citenamefont {Grahama},\ and\ \citenamefont
  {Vaccaa}}]{rajapakse2010sar}%
  \BibitemOpen
  \bibfield  {author} {\bibinfo {author} {\bibfnamefont {H.~A.}\ \bibnamefont
  {Rajapakse}}, \bibinfo {author} {\bibfnamefont {P.~G.}\ \bibnamefont
  {Nantermet}}, \bibinfo {author} {\bibfnamefont {H.~G.}\ \bibnamefont
  {Selnick}}, \bibinfo {author} {\bibfnamefont {J.~C.}\ \bibnamefont {Barrow}},
  \bibinfo {author} {\bibfnamefont {G.~B.}\ \bibnamefont {McGaughey}}, \bibinfo
  {author} {\bibfnamefont {S.}~\bibnamefont {Munshi}}, \bibinfo {author}
  {\bibfnamefont {S.~R.}\ \bibnamefont {Lindsley}}, \bibinfo {author}
  {\bibfnamefont {M.~B.}\ \bibnamefont {Young}}, \bibinfo {author}
  {\bibfnamefont {P.~L.}\ \bibnamefont {Ngo}}, \bibinfo {author} {\bibfnamefont
  {M.~K.}\ \bibnamefont {Holloway}}, \bibinfo {author} {\bibfnamefont {M.-T.}\
  \bibnamefont {Laid}}, \bibinfo {author} {\bibfnamefont {A.~S.}\ \bibnamefont
  {Espesethd}}, \bibinfo {author} {\bibfnamefont {X.-P.}\ \bibnamefont {Shid}},
  \bibinfo {author} {\bibfnamefont {D.}~\bibnamefont {Colussid}}, \bibinfo
  {author} {\bibfnamefont {B.}~\bibnamefont {Pietrakd}}, \bibinfo {author}
  {\bibfnamefont {M.-C.}\ \bibnamefont {Crouthameld}}, \bibinfo {author}
  {\bibfnamefont {K.}~\bibnamefont {Tugushevad}}, \bibinfo {author}
  {\bibfnamefont {Q.}~\bibnamefont {Huangd}}, \bibinfo {author} {\bibfnamefont
  {M.}~\bibnamefont {Xud}}, \bibinfo {author} {\bibfnamefont {A.~J.}\
  \bibnamefont {Simond}}, \bibinfo {author} {\bibfnamefont {L.}~\bibnamefont
  {Kuoc}}, \bibinfo {author} {\bibfnamefont {D.~J.}\ \bibnamefont {Hazudad}},
  \bibinfo {author} {\bibfnamefont {S.}~\bibnamefont {Grahama}}, \ and\
  \bibinfo {author} {\bibfnamefont {J.~P.}\ \bibnamefont {Vaccaa}},\
  }\href@noop {} {\bibfield  {journal} {\bibinfo  {journal} {Bioorg. Med. Chem.
  Lett.}\ }\textbf {\bibinfo {volume} {20}},\ \bibinfo {pages} {1885} (\bibinfo
  {year} {2010})}\BibitemShut {NoStop}%
\bibitem [{\citenamefont {Sprous}\ \emph {et~al.}(2010)\citenamefont {Sprous},
  \citenamefont {Palmer}, \citenamefont {Swanson},\ and\ \citenamefont
  {Lawless}}]{sprous2010qsar}%
  \BibitemOpen
  \bibfield  {author} {\bibinfo {author} {\bibfnamefont {D.}~\bibnamefont
  {Sprous}}, \bibinfo {author} {\bibfnamefont {R.}~\bibnamefont {Palmer}},
  \bibinfo {author} {\bibfnamefont {J.}~\bibnamefont {Swanson}}, \ and\
  \bibinfo {author} {\bibfnamefont {M.}~\bibnamefont {Lawless}},\ }\href@noop
  {} {\bibfield  {journal} {\bibinfo  {journal} {Current topics in medicinal
  chemistry}\ }\textbf {\bibinfo {volume} {10}},\ \bibinfo {pages} {619}
  (\bibinfo {year} {2010})}\BibitemShut {NoStop}%
\bibitem [{\citenamefont {Cheng}\ and\ \citenamefont
  {Sprik}(2010)}]{cheng2010acidity}%
  \BibitemOpen
  \bibfield  {author} {\bibinfo {author} {\bibfnamefont {J.}~\bibnamefont
  {Cheng}}\ and\ \bibinfo {author} {\bibfnamefont {M.}~\bibnamefont {Sprik}},\
  }\href@noop {} {\bibfield  {journal} {\bibinfo  {journal} {J. Chem. Theory
  Comput.}\ }\textbf {\bibinfo {volume} {6}},\ \bibinfo {pages} {880} (\bibinfo
  {year} {2010})}\BibitemShut {NoStop}%
\bibitem [{\citenamefont {Gallus}\ \emph {et~al.}(2015)\citenamefont {Gallus},
  \citenamefont {Wagner}, \citenamefont {Wiemers-Meyer}, \citenamefont
  {Winter},\ and\ \citenamefont {Cekic-Laskovic}}]{gallus2015new}%
  \BibitemOpen
  \bibfield  {author} {\bibinfo {author} {\bibfnamefont {D.~R.}\ \bibnamefont
  {Gallus}}, \bibinfo {author} {\bibfnamefont {R.}~\bibnamefont {Wagner}},
  \bibinfo {author} {\bibfnamefont {S.}~\bibnamefont {Wiemers-Meyer}}, \bibinfo
  {author} {\bibfnamefont {M.}~\bibnamefont {Winter}}, \ and\ \bibinfo {author}
  {\bibfnamefont {I.}~\bibnamefont {Cekic-Laskovic}},\ }\href@noop {}
  {\bibfield  {journal} {\bibinfo  {journal} {Electrochim. Acta}\ }\textbf
  {\bibinfo {volume} {184}},\ \bibinfo {pages} {410} (\bibinfo {year}
  {2015})}\BibitemShut {NoStop}%
\bibitem [{\citenamefont {Bryantsev}(2013)}]{bryantsev2013predicting}%
  \BibitemOpen
  \bibfield  {author} {\bibinfo {author} {\bibfnamefont {V.~S.}\ \bibnamefont
  {Bryantsev}},\ }\href@noop {} {\bibfield  {journal} {\bibinfo  {journal}
  {Chem. Phys. Lett.}\ }\textbf {\bibinfo {volume} {558}},\ \bibinfo {pages}
  {42} (\bibinfo {year} {2013})}\BibitemShut {NoStop}%
\bibitem [{\citenamefont {Ames}\ \emph {et~al.}(2011)\citenamefont {Ames},
  \citenamefont {Pantazis}, \citenamefont {Krewald}, \citenamefont {Cox},
  \citenamefont {Messinger}, \citenamefont {Lubitz},\ and\ \citenamefont
  {Neese}}]{ames2011theoretical}%
  \BibitemOpen
  \bibfield  {author} {\bibinfo {author} {\bibfnamefont {W.}~\bibnamefont
  {Ames}}, \bibinfo {author} {\bibfnamefont {D.~A.}\ \bibnamefont {Pantazis}},
  \bibinfo {author} {\bibfnamefont {V.}~\bibnamefont {Krewald}}, \bibinfo
  {author} {\bibfnamefont {N.}~\bibnamefont {Cox}}, \bibinfo {author}
  {\bibfnamefont {J.}~\bibnamefont {Messinger}}, \bibinfo {author}
  {\bibfnamefont {W.}~\bibnamefont {Lubitz}}, \ and\ \bibinfo {author}
  {\bibfnamefont {F.}~\bibnamefont {Neese}},\ }\href@noop {} {\bibfield
  {journal} {\bibinfo  {journal} {J. Am. Chem. Soc.}\ }\textbf {\bibinfo
  {volume} {133}},\ \bibinfo {pages} {19743} (\bibinfo {year}
  {2011})}\BibitemShut {NoStop}%
\bibitem [{\citenamefont {El~Yazal}\ \emph {et~al.}(2000)\citenamefont
  {El~Yazal}, \citenamefont {Prendergast}, \citenamefont {Shaw},\ and\
  \citenamefont {Pang}}]{el2000protonation}%
  \BibitemOpen
  \bibfield  {author} {\bibinfo {author} {\bibfnamefont {J.}~\bibnamefont
  {El~Yazal}}, \bibinfo {author} {\bibfnamefont {F.~G.}\ \bibnamefont
  {Prendergast}}, \bibinfo {author} {\bibfnamefont {D.~E.}\ \bibnamefont
  {Shaw}}, \ and\ \bibinfo {author} {\bibfnamefont {Y.-P.}\ \bibnamefont
  {Pang}},\ }\href@noop {} {\bibfield  {journal} {\bibinfo  {journal} {J. Am.
  Chem. Soc.}\ }\textbf {\bibinfo {volume} {122}},\ \bibinfo {pages} {11411}
  (\bibinfo {year} {2000})}\BibitemShut {NoStop}%
\bibitem [{\citenamefont {Chen}\ \emph {et~al.}(2013)\citenamefont {Chen},
  \citenamefont {Li}, \citenamefont {Sit},\ and\ \citenamefont
  {Selloni}}]{2013chemical}%
  \BibitemOpen
  \bibfield  {author} {\bibinfo {author} {\bibfnamefont {J.}~\bibnamefont
  {Chen}}, \bibinfo {author} {\bibfnamefont {Y.-F.}\ \bibnamefont {Li}},
  \bibinfo {author} {\bibfnamefont {P.}~\bibnamefont {Sit}}, \ and\ \bibinfo
  {author} {\bibfnamefont {A.}~\bibnamefont {Selloni}},\ }\href@noop {}
  {\bibfield  {journal} {\bibinfo  {journal} {J. Am. Chem. Soc.}\ }\textbf
  {\bibinfo {volume} {135}},\ \bibinfo {pages} {18774} (\bibinfo {year}
  {2013})}\BibitemShut {NoStop}%
\bibitem [{\citenamefont {Olah}\ and\ \citenamefont
  {Molnar}(2003)}]{olah2003hydrocarbon}%
  \BibitemOpen
  \bibfield  {author} {\bibinfo {author} {\bibfnamefont {G.~A.}\ \bibnamefont
  {Olah}}\ and\ \bibinfo {author} {\bibfnamefont {A.}~\bibnamefont {Molnar}},\
  }\href@noop {} {\emph {\bibinfo {title} {Hydrocarbon chemistry}}}\ (\bibinfo
  {publisher} {John Wiley \& Sons},\ \bibinfo {year} {2003})\BibitemShut
  {NoStop}%
\bibitem [{\citenamefont {Wang}\ and\ \citenamefont
  {Frenklach}(1994)}]{wang1994calculations}%
  \BibitemOpen
  \bibfield  {author} {\bibinfo {author} {\bibfnamefont {H.}~\bibnamefont
  {Wang}}\ and\ \bibinfo {author} {\bibfnamefont {M.}~\bibnamefont
  {Frenklach}},\ }\href@noop {} {\bibfield  {journal} {\bibinfo  {journal} {J.
  Phys. Chem.}\ }\textbf {\bibinfo {volume} {98}},\ \bibinfo {pages} {11465}
  (\bibinfo {year} {1994})}\BibitemShut {NoStop}%
\bibitem [{\citenamefont {Page}\ and\ \citenamefont
  {Brenner}(1991)}]{page1991hydrogen}%
  \BibitemOpen
  \bibfield  {author} {\bibinfo {author} {\bibfnamefont {M.}~\bibnamefont
  {Page}}\ and\ \bibinfo {author} {\bibfnamefont {D.~W.}\ \bibnamefont
  {Brenner}},\ }\href@noop {} {\bibfield  {journal} {\bibinfo  {journal} {J.
  Am. Chem. Soc.}\ }\textbf {\bibinfo {volume} {113}},\ \bibinfo {pages} {3270}
  (\bibinfo {year} {1991})}\BibitemShut {NoStop}%
\bibitem [{\citenamefont {Burton}\ and\ \citenamefont
  {Ingold}(1989)}]{burton1989vitamin}%
  \BibitemOpen
  \bibfield  {author} {\bibinfo {author} {\bibfnamefont {G.~W.}\ \bibnamefont
  {Burton}}\ and\ \bibinfo {author} {\bibfnamefont {K.~U.}\ \bibnamefont
  {Ingold}},\ }\href@noop {} {\bibfield  {journal} {\bibinfo  {journal} {Ann.
  N. Y. Acad. Sci.}\ }\textbf {\bibinfo {volume} {570}},\ \bibinfo {pages} {7}
  (\bibinfo {year} {1989})}\BibitemShut {NoStop}%
\bibitem [{\citenamefont {Mayer}(1998)}]{mayer1998hydrogen}%
  \BibitemOpen
  \bibfield  {author} {\bibinfo {author} {\bibfnamefont {J.~M.}\ \bibnamefont
  {Mayer}},\ }\href@noop {} {\bibfield  {journal} {\bibinfo  {journal} {Acc.
  Chem. Res.}\ }\textbf {\bibinfo {volume} {31}},\ \bibinfo {pages} {441}
  (\bibinfo {year} {1998})}\BibitemShut {NoStop}%
\bibitem [{\citenamefont {Blomberg}\ \emph {et~al.}(1997)\citenamefont
  {Blomberg}, \citenamefont {Siegbahn}, \citenamefont {Styring}, \citenamefont
  {Babcock}, \citenamefont {{\AA}kermark},\ and\ \citenamefont
  {Korall}}]{blomberg1997quantum}%
  \BibitemOpen
  \bibfield  {author} {\bibinfo {author} {\bibfnamefont {M.~R.}\ \bibnamefont
  {Blomberg}}, \bibinfo {author} {\bibfnamefont {P.~E.}\ \bibnamefont
  {Siegbahn}}, \bibinfo {author} {\bibfnamefont {S.}~\bibnamefont {Styring}},
  \bibinfo {author} {\bibfnamefont {G.~T.}\ \bibnamefont {Babcock}}, \bibinfo
  {author} {\bibfnamefont {B.}~\bibnamefont {{\AA}kermark}}, \ and\ \bibinfo
  {author} {\bibfnamefont {P.}~\bibnamefont {Korall}},\ }\href@noop {}
  {\bibfield  {journal} {\bibinfo  {journal} {J. Am. Chem. Soc.}\ }\textbf
  {\bibinfo {volume} {119}},\ \bibinfo {pages} {8285} (\bibinfo {year}
  {1997})}\BibitemShut {NoStop}%
\bibitem [{\citenamefont {Grasselli}(1999)}]{grasselli1999advances}%
  \BibitemOpen
  \bibfield  {author} {\bibinfo {author} {\bibfnamefont {R.~K.}\ \bibnamefont
  {Grasselli}},\ }\href@noop {} {\bibfield  {journal} {\bibinfo  {journal}
  {Catal. Today}\ }\textbf {\bibinfo {volume} {49}},\ \bibinfo {pages} {141}
  (\bibinfo {year} {1999})}\BibitemShut {NoStop}%
\bibitem [{\citenamefont {Snider}(1996)}]{snider1996manganese}%
  \BibitemOpen
  \bibfield  {author} {\bibinfo {author} {\bibfnamefont {B.~B.}\ \bibnamefont
  {Snider}},\ }\href@noop {} {\bibfield  {journal} {\bibinfo  {journal} {Chem.
  Rev.}\ }\textbf {\bibinfo {volume} {96}},\ \bibinfo {pages} {339} (\bibinfo
  {year} {1996})}\BibitemShut {NoStop}%
\bibitem [{\citenamefont {Basch}\ and\ \citenamefont
  {Hoz}(1997)}]{basch1997ab}%
  \BibitemOpen
  \bibfield  {author} {\bibinfo {author} {\bibfnamefont {H.}~\bibnamefont
  {Basch}}\ and\ \bibinfo {author} {\bibfnamefont {S.}~\bibnamefont {Hoz}},\
  }\href@noop {} {\bibfield  {journal} {\bibinfo  {journal} {J. Phys. Chem. A}\
  }\textbf {\bibinfo {volume} {101}},\ \bibinfo {pages} {4416} (\bibinfo {year}
  {1997})}\BibitemShut {NoStop}%
\bibitem [{\citenamefont {Coote}(2004)}]{coote2004reliable}%
  \BibitemOpen
  \bibfield  {author} {\bibinfo {author} {\bibfnamefont {M.~L.}\ \bibnamefont
  {Coote}},\ }\href@noop {} {\bibfield  {journal} {\bibinfo  {journal} {J.
  Phys. Chem. A}\ }\textbf {\bibinfo {volume} {108}},\ \bibinfo {pages} {3865}
  (\bibinfo {year} {2004})}\BibitemShut {NoStop}%
\bibitem [{\citenamefont {Pu}\ and\ \citenamefont
  {Truhlar}(2005)}]{pu2005benchmark}%
  \BibitemOpen
  \bibfield  {author} {\bibinfo {author} {\bibfnamefont {J.}~\bibnamefont
  {Pu}}\ and\ \bibinfo {author} {\bibfnamefont {D.~G.}\ \bibnamefont
  {Truhlar}},\ }\href@noop {} {\bibfield  {journal} {\bibinfo  {journal} {J.
  Phys. Chem. A}\ }\textbf {\bibinfo {volume} {109}},\ \bibinfo {pages} {773}
  (\bibinfo {year} {2005})}\BibitemShut {NoStop}%
\bibitem [{\citenamefont {Carvalho}\ \emph
  {et~al.}(2008{\natexlab{a}})\citenamefont {Carvalho}, \citenamefont
  {Barauna}, \citenamefont {Machado},\ and\ \citenamefont
  {Roberto-Neto}}]{carvalho2008dft}%
  \BibitemOpen
  \bibfield  {author} {\bibinfo {author} {\bibfnamefont {E.~F.}\ \bibnamefont
  {Carvalho}}, \bibinfo {author} {\bibfnamefont {A.~N.}\ \bibnamefont
  {Barauna}}, \bibinfo {author} {\bibfnamefont {F.~B.}\ \bibnamefont
  {Machado}}, \ and\ \bibinfo {author} {\bibfnamefont {O.}~\bibnamefont
  {Roberto-Neto}},\ }\href@noop {} {\bibfield  {journal} {\bibinfo  {journal}
  {Int. J. Quantum Chem.}\ }\textbf {\bibinfo {volume} {108}},\ \bibinfo
  {pages} {2476} (\bibinfo {year} {2008}{\natexlab{a}})}\BibitemShut {NoStop}%
\bibitem [{\citenamefont {Carvalho}\ \emph
  {et~al.}(2008{\natexlab{b}})\citenamefont {Carvalho}, \citenamefont
  {Barauna}, \citenamefont {Machado},\ and\ \citenamefont
  {Roberto-Neto}}]{carvalho2008theoretical}%
  \BibitemOpen
  \bibfield  {author} {\bibinfo {author} {\bibfnamefont {E.}~\bibnamefont
  {Carvalho}}, \bibinfo {author} {\bibfnamefont {A.~N.}\ \bibnamefont
  {Barauna}}, \bibinfo {author} {\bibfnamefont {F.~B.}\ \bibnamefont
  {Machado}}, \ and\ \bibinfo {author} {\bibfnamefont {O.}~\bibnamefont
  {Roberto-Neto}},\ }\href@noop {} {\bibfield  {journal} {\bibinfo  {journal}
  {Chem. Phys. Lett.}\ }\textbf {\bibinfo {volume} {463}},\ \bibinfo {pages}
  {33} (\bibinfo {year} {2008}{\natexlab{b}})}\BibitemShut {NoStop}%
\bibitem [{\citenamefont {Fracchia}\ \emph {et~al.}(2013)\citenamefont
  {Fracchia}, \citenamefont {Filippi},\ and\ \citenamefont
  {Amovilli}}]{fracchia2013barrier}%
  \BibitemOpen
  \bibfield  {author} {\bibinfo {author} {\bibfnamefont {F.}~\bibnamefont
  {Fracchia}}, \bibinfo {author} {\bibfnamefont {C.}~\bibnamefont {Filippi}}, \
  and\ \bibinfo {author} {\bibfnamefont {C.}~\bibnamefont {Amovilli}},\
  }\href@noop {} {\bibfield  {journal} {\bibinfo  {journal} {J. Chem. Theory
  Comput.}\ }\textbf {\bibinfo {volume} {9}},\ \bibinfo {pages} {3453}
  (\bibinfo {year} {2013})}\BibitemShut {NoStop}%
\bibitem [{\citenamefont {Kanai}\ and\ \citenamefont
  {Takeuchi}(2009)}]{kanai2009toward}%
  \BibitemOpen
  \bibfield  {author} {\bibinfo {author} {\bibfnamefont {Y.}~\bibnamefont
  {Kanai}}\ and\ \bibinfo {author} {\bibfnamefont {N.}~\bibnamefont
  {Takeuchi}},\ }\href@noop {} {\bibfield  {journal} {\bibinfo  {journal} {J.
  Chem. Phys.}\ }\textbf {\bibinfo {volume} {131}},\ \bibinfo {pages} {214708}
  (\bibinfo {year} {2009})}\BibitemShut {NoStop}%
\bibitem [{\citenamefont {Kollias}\ \emph {et~al.}(2004)\citenamefont
  {Kollias}, \citenamefont {Couronne},\ and\ \citenamefont
  {Lester~Jr}}]{kollias2004quantum}%
  \BibitemOpen
  \bibfield  {author} {\bibinfo {author} {\bibfnamefont {A.}~\bibnamefont
  {Kollias}}, \bibinfo {author} {\bibfnamefont {O.}~\bibnamefont {Couronne}}, \
  and\ \bibinfo {author} {\bibfnamefont {W.}~\bibnamefont {Lester~Jr}},\
  }\href@noop {} {\bibfield  {journal} {\bibinfo  {journal} {J. Chem. Phys.}\
  }\textbf {\bibinfo {volume} {121}},\ \bibinfo {pages} {1357} (\bibinfo {year}
  {2004})}\BibitemShut {NoStop}%
\bibitem [{\citenamefont {Curtiss}\ \emph {et~al.}(1998)\citenamefont
  {Curtiss}, \citenamefont {Redfern}, \citenamefont {Raghavachari},\ and\
  \citenamefont {Pople}}]{curtiss1998assessment}%
  \BibitemOpen
  \bibfield  {author} {\bibinfo {author} {\bibfnamefont {L.~A.}\ \bibnamefont
  {Curtiss}}, \bibinfo {author} {\bibfnamefont {P.~C.}\ \bibnamefont
  {Redfern}}, \bibinfo {author} {\bibfnamefont {K.}~\bibnamefont
  {Raghavachari}}, \ and\ \bibinfo {author} {\bibfnamefont {J.~A.}\
  \bibnamefont {Pople}},\ }\href@noop {} {\bibfield  {journal} {\bibinfo
  {journal} {J. Chem. Phys.}\ }\textbf {\bibinfo {volume} {109}},\ \bibinfo
  {pages} {42} (\bibinfo {year} {1998})}\BibitemShut {NoStop}%
\bibitem [{\citenamefont {Purwanto}\ \emph {et~al.}(2011)\citenamefont
  {Purwanto}, \citenamefont {Krakauer}, \citenamefont {Virgus},\ and\
  \citenamefont {Zhang}}]{purwanto2011Ca}%
  \BibitemOpen
  \bibfield  {author} {\bibinfo {author} {\bibfnamefont {W.}~\bibnamefont
  {Purwanto}}, \bibinfo {author} {\bibfnamefont {H.}~\bibnamefont {Krakauer}},
  \bibinfo {author} {\bibfnamefont {Y.}~\bibnamefont {Virgus}}, \ and\ \bibinfo
  {author} {\bibfnamefont {S.}~\bibnamefont {Zhang}},\ }\href@noop {}
  {\bibfield  {journal} {\bibinfo  {journal} {J. Chem. Phys.}\ }\textbf
  {\bibinfo {volume} {135}},\ \bibinfo {pages} {164105} (\bibinfo {year}
  {2011})}\BibitemShut {NoStop}%
\bibitem [{\citenamefont {Trotter}(1959)}]{trotter1959product}%
  \BibitemOpen
  \bibfield  {author} {\bibinfo {author} {\bibfnamefont {H.~F.}\ \bibnamefont
  {Trotter}},\ }\href@noop {} {\bibfield  {journal} {\bibinfo  {journal} {Proc.
  Am. Math. Soc.}\ }\textbf {\bibinfo {volume} {10}},\ \bibinfo {pages} {545}
  (\bibinfo {year} {1959})}\BibitemShut {NoStop}%
\bibitem [{\citenamefont {Suzuki}(1976)}]{suzuki1976generalized}%
  \BibitemOpen
  \bibfield  {author} {\bibinfo {author} {\bibfnamefont {M.}~\bibnamefont
  {Suzuki}},\ }\href@noop {} {\bibfield  {journal} {\bibinfo  {journal}
  {Commun. Math. Phys.}\ }\textbf {\bibinfo {volume} {51}},\ \bibinfo {pages}
  {183} (\bibinfo {year} {1976})}\BibitemShut {NoStop}%
\bibitem [{\citenamefont {Stratonovich}(1957)}]{stratonovich1957method}%
  \BibitemOpen
  \bibfield  {author} {\bibinfo {author} {\bibfnamefont {R.}~\bibnamefont
  {Stratonovich}},\ }\href@noop {} {\bibfield  {journal} {\bibinfo  {journal}
  {Dokl. Akad. Nauk SSSR}\ }\textbf {\bibinfo {volume} {115}},\ \bibinfo
  {pages} {1097} (\bibinfo {year} {1957})}\BibitemShut {NoStop}%
\bibitem [{\citenamefont {Hubbard}(1959)}]{hubbard1959calculation}%
  \BibitemOpen
  \bibfield  {author} {\bibinfo {author} {\bibfnamefont {J.}~\bibnamefont
  {Hubbard}},\ }\href {\doibase 10.1103/PhysRevLett.3.77} {\bibfield  {journal}
  {\bibinfo  {journal} {Phys. Rev. Lett.}\ }\textbf {\bibinfo {volume} {3}},\
  \bibinfo {pages} {77} (\bibinfo {year} {1959})}\BibitemShut {NoStop}%
\bibitem [{\citenamefont {Hamann}\ and\ \citenamefont
  {Fahy}(1990)}]{hamann1990energy}%
  \BibitemOpen
  \bibfield  {author} {\bibinfo {author} {\bibfnamefont {D.~R.}\ \bibnamefont
  {Hamann}}\ and\ \bibinfo {author} {\bibfnamefont {S.~B.}\ \bibnamefont
  {Fahy}},\ }\href {\doibase 10.1103/PhysRevB.41.11352} {\bibfield  {journal}
  {\bibinfo  {journal} {Phys. Rev. B}\ }\textbf {\bibinfo {volume} {41}},\
  \bibinfo {pages} {11352} (\bibinfo {year} {1990})}\BibitemShut {NoStop}%
\bibitem [{\citenamefont {Nguyen}(2014)}]{HuyThesis}%
  \BibitemOpen
  \bibfield  {author} {\bibinfo {author} {\bibfnamefont {H.}~\bibnamefont
  {Nguyen}},\ }\emph {\bibinfo {title} {Quantum Monte Carlo Calculation of the
  Imaginary-Time Green’s Function in the Hubbard Model}},\ \href@noop {}
  {\bibinfo {type} {Undergraduate thesis}},\ \bibinfo  {school} {Reed College}
  (\bibinfo {year} {2014})\BibitemShut {NoStop}%
\bibitem [{\citenamefont {Rubenstein}\ \emph {et~al.}(2012)\citenamefont
  {Rubenstein}, \citenamefont {Zhang},\ and\ \citenamefont
  {Reichman}}]{rubenstein2012finite}%
  \BibitemOpen
  \bibfield  {author} {\bibinfo {author} {\bibfnamefont {B.~M.}\ \bibnamefont
  {Rubenstein}}, \bibinfo {author} {\bibfnamefont {S.}~\bibnamefont {Zhang}}, \
  and\ \bibinfo {author} {\bibfnamefont {D.~R.}\ \bibnamefont {Reichman}},\
  }\href {\doibase 10.1103/PhysRevA.86.053606} {\bibfield  {journal} {\bibinfo
  {journal} {Phys. Rev. A}\ }\textbf {\bibinfo {volume} {86}},\ \bibinfo
  {pages} {053606} (\bibinfo {year} {2012})}\BibitemShut {NoStop}%
\bibitem [{\citenamefont {Reynolds}\ \emph {et~al.}(1982)\citenamefont
  {Reynolds}, \citenamefont {Ceperley}, \citenamefont {Alder},\ and\
  \citenamefont {Lester~Jr}}]{reynolds1982fixed}%
  \BibitemOpen
  \bibfield  {author} {\bibinfo {author} {\bibfnamefont {P.~J.}\ \bibnamefont
  {Reynolds}}, \bibinfo {author} {\bibfnamefont {D.~M.}\ \bibnamefont
  {Ceperley}}, \bibinfo {author} {\bibfnamefont {B.~J.}\ \bibnamefont {Alder}},
  \ and\ \bibinfo {author} {\bibfnamefont {W.~A.}\ \bibnamefont {Lester~Jr}},\
  }\href@noop {} {\bibfield  {journal} {\bibinfo  {journal} {J. Chem. Phys.}\
  }\textbf {\bibinfo {volume} {77}},\ \bibinfo {pages} {5593} (\bibinfo {year}
  {1982})}\BibitemShut {NoStop}%
\bibitem [{\citenamefont {Shi}\ and\ \citenamefont
  {Zhang}(2013)}]{shi2013symmetry}%
  \BibitemOpen
  \bibfield  {author} {\bibinfo {author} {\bibfnamefont {H.}~\bibnamefont
  {Shi}}\ and\ \bibinfo {author} {\bibfnamefont {S.}~\bibnamefont {Zhang}},\
  }\href {\doibase 10.1103/PhysRevB.88.125132} {\bibfield  {journal} {\bibinfo
  {journal} {Phys. Rev. B}\ }\textbf {\bibinfo {volume} {88}},\ \bibinfo
  {pages} {125132} (\bibinfo {year} {2013})}\BibitemShut {NoStop}%
\bibitem [{\citenamefont {Zhang}\ \emph {et~al.}(1997)\citenamefont {Zhang},
  \citenamefont {Carlson},\ and\ \citenamefont
  {Gubernatis}}]{zhang1997constrained}%
  \BibitemOpen
  \bibfield  {author} {\bibinfo {author} {\bibfnamefont {S.}~\bibnamefont
  {Zhang}}, \bibinfo {author} {\bibfnamefont {J.}~\bibnamefont {Carlson}}, \
  and\ \bibinfo {author} {\bibfnamefont {J.~E.}\ \bibnamefont {Gubernatis}},\
  }\href {\doibase 10.1103/PhysRevB.55.7464} {\bibfield  {journal} {\bibinfo
  {journal} {Phys. Rev. B}\ }\textbf {\bibinfo {volume} {55}},\ \bibinfo
  {pages} {7464} (\bibinfo {year} {1997})}\BibitemShut {NoStop}%
\bibitem [{\citenamefont {Loh}\ \emph {et~al.}(1990)\citenamefont {Loh},
  \citenamefont {Gubernatis}, \citenamefont {Scalettar}, \citenamefont {White},
  \citenamefont {Scalapino},\ and\ \citenamefont {Sugar}}]{loh1990sign}%
  \BibitemOpen
  \bibfield  {author} {\bibinfo {author} {\bibfnamefont {E.~Y.}\ \bibnamefont
  {Loh}}, \bibinfo {author} {\bibfnamefont {J.~E.}\ \bibnamefont {Gubernatis}},
  \bibinfo {author} {\bibfnamefont {R.~T.}\ \bibnamefont {Scalettar}}, \bibinfo
  {author} {\bibfnamefont {S.~R.}\ \bibnamefont {White}}, \bibinfo {author}
  {\bibfnamefont {D.~J.}\ \bibnamefont {Scalapino}}, \ and\ \bibinfo {author}
  {\bibfnamefont {R.~L.}\ \bibnamefont {Sugar}},\ }\href {\doibase
  10.1103/PhysRevB.41.9301} {\bibfield  {journal} {\bibinfo  {journal} {Phys.
  Rev. B}\ }\textbf {\bibinfo {volume} {41}},\ \bibinfo {pages} {9301}
  (\bibinfo {year} {1990})}\BibitemShut {NoStop}%
\bibitem [{\citenamefont {Troyer}\ and\ \citenamefont
  {Wiese}(2005)}]{troyer2005computational}%
  \BibitemOpen
  \bibfield  {author} {\bibinfo {author} {\bibfnamefont {M.}~\bibnamefont
  {Troyer}}\ and\ \bibinfo {author} {\bibfnamefont {U.-J.}\ \bibnamefont
  {Wiese}},\ }\href {\doibase 10.1103/PhysRevLett.94.170201} {\bibfield
  {journal} {\bibinfo  {journal} {Phys. Rev. Lett.}\ }\textbf {\bibinfo
  {volume} {94}},\ \bibinfo {pages} {170201} (\bibinfo {year}
  {2005})}\BibitemShut {NoStop}%
\bibitem [{\citenamefont {Motta}\ \emph {et~al.}(2014)\citenamefont {Motta},
  \citenamefont {Galli}, \citenamefont {Moroni},\ and\ \citenamefont
  {Vitali}}]{motta2014imaginary}%
  \BibitemOpen
  \bibfield  {author} {\bibinfo {author} {\bibfnamefont {M.}~\bibnamefont
  {Motta}}, \bibinfo {author} {\bibfnamefont {D.~E.}\ \bibnamefont {Galli}},
  \bibinfo {author} {\bibfnamefont {S.}~\bibnamefont {Moroni}}, \ and\ \bibinfo
  {author} {\bibfnamefont {E.}~\bibnamefont {Vitali}},\ }\href@noop {}
  {\bibfield  {journal} {\bibinfo  {journal} {J. Chem. Phys.}\ }\textbf
  {\bibinfo {volume} {140}},\ \bibinfo {pages} {024107} (\bibinfo {year}
  {2014})}\BibitemShut {NoStop}%
\bibitem [{\citenamefont {Purwanto}\ and\ \citenamefont
  {Zhang}(2004)}]{purwanto2004quantum}%
  \BibitemOpen
  \bibfield  {author} {\bibinfo {author} {\bibfnamefont {W.}~\bibnamefont
  {Purwanto}}\ and\ \bibinfo {author} {\bibfnamefont {S.}~\bibnamefont
  {Zhang}},\ }\href {\doibase 10.1103/PhysRevE.70.056702} {\bibfield  {journal}
  {\bibinfo  {journal} {Phys. Rev. E}\ }\textbf {\bibinfo {volume} {70}},\
  \bibinfo {pages} {056702} (\bibinfo {year} {2004})}\BibitemShut {NoStop}%
\bibitem [{\citenamefont {Al-Saidi}\ \emph
  {et~al.}(2006{\natexlab{c}})\citenamefont {Al-Saidi}, \citenamefont {Zhang},\
  and\ \citenamefont {Krakauer}}]{al2006gaussian}%
  \BibitemOpen
  \bibfield  {author} {\bibinfo {author} {\bibfnamefont {W.}~\bibnamefont
  {Al-Saidi}}, \bibinfo {author} {\bibfnamefont {S.}~\bibnamefont {Zhang}}, \
  and\ \bibinfo {author} {\bibfnamefont {H.}~\bibnamefont {Krakauer}},\
  }\href@noop {} {\bibfield  {journal} {\bibinfo  {journal} {J. Chem. Phys.}\
  }\textbf {\bibinfo {volume} {124}},\ \bibinfo {pages} {224101} (\bibinfo
  {year} {2006}{\natexlab{c}})}\BibitemShut {NoStop}%
\bibitem [{\citenamefont {Flyvbjerg}\ and\ \citenamefont
  {Petersen}(1989)}]{flyvbjerg1989error}%
  \BibitemOpen
  \bibfield  {author} {\bibinfo {author} {\bibfnamefont {H.}~\bibnamefont
  {Flyvbjerg}}\ and\ \bibinfo {author} {\bibfnamefont {H.~G.}\ \bibnamefont
  {Petersen}},\ }\href@noop {} {\bibfield  {journal} {\bibinfo  {journal} {J.
  Chem. Phys.}\ }\textbf {\bibinfo {volume} {91}},\ \bibinfo {pages} {461}
  (\bibinfo {year} {1989})}\BibitemShut {NoStop}%
\bibitem [{\citenamefont {Nguyen}\ \emph {et~al.}(2014)\citenamefont {Nguyen},
  \citenamefont {Shi}, \citenamefont {Xu},\ and\ \citenamefont
  {Zhang}}]{nguyen2014cpmc}%
  \BibitemOpen
  \bibfield  {author} {\bibinfo {author} {\bibfnamefont {H.}~\bibnamefont
  {Nguyen}}, \bibinfo {author} {\bibfnamefont {H.}~\bibnamefont {Shi}},
  \bibinfo {author} {\bibfnamefont {J.}~\bibnamefont {Xu}}, \ and\ \bibinfo
  {author} {\bibfnamefont {S.}~\bibnamefont {Zhang}},\ }\href@noop {}
  {\bibfield  {journal} {\bibinfo  {journal} {Comput. Phys. Commun.}\ }\textbf
  {\bibinfo {volume} {185}},\ \bibinfo {pages} {3344} (\bibinfo {year}
  {2014})}\BibitemShut {NoStop}%
\bibitem [{\citenamefont {Valiev}\ \emph {et~al.}(2010)\citenamefont {Valiev},
  \citenamefont {Bylaska}, \citenamefont {Govind}, \citenamefont {Kowalski},
  \citenamefont {Straatsma}, \citenamefont {Van~Dam}, \citenamefont {Wang},
  \citenamefont {Nieplocha}, \citenamefont {Apra}, \citenamefont {Windus},\
  and\ \citenamefont {de~Jong}}]{valiev2010nwchem}%
  \BibitemOpen
  \bibfield  {author} {\bibinfo {author} {\bibfnamefont {M.}~\bibnamefont
  {Valiev}}, \bibinfo {author} {\bibfnamefont {E.~J.}\ \bibnamefont {Bylaska}},
  \bibinfo {author} {\bibfnamefont {N.}~\bibnamefont {Govind}}, \bibinfo
  {author} {\bibfnamefont {K.}~\bibnamefont {Kowalski}}, \bibinfo {author}
  {\bibfnamefont {T.~P.}\ \bibnamefont {Straatsma}}, \bibinfo {author}
  {\bibfnamefont {H.~J.}\ \bibnamefont {Van~Dam}}, \bibinfo {author}
  {\bibfnamefont {D.}~\bibnamefont {Wang}}, \bibinfo {author} {\bibfnamefont
  {J.}~\bibnamefont {Nieplocha}}, \bibinfo {author} {\bibfnamefont
  {E.}~\bibnamefont {Apra}}, \bibinfo {author} {\bibfnamefont {T.~L.}\
  \bibnamefont {Windus}}, \ and\ \bibinfo {author} {\bibfnamefont {W.~A.}\
  \bibnamefont {de~Jong}},\ }\href@noop {} {\bibfield  {journal} {\bibinfo
  {journal} {Comput. Phys. Commun.}\ }\textbf {\bibinfo {volume} {181}},\
  \bibinfo {pages} {1477} (\bibinfo {year} {2010})}\BibitemShut {NoStop}%
\bibitem [{\citenamefont {Purwanto}\ \emph {et~al.}(2008)\citenamefont
  {Purwanto}, \citenamefont {Al-Saidi}, \citenamefont {Krakauer},\ and\
  \citenamefont {Zhang}}]{purwanto2008eliminating}%
  \BibitemOpen
  \bibfield  {author} {\bibinfo {author} {\bibfnamefont {W.}~\bibnamefont
  {Purwanto}}, \bibinfo {author} {\bibfnamefont {W.}~\bibnamefont {Al-Saidi}},
  \bibinfo {author} {\bibfnamefont {H.}~\bibnamefont {Krakauer}}, \ and\
  \bibinfo {author} {\bibfnamefont {S.}~\bibnamefont {Zhang}},\ }\href@noop {}
  {\bibfield  {journal} {\bibinfo  {journal} {J. Chem. Phys.}\ }\textbf
  {\bibinfo {volume} {128}},\ \bibinfo {pages} {114309} (\bibinfo {year}
  {2008})}\BibitemShut {NoStop}%
\bibitem [{pys()}]{pyscf}%
  \BibitemOpen
  \href@noop {} {\enquote {\bibinfo {title} {{PySCF}},}\ }\bibinfo
  {howpublished} {https://github.com/sunqm/pyscf},\ \bibinfo {note}
  {2014}\BibitemShut {NoStop}%
\bibitem [{\citenamefont {White}\ \emph {et~al.}(1989)\citenamefont {White},
  \citenamefont {Scalapino}, \citenamefont {Sugar}, \citenamefont {Loh},
  \citenamefont {Gubernatis},\ and\ \citenamefont
  {Scalettar}}]{white1989numerical}%
  \BibitemOpen
  \bibfield  {author} {\bibinfo {author} {\bibfnamefont {S.~R.}\ \bibnamefont
  {White}}, \bibinfo {author} {\bibfnamefont {D.~J.}\ \bibnamefont
  {Scalapino}}, \bibinfo {author} {\bibfnamefont {R.~L.}\ \bibnamefont
  {Sugar}}, \bibinfo {author} {\bibfnamefont {E.~Y.}\ \bibnamefont {Loh}},
  \bibinfo {author} {\bibfnamefont {J.~E.}\ \bibnamefont {Gubernatis}}, \ and\
  \bibinfo {author} {\bibfnamefont {R.~T.}\ \bibnamefont {Scalettar}},\ }\href
  {\doibase 10.1103/PhysRevB.40.506} {\bibfield  {journal} {\bibinfo  {journal}
  {Phys. Rev. B}\ }\textbf {\bibinfo {volume} {40}},\ \bibinfo {pages} {506}
  (\bibinfo {year} {1989})}\BibitemShut {NoStop}%
\bibitem [{\citenamefont {Purwanto}\ \emph {et~al.}(2009)\citenamefont
  {Purwanto}, \citenamefont {Krakauer},\ and\ \citenamefont
  {Zhang}}]{purwanto2009pressure}%
  \BibitemOpen
  \bibfield  {author} {\bibinfo {author} {\bibfnamefont {W.}~\bibnamefont
  {Purwanto}}, \bibinfo {author} {\bibfnamefont {H.}~\bibnamefont {Krakauer}},
  \ and\ \bibinfo {author} {\bibfnamefont {S.}~\bibnamefont {Zhang}},\ }\href
  {\doibase 10.1103/PhysRevB.80.214116} {\bibfield  {journal} {\bibinfo
  {journal} {Phys. Rev. B}\ }\textbf {\bibinfo {volume} {80}},\ \bibinfo
  {pages} {214116} (\bibinfo {year} {2009})}\BibitemShut {NoStop}%
\bibitem [{\citenamefont {Schmidt}\ \emph {et~al.}(1993)\citenamefont
  {Schmidt}, \citenamefont {Baldridge}, \citenamefont {Boatz}, \citenamefont
  {Elbert}, \citenamefont {Gordon}, \citenamefont {Jensen}, \citenamefont
  {Koseki}, \citenamefont {Matsunaga}, \citenamefont {Nguyen}, \citenamefont
  {Su}, \citenamefont {Windus}, \citenamefont {Dupuis},\ and\ \citenamefont
  {Montgomery~Jr}}]{schmidt1993general}%
  \BibitemOpen
  \bibfield  {author} {\bibinfo {author} {\bibfnamefont {M.~W.}\ \bibnamefont
  {Schmidt}}, \bibinfo {author} {\bibfnamefont {K.~K.}\ \bibnamefont
  {Baldridge}}, \bibinfo {author} {\bibfnamefont {J.~A.}\ \bibnamefont
  {Boatz}}, \bibinfo {author} {\bibfnamefont {S.~T.}\ \bibnamefont {Elbert}},
  \bibinfo {author} {\bibfnamefont {M.~S.}\ \bibnamefont {Gordon}}, \bibinfo
  {author} {\bibfnamefont {J.~H.}\ \bibnamefont {Jensen}}, \bibinfo {author}
  {\bibfnamefont {S.}~\bibnamefont {Koseki}}, \bibinfo {author} {\bibfnamefont
  {N.}~\bibnamefont {Matsunaga}}, \bibinfo {author} {\bibfnamefont {K.~A.}\
  \bibnamefont {Nguyen}}, \bibinfo {author} {\bibfnamefont {S.}~\bibnamefont
  {Su}}, \bibinfo {author} {\bibfnamefont {T.~L.}\ \bibnamefont {Windus}},
  \bibinfo {author} {\bibfnamefont {M.}~\bibnamefont {Dupuis}}, \ and\ \bibinfo
  {author} {\bibfnamefont {J.~A.}\ \bibnamefont {Montgomery~Jr}},\ }\href@noop
  {} {\bibfield  {journal} {\bibinfo  {journal} {J. Comput. Chem.}\ }\textbf
  {\bibinfo {volume} {14}},\ \bibinfo {pages} {1347} (\bibinfo {year}
  {1993})}\BibitemShut {NoStop}%
\bibitem [{\citenamefont {Dunning~Jr}(1989)}]{dunning1989gaussian}%
  \BibitemOpen
  \bibfield  {author} {\bibinfo {author} {\bibfnamefont {T.~H.}\ \bibnamefont
  {Dunning~Jr}},\ }\href@noop {} {\bibfield  {journal} {\bibinfo  {journal} {J.
  Chem. Phys.}\ }\textbf {\bibinfo {volume} {90}},\ \bibinfo {pages} {1007}
  (\bibinfo {year} {1989})}\BibitemShut {NoStop}%
\bibitem [{\citenamefont {Kendall}\ \emph {et~al.}(1992)\citenamefont
  {Kendall}, \citenamefont {Dunning~Jr},\ and\ \citenamefont
  {Harrison}}]{kendall1992electron}%
  \BibitemOpen
  \bibfield  {author} {\bibinfo {author} {\bibfnamefont {R.~A.}\ \bibnamefont
  {Kendall}}, \bibinfo {author} {\bibfnamefont {T.~H.}\ \bibnamefont
  {Dunning~Jr}}, \ and\ \bibinfo {author} {\bibfnamefont {R.~J.}\ \bibnamefont
  {Harrison}},\ }\href@noop {} {\bibfield  {journal} {\bibinfo  {journal} {J.
  Chem. Phys.}\ }\textbf {\bibinfo {volume} {96}},\ \bibinfo {pages} {6796}
  (\bibinfo {year} {1992})}\BibitemShut {NoStop}%
\bibitem [{\citenamefont {Johnson~III}()}]{johnsonnist}%
  \BibitemOpen
  \bibfield  {author} {\bibinfo {author} {\bibfnamefont {R.~J.}\ \bibnamefont
  {Johnson~III}},\ }\href@noop {} {\enquote {\bibinfo {title} {{NIST}
  {C}omputational {C}hemistry {C}omparison and {B}enchmark {D}atabase},}\
  }\bibinfo {howpublished} {http://cccbdb.nist.gov/},\ \bibinfo {note} {{NIST}
  {S}tandard {R}eference {D}atabase {N}umber 101, {R}elease 17b, {S}eptember
  2015}\BibitemShut {NoStop}%
\bibitem [{\citenamefont {Wang}\ \emph {et~al.}(2013)\citenamefont {Wang},
  \citenamefont {Zhang}, \citenamefont {Zhang},\ and\ \citenamefont
  {Yi}}]{Wang_OpenBLAS}%
  \BibitemOpen
  \bibfield  {author} {\bibinfo {author} {\bibfnamefont {Q.}~\bibnamefont
  {Wang}}, \bibinfo {author} {\bibfnamefont {X.}~\bibnamefont {Zhang}},
  \bibinfo {author} {\bibfnamefont {Y.}~\bibnamefont {Zhang}}, \ and\ \bibinfo
  {author} {\bibfnamefont {Q.}~\bibnamefont {Yi}},\ }in\ \href {\doibase
  10.1145/2503210.2503219} {\emph {\bibinfo {booktitle} {Proceedings of the
  International Conference on High Performance Computing, Networking, Storage
  and Analysis}}},\ \bibinfo {series and number} {SC '13}\ (\bibinfo
  {publisher} {ACM},\ \bibinfo {address} {New York, NY, USA},\ \bibinfo {year}
  {2013})\ pp.\ \bibinfo {pages} {25:1--25:12}\BibitemShut {NoStop}%
\bibitem [{\citenamefont {Xianyi}\ \emph {et~al.}(2012)\citenamefont {Xianyi},
  \citenamefont {Qian},\ and\ \citenamefont {Yunquan}}]{Xianyi_OpenBLAS}%
  \BibitemOpen
  \bibfield  {author} {\bibinfo {author} {\bibfnamefont {Z.}~\bibnamefont
  {Xianyi}}, \bibinfo {author} {\bibfnamefont {W.}~\bibnamefont {Qian}}, \ and\
  \bibinfo {author} {\bibfnamefont {Z.}~\bibnamefont {Yunquan}},\ }in\ \href
  {\doibase 10.1109/ICPADS.2012.97} {\emph {\bibinfo {booktitle} {2012 IEEE
  18th International Conference on Parallel and Distributed Systems}}}\
  (\bibinfo {year} {2012})\ pp.\ \bibinfo {pages} {684--691}\BibitemShut
  {NoStop}%
\bibitem [{\citenamefont {Sidje}(1998)}]{sidje1998expokit}%
  \BibitemOpen
  \bibfield  {author} {\bibinfo {author} {\bibfnamefont {R.~B.}\ \bibnamefont
  {Sidje}},\ }\href@noop {} {\bibfield  {journal} {\bibinfo  {journal} {ACM
  Transactions on Mathematical Software (TOMS)}\ }\textbf {\bibinfo {volume}
  {24}},\ \bibinfo {pages} {130} (\bibinfo {year} {1998})}\BibitemShut
  {NoStop}%
\bibitem [{\citenamefont {Mascagni}\ and\ \citenamefont
  {Srinivasan}(2000)}]{mascagni2000algorithm}%
  \BibitemOpen
  \bibfield  {author} {\bibinfo {author} {\bibfnamefont {M.}~\bibnamefont
  {Mascagni}}\ and\ \bibinfo {author} {\bibfnamefont {A.}~\bibnamefont
  {Srinivasan}},\ }\href@noop {} {\bibfield  {journal} {\bibinfo  {journal}
  {ACM Transactions on Mathematical Software (TOMS)}\ }\textbf {\bibinfo
  {volume} {26}},\ \bibinfo {pages} {436} (\bibinfo {year} {2000})}\BibitemShut
  {NoStop}%
\bibitem [{\citenamefont {Curtiss}\ \emph
  {et~al.}(1991{\natexlab{a}})\citenamefont {Curtiss}, \citenamefont
  {Raghavachari}, \citenamefont {Trucks},\ and\ \citenamefont
  {Pople}}]{curtiss1991gaussian}%
  \BibitemOpen
  \bibfield  {author} {\bibinfo {author} {\bibfnamefont {L.~A.}\ \bibnamefont
  {Curtiss}}, \bibinfo {author} {\bibfnamefont {K.}~\bibnamefont
  {Raghavachari}}, \bibinfo {author} {\bibfnamefont {G.~W.}\ \bibnamefont
  {Trucks}}, \ and\ \bibinfo {author} {\bibfnamefont {J.~A.}\ \bibnamefont
  {Pople}},\ }\href@noop {} {\bibfield  {journal} {\bibinfo  {journal} {J.
  Chem. Phys.}\ }\textbf {\bibinfo {volume} {94}},\ \bibinfo {pages} {7221}
  (\bibinfo {year} {1991}{\natexlab{a}})}\BibitemShut {NoStop}%
\bibitem [{\citenamefont {Becke}(1993)}]{becke1993density}%
  \BibitemOpen
  \bibfield  {author} {\bibinfo {author} {\bibfnamefont {A.~D.}\ \bibnamefont
  {Becke}},\ }\href@noop {} {\bibfield  {journal} {\bibinfo  {journal} {J.
  Chem. Phys.}\ }\textbf {\bibinfo {volume} {98}},\ \bibinfo {pages} {5648}
  (\bibinfo {year} {1993})}\BibitemShut {NoStop}%
\bibitem [{\citenamefont {Lee}\ \emph {et~al.}(1988)\citenamefont {Lee},
  \citenamefont {Yang},\ and\ \citenamefont {Parr}}]{lee1988development}%
  \BibitemOpen
  \bibfield  {author} {\bibinfo {author} {\bibfnamefont {C.}~\bibnamefont
  {Lee}}, \bibinfo {author} {\bibfnamefont {W.}~\bibnamefont {Yang}}, \ and\
  \bibinfo {author} {\bibfnamefont {R.~G.}\ \bibnamefont {Parr}},\ }\href
  {\doibase 10.1103/PhysRevB.37.785} {\bibfield  {journal} {\bibinfo  {journal}
  {Phys. Rev. B}\ }\textbf {\bibinfo {volume} {37}},\ \bibinfo {pages} {785}
  (\bibinfo {year} {1988})}\BibitemShut {NoStop}%
\bibitem [{\citenamefont {Tawa}\ \emph {et~al.}(1998)\citenamefont {Tawa},
  \citenamefont {Topol}, \citenamefont {Burt}, \citenamefont {Caldwell},\ and\
  \citenamefont {Rashin}}]{tawa1998calculation}%
  \BibitemOpen
  \bibfield  {author} {\bibinfo {author} {\bibfnamefont {G.}~\bibnamefont
  {Tawa}}, \bibinfo {author} {\bibfnamefont {I.}~\bibnamefont {Topol}},
  \bibinfo {author} {\bibfnamefont {S.}~\bibnamefont {Burt}}, \bibinfo {author}
  {\bibfnamefont {R.}~\bibnamefont {Caldwell}}, \ and\ \bibinfo {author}
  {\bibfnamefont {A.}~\bibnamefont {Rashin}},\ }\href@noop {} {\bibfield
  {journal} {\bibinfo  {journal} {J. Chem. Phys.}\ }\textbf {\bibinfo {volume}
  {109}},\ \bibinfo {pages} {4852} (\bibinfo {year} {1998})}\BibitemShut
  {NoStop}%
\bibitem [{\citenamefont {Merrill}\ and\ \citenamefont
  {Kass}(1996)}]{merrill1996calculated}%
  \BibitemOpen
  \bibfield  {author} {\bibinfo {author} {\bibfnamefont {G.~N.}\ \bibnamefont
  {Merrill}}\ and\ \bibinfo {author} {\bibfnamefont {S.~R.}\ \bibnamefont
  {Kass}},\ }\href@noop {} {\bibfield  {journal} {\bibinfo  {journal} {J. Phys.
  Chem.}\ }\textbf {\bibinfo {volume} {100}},\ \bibinfo {pages} {17465}
  (\bibinfo {year} {1996})}\BibitemShut {NoStop}%
\bibitem [{\citenamefont {Curtiss}\ \emph
  {et~al.}(1991{\natexlab{b}})\citenamefont {Curtiss}, \citenamefont {Kock},\
  and\ \citenamefont {Pople}}]{curtiss1991energies}%
  \BibitemOpen
  \bibfield  {author} {\bibinfo {author} {\bibfnamefont {L.~A.}\ \bibnamefont
  {Curtiss}}, \bibinfo {author} {\bibfnamefont {L.~D.}\ \bibnamefont {Kock}}, \
  and\ \bibinfo {author} {\bibfnamefont {J.~A.}\ \bibnamefont {Pople}},\
  }\href@noop {} {\bibfield  {journal} {\bibinfo  {journal} {J. Chem. Phys.}\
  }\textbf {\bibinfo {volume} {95}},\ \bibinfo {pages} {4040} (\bibinfo {year}
  {1991}{\natexlab{b}})}\BibitemShut {NoStop}%
\bibitem [{\citenamefont {Suewattana}\ \emph {et~al.}(2007)\citenamefont
  {Suewattana}, \citenamefont {Purwanto}, \citenamefont {Zhang}, \citenamefont
  {Krakauer},\ and\ \citenamefont {Walter}}]{suewattana2007phaseless}%
  \BibitemOpen
  \bibfield  {author} {\bibinfo {author} {\bibfnamefont {M.}~\bibnamefont
  {Suewattana}}, \bibinfo {author} {\bibfnamefont {W.}~\bibnamefont
  {Purwanto}}, \bibinfo {author} {\bibfnamefont {S.}~\bibnamefont {Zhang}},
  \bibinfo {author} {\bibfnamefont {H.}~\bibnamefont {Krakauer}}, \ and\
  \bibinfo {author} {\bibfnamefont {E.~J.}\ \bibnamefont {Walter}},\ }\href
  {\doibase 10.1103/PhysRevB.75.245123} {\bibfield  {journal} {\bibinfo
  {journal} {Phys. Rev. B}\ }\textbf {\bibinfo {volume} {75}},\ \bibinfo
  {pages} {245123} (\bibinfo {year} {2007})}\BibitemShut {NoStop}%
\bibitem [{\citenamefont {Chan}\ \emph {et~al.}(2012)\citenamefont {Chan},
  \citenamefont {Trevitt}, \citenamefont {Blanksby},\ and\ \citenamefont
  {Radom}}]{chan2012comment}%
  \BibitemOpen
  \bibfield  {author} {\bibinfo {author} {\bibfnamefont {B.}~\bibnamefont
  {Chan}}, \bibinfo {author} {\bibfnamefont {A.~J.}\ \bibnamefont {Trevitt}},
  \bibinfo {author} {\bibfnamefont {S.~J.}\ \bibnamefont {Blanksby}}, \ and\
  \bibinfo {author} {\bibfnamefont {L.}~\bibnamefont {Radom}},\ }\href@noop {}
  {\bibfield  {journal} {\bibinfo  {journal} {J. Phys. Chem. A}\ }\textbf
  {\bibinfo {volume} {116}},\ \bibinfo {pages} {9214} (\bibinfo {year}
  {2012})}\BibitemShut {NoStop}%
\bibitem [{\citenamefont {Li}\ and\ \citenamefont {Fan}(2002)}]{li2002b2f4}%
  \BibitemOpen
  \bibfield  {author} {\bibinfo {author} {\bibfnamefont {Z.-H.}\ \bibnamefont
  {Li}}\ and\ \bibinfo {author} {\bibfnamefont {K.-N.}\ \bibnamefont {Fan}},\
  }\href@noop {} {\bibfield  {journal} {\bibinfo  {journal} {J. Phys. Chem. A}\
  }\textbf {\bibinfo {volume} {106}},\ \bibinfo {pages} {6659} (\bibinfo {year}
  {2002})}\BibitemShut {NoStop}%
\bibitem [{\citenamefont {Coskun}\ \emph {et~al.}(2016)\citenamefont {Coskun},
  \citenamefont {Jerome},\ and\ \citenamefont
  {Friesner}}]{coskun2016evaluation}%
  \BibitemOpen
  \bibfield  {author} {\bibinfo {author} {\bibfnamefont {D.}~\bibnamefont
  {Coskun}}, \bibinfo {author} {\bibfnamefont {S.~V.}\ \bibnamefont {Jerome}},
  \ and\ \bibinfo {author} {\bibfnamefont {R.~A.}\ \bibnamefont {Friesner}},\
  }\href@noop {} {\bibfield  {journal} {\bibinfo  {journal} {J. Chem. Theory
  Comput.}\ }\textbf {\bibinfo {volume} {12}},\ \bibinfo {pages} {1121}
  (\bibinfo {year} {2016})}\BibitemShut {NoStop}%
\bibitem [{\citenamefont {Roy}\ \emph {et~al.}(2009)\citenamefont {Roy},
  \citenamefont {Jakubikova}, \citenamefont {Guthrie},\ and\ \citenamefont
  {Batista}}]{roy2009calculation}%
  \BibitemOpen
  \bibfield  {author} {\bibinfo {author} {\bibfnamefont {L.~E.}\ \bibnamefont
  {Roy}}, \bibinfo {author} {\bibfnamefont {E.}~\bibnamefont {Jakubikova}},
  \bibinfo {author} {\bibfnamefont {M.~G.}\ \bibnamefont {Guthrie}}, \ and\
  \bibinfo {author} {\bibfnamefont {E.~R.}\ \bibnamefont {Batista}},\
  }\href@noop {} {\bibfield  {journal} {\bibinfo  {journal} {J. Phys. Chem. A}\
  }\textbf {\bibinfo {volume} {113}},\ \bibinfo {pages} {6745} (\bibinfo {year}
  {2009})}\BibitemShut {NoStop}%
\bibitem [{\citenamefont {Konezny}\ \emph {et~al.}(2012)\citenamefont
  {Konezny}, \citenamefont {Doherty}, \citenamefont {Luca}, \citenamefont
  {Crabtree}, \citenamefont {Soloveichik},\ and\ \citenamefont
  {Batista}}]{konezny2012reduction}%
  \BibitemOpen
  \bibfield  {author} {\bibinfo {author} {\bibfnamefont {S.~J.}\ \bibnamefont
  {Konezny}}, \bibinfo {author} {\bibfnamefont {M.~D.}\ \bibnamefont
  {Doherty}}, \bibinfo {author} {\bibfnamefont {O.~R.}\ \bibnamefont {Luca}},
  \bibinfo {author} {\bibfnamefont {R.~H.}\ \bibnamefont {Crabtree}}, \bibinfo
  {author} {\bibfnamefont {G.~L.}\ \bibnamefont {Soloveichik}}, \ and\ \bibinfo
  {author} {\bibfnamefont {V.~S.}\ \bibnamefont {Batista}},\ }\href@noop {}
  {\bibfield  {journal} {\bibinfo  {journal} {J. Phys. Chem. C}\ }\textbf
  {\bibinfo {volume} {116}},\ \bibinfo {pages} {6349} (\bibinfo {year}
  {2012})}\BibitemShut {NoStop}%
\bibitem [{\citenamefont {Baik}\ and\ \citenamefont
  {Friesner}(2002)}]{baik2002computing}%
  \BibitemOpen
  \bibfield  {author} {\bibinfo {author} {\bibfnamefont {M.-H.}\ \bibnamefont
  {Baik}}\ and\ \bibinfo {author} {\bibfnamefont {R.~A.}\ \bibnamefont
  {Friesner}},\ }\href@noop {} {\bibfield  {journal} {\bibinfo  {journal} {J.
  Phys. Chem. A}\ }\textbf {\bibinfo {volume} {106}},\ \bibinfo {pages} {7407}
  (\bibinfo {year} {2002})}\BibitemShut {NoStop}%
\bibitem [{\citenamefont {Borodin}\ \emph {et~al.}(2013)\citenamefont
  {Borodin}, \citenamefont {Behl},\ and\ \citenamefont
  {Jow}}]{borodin2013oxidative}%
  \BibitemOpen
  \bibfield  {author} {\bibinfo {author} {\bibfnamefont {O.}~\bibnamefont
  {Borodin}}, \bibinfo {author} {\bibfnamefont {W.}~\bibnamefont {Behl}}, \
  and\ \bibinfo {author} {\bibfnamefont {T.~R.}\ \bibnamefont {Jow}},\
  }\href@noop {} {\bibfield  {journal} {\bibinfo  {journal} {J. Phys. Chem. C}\
  }\textbf {\bibinfo {volume} {117}},\ \bibinfo {pages} {8661} (\bibinfo {year}
  {2013})}\BibitemShut {NoStop}%
\bibitem [{\citenamefont {Shao}\ \emph {et~al.}(2012)\citenamefont {Shao},
  \citenamefont {Sun}, \citenamefont {Dai},\ and\ \citenamefont
  {Jiang}}]{shao2012oxidation}%
  \BibitemOpen
  \bibfield  {author} {\bibinfo {author} {\bibfnamefont {N.}~\bibnamefont
  {Shao}}, \bibinfo {author} {\bibfnamefont {X.-G.}\ \bibnamefont {Sun}},
  \bibinfo {author} {\bibfnamefont {S.}~\bibnamefont {Dai}}, \ and\ \bibinfo
  {author} {\bibfnamefont {D.-e.}\ \bibnamefont {Jiang}},\ }\href@noop {}
  {\bibfield  {journal} {\bibinfo  {journal} {J. Phys. Chem. B}\ }\textbf
  {\bibinfo {volume} {116}},\ \bibinfo {pages} {3235} (\bibinfo {year}
  {2012})}\BibitemShut {NoStop}%
\bibitem [{\citenamefont {Qu}\ \emph {et~al.}(2015)\citenamefont {Qu},
  \citenamefont {Jain}, \citenamefont {Rajput}, \citenamefont {Cheng},
  \citenamefont {Zhang}, \citenamefont {Ong}, \citenamefont {Brafman},
  \citenamefont {Maginn}, \citenamefont {Curtiss},\ and\ \citenamefont
  {Persson}}]{qu2015electrolyte}%
  \BibitemOpen
  \bibfield  {author} {\bibinfo {author} {\bibfnamefont {X.}~\bibnamefont
  {Qu}}, \bibinfo {author} {\bibfnamefont {A.}~\bibnamefont {Jain}}, \bibinfo
  {author} {\bibfnamefont {N.~N.}\ \bibnamefont {Rajput}}, \bibinfo {author}
  {\bibfnamefont {L.}~\bibnamefont {Cheng}}, \bibinfo {author} {\bibfnamefont
  {Y.}~\bibnamefont {Zhang}}, \bibinfo {author} {\bibfnamefont {S.~P.}\
  \bibnamefont {Ong}}, \bibinfo {author} {\bibfnamefont {M.}~\bibnamefont
  {Brafman}}, \bibinfo {author} {\bibfnamefont {E.}~\bibnamefont {Maginn}},
  \bibinfo {author} {\bibfnamefont {L.~A.}\ \bibnamefont {Curtiss}}, \ and\
  \bibinfo {author} {\bibfnamefont {K.~A.}\ \bibnamefont {Persson}},\
  }\href@noop {} {\bibfield  {journal} {\bibinfo  {journal} {Comput. Mater.
  Sci.}\ }\textbf {\bibinfo {volume} {103}},\ \bibinfo {pages} {56} (\bibinfo
  {year} {2015})}\BibitemShut {NoStop}%
\end{thebibliography}%

\end{document}